\documentclass[final,3p,times]{elsarticle}
\usepackage{dcolumn}
\usepackage{bm}
\usepackage{caption}
\usepackage{subcaption}
\usepackage{textcomp}
\usepackage{gensymb}
\usepackage{pifont}
\usepackage{amsmath,amssymb,amsfonts}
\usepackage{graphicx}
\usepackage{color}
\usepackage{longtable}
\usepackage{xcolor}
\usepackage{colortbl}
\usepackage{longtable}
\definecolor{rev1}{rgb}{0,0,1}
\usepackage{comment}
\usepackage{booktabs}
\usepackage[hidelinks,colorlinks = true,urlcolor=blue,linkcolor = blue,citecolor = blue]{hyperref}
\usepackage{lipsum}
\usepackage{tabularx}
\usepackage{stackengine}
\usepackage[switch]{lineno} 
\usepackage{makecell}
\usepackage{float}
\journal{Elsevier}

\let\oldalign\align
\let\oldendalign\endalign

\renewenvironment{align}
  {\linenomathNonumbers\oldalign}
  {\oldendalign\endlinenomath}

\begin{document}
%\linenumbers

\begin{frontmatter}
    \title{Enhancing wind field resolution in complex terrain through a knowledge-driven machine learning approach}

\author[label1]{Jacob Wulff Wold}
\author[label2]{Florian Stadtmann}
\author[label2,label3]{Adil Rasheed\corref{mycorrespondingauthor}}
\cortext[mycorrespondingauthor]{Corresponding author}
\ead{adil.rasheed@ntnu.no}
\author[label3]{Mandar Tabib}
\author[label4]{Omer San}
\author[label5]{Jan-Tore Horn}

\affiliation[label1]{
    organization={Department of Physics, NTNU},
            addressline={Høgskoleringen 5}, 
            city={Trondheim},
            postcode={7034},
            country={Norway}
            }
   
    \affiliation[label2]{
    organization={Department of Engineering Cybernetics, NTNU},
            addressline={O.S.Bragstads plass 2}, 
            city={Trondheim},
            postcode={7034},
            country={Norway}
            }
    \affiliation[label3]{
    organization={Mathematics and Cybernetics, SINTEF Digital},
            addressline={Klæbuveien 153}, 
            city={Trondheim},
            postcode={7031},
            country={Norway}
            }
    
    \affiliation[label4]{
    organization={Department of Mechanical, Aerospace and Biomedical Engineering, UTK},
            addressline={1512 Middle Dr}, 
            city={Knoxville, Tennessee},
            postcode={37996},
            country={USA}
            }

    \affiliation[label5]{
    organization={Vind Technologies AS},
            addressline={Grundingen 2}, 
            city={Oslo},
            postcode={0250},
            country={Norway}
            }
            
    \begin{abstract}
 
Wind energy, an essential component in the fight against climate change, relies heavily on precise, detailed wind field simulations to optimize wind farms, particularly in complex terrain with intricate wind patterns. However, conventional high-resolution simulations come with a hefty computational cost, limiting their applicability in real-time decision-making. This research addresses this challenge by proposing a machine learning-driven, computationally efficient approach utilizing a modified Generative Adversarial Network.
The contribution of this work is threefold. Firstly, we provide access to a unique high-resolution dataset of wind fields in complex terrain. Secondly, we introduce a knowledge-based modification to the loss function, ensuring that the algorithm captures crucial characteristics of the flow within complex terrains. Finally, we demonstrate the potential of our approach to enhance wind flow resolution in real-life wind farms. Through this, our method delivers comparable accuracy to high-resolution simulations while substantially reducing computational demands. This advancement greatly enhances the accessibility and efficiency of high-resolution wind field simulations, facilitating real-time optimization of wind farms. Moreover, we illustrate that by designing an appropriate loss function informed by domain knowledge, we can mitigate the need for adversarial training.
    \end{abstract}
\end{frontmatter}

\section{Introduction}
\label{sec:intro}
Wind energy, a pivotal technology for addressing climate change, is poised to play a significant role in the global shift towards sustainable energy sources. With the increasing demand for clean energy and the urgent need to reduce carbon emissions, wind power has emerged as a cornerstone in our pursuit of a more sustainable future. Projections suggest that by 2050, wind energy could contribute to more than 50\% of total power production, marking a remarkable milestone in our journey toward a low carbon energy landscape \cite{Gibon2022cni, 2018ida}. However, this growth is not without its challenges. The wind energy industry is experiencing substantial expansion, but faces hurdles in achieving and maintaining ambitious goals \cite{Lee2021gwr}. Meeting these goals requires not only increasing the number of wind farms, but also enhancing their efficiency and reducing operational and maintenance costs. To address these challenges, there has been an increasing focus on digitalization, with digital twins \cite{Rasheed2020dtv} as a vital enabler. In a recent survey \cite{Stadtmann2023dti} conducted in collaboration with key industry partners active in the digitalization of the wind energy sector, a common understanding of the concept of digital twins was developed, categorizing digital twins on a scale of 0 to 5, which ranged from standalone (0), descriptive (1), diagnostic (2), predictive (3), prescriptive (4) to autonomous (5). Various challenges facing the development of digital twin technologies were also identified and categorized into data-related, model-related, and industrial acceptance-related issues. Data-related challenges include issues such as the lack of high-quality data, intellectual property rights, data security, and privacy. On the other hand, model-related challenges pertain to the need for models that are accurate, computationally efficient, generalizable, and capable of self-evolution. 

In the context of our current research, our primary focus is addressing one each of the data- and model-related issues by providing open-source data and a computational approach for efficiently modeling high-resolution wind fields in complex terrain. Wind flows in complex terrain are dominated by distinct features, including rotors, hydraulic jumps, and substantial spatiotemporal variations. Achieving an accurate and computationally efficient modeling of these phenomena is of utmost importance for optimizing wind farm layouts. Accurate predictions of power production are also crucial for energy producers, as they are accountable for any deviations from the predicted power output \cite{Skajaa2015ito}. Furthermore, high-resolution wind fields serve as inputs for wake models, allowing for the precise prediction of how one turbine's wake affects others. This capability empowers operators to make real-time adjustments, such as modifying the yaw misalignment of individual turbines, to maximize power production and minimize stress on the turbines~\cite{Knudsen2015sow, VanDijk2017wfm}. A good understanding of load distribution on wind turbine components is critical for assessing structural integrity and ensuring the safety of both the turbines and their surrounding environment. Precise load estimations facilitate proactive maintenance and repair scheduling, thus reducing costly downtime \cite{Stanley2020wfl} and extending the lifespan of the turbines~\cite{Ernst2012ios}.

Although the importance of high-resolution wind field simulations is evident, they are not without their challenges, especially in complex terrains. The computational demands of high-resolution simulations can be prohibitive, limiting their practicality for real-time applications. We have been operating a terrain-induced turbulence alert system for 19 Norwegian airports (www.ippc.no) at a horizontal resolution of approximately $400m\times400m$ since 2010 on the national supercomputing infrastructure. However, due to computational constraints, we could only afford steady-state simulations for each hour for the next 12 hours. More details of the model and forecast system can be found in (\url{https://www.sesarju.eu/sites/default/files/sids2013_submission_23.pdf}) and \cite{Midjiyawa2023ncf}. The sheer number of grid elements required to represent complex terrains accurately, along with the need for substantial computational resources, can result in simulations that are unfeasible to perform instantaneously~\cite{Rasheed2017wfm}. This poses a significant challenge for operational wind farms, where real-time decision-making is essential for efficient energy generation and load management.

To address the challenges associated with high-resolution wind field simulations, machine learning approaches have gained prominence. Machine learning techniques, particularly Deep Neural Networks (DNNs) and Generative Adversarial Networks (GANs), have demonstrated promise in super-resolving wind fields from lower resolutions to higher resolutions. These approaches hold the potential to accelerate wind field simulations and make them more accessible for practical applications. In a comprehensive survey~\cite{Fukami2023sra}, various super-resolution neural networks for fluid flows, including wind fields, are reviewed. Some of these networks incorporate physics-informed losses during training, aligning them more closely with the underlying physical principles governing wind behavior. Likewise, Convolutional Neural Networks (CNNs) have been used to super-resolve 3D wind fields, effectively increasing their resolution from lower resolutions to higher resolutions. Such models have demonstrated success in capturing detailed wind behavior at various heights within the atmosphere~\cite{Hohlein2020acs}. The authors employed CNNs to super-resolve wind fields from $31km\times31km$ resolution to $9km\times9km$ resolution at a height of 100 meters. In another study~\cite{Stengel2020asr} the authors applied a two-stage GANs approach to super-resolve wind fields from $100km\times100km$ resolution to $10km\times10km$ and further to $2km\times2km$ resolution at a height of 100 meters. GANs employ a generative approach where they create high-resolution representations of wind fields from lower-resolution inputs~\cite{Xie2018tat, Werhahn2019amp}. An enhanced super-resolution GAN structure with an RGB-image-based perceptual loss component was used to reconstruct simulated high-resolution wind fields over real terrain in two dimensions~\cite{Tran2020ges}. Nonetheless, it's worth noting that the model setup proposed in the study was found to be inadequate for complex terrain~\cite{Larsen2020ota}. The results were reproducible only for smooth airflow conditions, but when dealing with complex flow, the model consistently failed to replicate the ground truth. The instability in the model was attributed to the perceptual loss. Additionally, there continues to be a scarcity of open-source, high-quality 3D wind flow data in complex terrain, which is crucial for facilitating more realistic investigations. To this end, the objective of the present work is to: 

\begin{itemize}
    \item Simulate and disseminate comprehensive high-resolution wind field data for complex terrains to a wider community of wind modeling experts.
    \item Develop a data-driven approach that can achieve the same level of accuracy as a high-resolution simulation, but with a significantly reduced computational cost by running the simulation at a much coarser resolution. 
    \item Explore techniques for improving the accuracy of the data-driven model by incorporating physics-based loss functions to enhance the super-resolution of 3D wind fields within realistic terrains.
\end{itemize}

In order to achieve the objectives, the ESRGAN model has been modified to ensure reliable results in reconstructing simulated three-dimensional high-resolution wind fields over complex terrain. The ESRGAN structure has been adjusted to handle the reconstruction of 3D vector fields, and additional information on pressure, height, and terrain has been incorporated into the early and intermediate layers of the original network. Additionally, the previous perceptual loss has been replaced with several physically motivated loss components. Furthermore, new hyperparameters have been introduced and optimized. 

This article is structured as follows. Section~\ref{sec:theory} presents the essential theory for the meso and microscale physics-based models used for data generation and for the Machine Learning (ML) models used in this work. Section~\ref{sec:methodandsetup} provides information on the data and its generation and preprocessing. It also presents the modified GAN architecture, hyperparameters, and loss function. Furthermore, the methodology of three experiments is presented throughout which the model setup is optimized for the task of super-resolving the wind field around the wind farm. Section~\ref{sec:resultsanddiscussions} analyzes the results obtained from the experiments, presents and evaluates the optimal model setup, and discusses the results in more detail. Finally, Section~\ref{sec:conclusionandfuturework} concludes the work and offers recommendations for areas of future research.

\section{Theory}
\label{sec:theory}
This section provides an overview of the theory necessary for simulating high-resolution wind fields in complex terrain, developing generative adversarial models, and establishing performance metrics for a quantitative evaluation of the proposed enhanced wind field resolution approach.

    \subsection{Meso- and microscale simulation model}
    \label{subsec:microscale}
    Atmospheric flows are governed by mass, momentum, and energy conservation principles given by Equation~\ref{eqn:mass}, Equation~\ref{eqn:momentum}, and Equation~\ref{eqn:energy} respectively.  
    \begin{equation}
    \nabla\cdot(\rho_{s} \mathbf u)=0
    \label{eqn:mass}
    \end{equation}
    \begin{equation}
    \frac{D \mathbf u}{Dt}=-\nabla \left(\frac{p_{d}}{\rho_{s}} \right)+\mathbf{g}\frac{\theta_{d}}{\theta_{s}}+\frac{1}{\rho_{s}}\nabla\cdot\mathbf R + \mathbf f	
    \label{eqn:momentum}
    \end{equation}
    \begin{equation}
    \frac{D \theta}{Dt}=\nabla\cdot(\gamma_{T} \nabla \theta)+q
    \label{eqn:energy}
    \end{equation}
    Here $\textbf{u},\rho, p,\theta, \textbf{R}, \textbf{f}$ represent velocity, density, pressure, potential temperature, stress tensor, and sink/source term (eg. Coriolis force) respectively. Furthermore, $\textbf{g}$, $\gamma_{T}$, and $q$ denote acceleration due to gravity, thermal diffusivity, and temperature source term. $\gamma_{T}$  can be used to model radiative heating of the atmosphere in a mesoscale modeling context. As for the subscripts, $s$ signifies hydrostatic values, while subscript $d$ indicates the deviation from this value. In mathematical terms this equals to $p=p_{s}+p_{d}$, $\theta=\theta_{s}+\theta_{d}$, $\rho=\rho_{s}+\rho_{d}$ where the hydrostatic relation is given by $\partial p_{s}/\partial z=-g\rho_{s}$ and $\rho_{s}=p_{s}/R\theta(p_{o}/p_{s})^{R_{g}/C_{p}}$ with $C_{p}$ representing the specific heat at constant pressure while $R_{g}$ is the gas constant. Again from \cite{Rasheed2017wfm}, $\textbf{R},P_{k},G_{\theta}$ are given by Equation~\ref{eqn:rst} and Equation~\ref{eqn:prod}.
    \begin{equation}
    R_{ij}=\nu_{T} \left( \frac{\partial u_{i}}{\partial x_{j}}+ \frac{\partial u_{j}}{\partial x_{i}} \right) - \frac{2}{3}k \delta_{ij}
    \label{eqn:rst}
    \end{equation}
    \begin{equation}
    P_{k}=\nu_{T}  \left( \frac{\partial u_{i}}{\partial x_{j}}+ \frac{\partial u_{j}}{\partial x_{i}} \right) \frac{\partial u_{i}}{\partial x_{j}},\quad G_{\theta} = -\frac{g}{\theta}\frac{\nu_{T}}{\sigma_{T}}\frac{\partial \theta}{\partial z}
    \label{eqn:prod}
    \end{equation}
    \begin{equation}
    \nu_{T}=C_{\mu}\frac{k^{2}}{\epsilon}
    \label{eqn:nuT}
    \end{equation}
    The turbulent viscosity $\nu_{T}$ given by Equation~\ref{eqn:nuT} is computed from the turbulent kinetic energy ($k$) and dissipation ($\epsilon$) given by Equation~\ref{eqn:tke} and Equation~\ref{eqn:epsilon}.  
    \begin{equation}
    \frac{D k}{Dt}=\nabla\cdot(\nu_{T} \nabla k)+P_{k}+G_{\theta}-\epsilon
    \label{eqn:tke}
    \end{equation}
    \begin{equation}
    \frac{D \epsilon}{Dt}=\nabla\cdot\left( \frac{\nu_{T}}{\sigma_{e}} \nabla \epsilon \right) +(C_{1}P_{k}+C_{3}G_{\theta})\frac{\epsilon}{k}-C_{2}\frac{\epsilon^{2}}{k}
    \label{eqn:epsilon}
    \end{equation}
    
    In the current work, two different models operating at different spatial resolutions are coupled together to make this realistic wind flow modeling computationally tractable. The large-scale model is called HARMONIE (Hirlam Aladin Regional Mesoscale Operational
    Numerical prediction in Europe) and is used as a weather forecast model in Norway. The wind field available from this model is at a horizontal resolution of $2.5 \, \mbox{km} \times 2.5 \, \mbox{km}$. The resolution of the wind field is improved to $200  \, \mbox{m} \times 200  \, \mbox{m}$ using a microscale model called SIMRA (Semi Implicit Method for Reynolds Averaged Navier Stokes Equations)\cite{Midjiyawa2023ncf,Rasheed2017wfm}. Both these models are essentially based on the equations presented above. One major difference between the two models is in the way turbulence is modeled. In SIMRA a two-equation turbulence model (one for turbulent kinetic energy, i.e. Equation~\ref{eqn:tke} and another for dissipation i.e. Equation~\ref{eqn:epsilon}) is used, while in HARMONIE, a one equation model given by Equation~\ref{eqn:tke} is employed. Further,  the turbulent dissipation is estimated from $\epsilon=(C_{\mu}^{1/2}K)^{3/2}/\ell_{t}$, with $\ell_{t}$ computed by applying the relationship
    \begin{equation}
    \ell_{t}\approx \frac{\min(\kappa z, 200 \, \mbox{m})}{1+5Ri}
    \end{equation}
    where
    \begin{equation}
    Ri=\frac{(g/\theta)\partial \theta / \partial z}{(\partial u / \partial
        z)^{2}}\approx -\frac{G}{P}
    \end{equation}
    The stability correction $(1+5Ri)$ is replaced by $(1-40Ri)^{-1/3}$ in convective conditions and the gradient Richardson number $Ri$ is expected to be less than~$1/4$. At last, the coefficients are $(C_{\mu}, C_{1}, C_{2}, C_{3}) = (0.09, 1.92, 1.43, 1)$ and the coefficients ($\kappa, \sigma_{K}, \sigma_{\epsilon}$) are ($0.4, 1, 1.3$), respectively.~\cite{Gatski1996sam}. For a more detailed description of the multiscale model discussed above, the reader is referred to the work of Zakari et al.~\cite{Midjiyawa2023ncf}.

    \subsection{Generative adversarial networks} % TODO directly introduce the esrgan here
    \label{subsec:GAN}
    In this work, the enhanced super-resolution generative adversarial network (ESRGAN) is used as the basic model. it builds on the generative adversarial network (GAN), which is an unsupervised generative model. A GAN model consists of a generator network and a discriminator network.
    The generator creates fake samples $\mathbf{f}=\mathbf{G}(\mathbf{z})$ from an input $\mathbf{z}$ which the discriminator should distinguish from real samples $\mathbf{r}$.
    The discriminator outputs a number between 1 and 0 representing the certainty of the input being a real sample (1.0) or a false sample (0.0). The generator is rewarded for creating images that the discriminator classifies as real, resulting in adversarial training. The loss functions of the discriminator and generator are defined as:
    \begin{equation} \label{eq:std_D_loss}
    L_D^{GAN} = \log\left(\mathbf{D}(\mathbf{r})\right) - \log\left(1-\mathbf{D}(\mathbf{G(\mathbf{z})})\right)
    \end{equation}
    \begin{equation} \label{eq:std_G_loss}
        L_G^{GAN} = \log\left(1-\mathbf{D}(\mathbf{G(\mathbf{z})})\right).
    \end{equation}
    The generator is penalized for the discriminator labeling its generated data as false, while the discriminator is penalized both for not correctly identifying true samples and for being fooled by the generator.

    \subsubsection{Relativistic Average GANs}
    Since its introduction in 2014, multiple adjustments to the original GAN model~\cite{Goodfellow2014gan} have been introduced. The relativistic average GAN (RaGAN) model\cite{Jolicoeur_Martineau2019trd} was introduced by Jolicoeur-Martineau in 2018:
    \begin{equation} \label{eq:RaGAN_D_loss}
        L_D^{RaGAN} = -\Bigl \langle \log\left(\mathbf{\Bar{D}}(\mathbf{r})\right)\Bigr \rangle - \Bigl \langle \log\left(1-\mathbf{\Bar{D}}(\mathbf{f})\right)\Bigr \rangle
    \end{equation}
    \begin{equation} \label{eq:RaGAN_G_loss}
        L_G^{RaGAN} = -\Bigl \langle\log\left(\mathbf{\Bar{D}}(\mathbf{f})\right)\Bigr \rangle - \Bigl \langle\log\left(1-\mathbf{\Bar{D}}(\mathbf{r})\right)\Bigr \rangle ,
    \end{equation}
    with
    \begin{equation}
    \begin{split}
        \mathbf{\Bar{D}}(\mathbf{r}) &= \sigma \Bigl(C\left(\mathbf{r}\right)- \Bigl \langle C(\mathbf{f})\Bigl \rangle \Bigr) \\
        \mathbf{\Bar{D}}(\mathbf{f}) &= \sigma \Bigl(C\left(\mathbf{f}\right)- \Bigl \langle C(\mathbf{r})\Bigl \rangle \Bigr).
    \end{split}
    \end{equation}
    Here $\sigma$ is the activation function of the discriminator, $C(\mathbf{x})$ is the discriminator output before the activation, $\mathbf{D}(\mathbf{x}) = \sigma \bigl( C(\mathbf{x}) \bigr)$, and the $\langle \rangle$ represents expected value, in practice meaning the batch average value. Instead of punishing the discriminator for absolute classification error, it is rewarded for its classification of true samples \textit{relative} to how it classifies false samples. Similarly for the generator. This exploits the knowledge that half of the samples presented to the discriminator are fake. Whereas in the original GAN, a perfectly calibrated model would classify all samples as real, a perfectly calibrated RaGAN would classify half of all samples as fake, and half as real. Both the generator and the discriminator are pushed to specifically target the recurring differences between the real and generated data, rather than just the result of each single sample.

    \subsubsection{Single Image Super-Resolution GANs} \label{sec:SISRGAN}
    A Single Image Super-Resolution GAN (SISRGAN) is a GAN network trained to approximate a high-resolution image $\mathbf{I}_\mathrm{HR}$ from a single low-resolution image $\mathbf{I}_\mathrm{LR}$ by generating a \textit{super-resolved} image $\mathbf{I}_\mathrm{SR}$. SISRGANs using deep residual networks have proven very effective at super-resolving images. Of the most effective and well-known of these architectures is the Enhanced Super-Resolution GAN (ESRGAN), introduced by Wang et. al\cite{Wang2019ees}. The ``enhance'' part refers to ESRGAN's inspiration from Super-Resolution GAN (SRGAN) \cite{Ledig2017prs}. In SRGAN, the generator consists of a number of residual dense blocks (RDBs) of Conv-\href{https://pytorch.org/docs/stable/generated/torch.nn.BatchNorm2d.html}{BatchNorm}-\href{https://pytorch.org/docs/stable/generated/torch.nn.PReLU.html}{PRelu}-Conv-BatchNorm, and the same discriminator. SRGAN combines the adversarial loss (slightly modified from Equation~\ref{eq:std_D_loss}-\ref{eq:std_G_loss}) with a perceptual loss and a pixel loss:
    \begin{equation}
        L_G^{pix} = MSE(I_\mathrm{HR},\mathbf{G}(I_\mathrm{LR}))
    \end{equation}
    \begin{equation}
        L_G^{perceptual_{ij}} = MSE(\phi_{i,j}(I_\mathrm{HR}),\phi_{i,j}(\mathbf{G}(I_\mathrm{LR}))
    \end{equation}
    \begin{equation} \label{eq:SISRGAN_L_G}
        L_G = \eta_1 L_G^{pix} + \eta_2 L_G^{perceptual_{ij}} - \eta_3\log\left(\mathbf{D}(\mathbf{G(\mathbf{z})})\right)
    \end{equation}
    \begin{equation} \label{eq:SISRGAN_L_D}
        L_D = L_D^{GAN}
    \end{equation}
    
    Here, $MSE$ is the mean square error, $\eta_1, \eta_2, \eta_3$ are constants and $\phi_{i,j}$ is the feature map obtained by the $j$-th convolution after activation, before the i-th max-pooling layer within the pre-trained VGG19 \cite{Simonyan2015vdc} network. The perceptual loss, therefore, penalizes the generator according to how differently the generated SR images and the HR images activate features in a network \textit{trained to detect important perceptual features}. The SRGAN generator is first trained on only $L_G^{pix}$. Then this pre-trained model is trained further using $L_G$ with, after some testing, $\eta_1, \eta_2, \eta_3=$  0.0, 0.06 (effectively as a result of feature scaling), 0.001 and $\phi_{i,j}=\phi_{5,4}$.
    
    In ESRGAN the perceptual loss is modified to use the feature output from the pre-trained image classifier before activation instead of after, thereby gaining more granular feature discrepancies. ESRGAN uses RaGAN losses instead of traditional adversarial loss, and $\eta_1, \eta_2, \eta_3=0.01, 1.0, 0.005$. ESRGAN also uses mean square error for pretraining the generator, but average absolute error instead of MSE for $L_G^{pix}$ when used in \eqref{eq:SISRGAN_L_G}. Having the pre-trained MSE generator and the trained perceptual GAN generator, they use weight interpolation between the two generators, allowing the choice of how much one wants to weigh pure content error versus perceptual features. Finally, they remove batch normalization from the generator, seeing that this further increased performance. A more detailed description of the ESRGAN architecture on which the current work is based can be found in the work by Tran et al.\cite{Tran2020ges}.

    \subsubsection{Loss modification in this work}
    \label{subsec:loss}
    
    In this work, the generator loss function $L_G$ is modified from the default setup and consists of several components, which are explained here. The first component is the pixel loss, which is calculated between each directional component of the vector in a certain position. Let $V^{HR}$ and $V^{SR}$ be the original high-resolution vector field and the reconstructed super-resolution vector field with horizontal components $V_1, V_2$ and vertical component $V_3$. The pixel loss can then be written as a matrix
    \begin{equation}
        L_G^{pix} = \frac{1}{3} \sum_{i}^{3} \left| V_{i}^{HR} - V_{i}^{SR} \right|
    \end{equation}
    with one value per vector in the (discrete) vector field.
    
    In Section~\ref{subsec:microscale} it is shown that not only the vectors themselves but also the local partial derivatives are of importance. Therefore, a gradient-like loss matrix is calculated in the horizontal and vertical directions with
    \begin{align}
        L_G^{grad_{x_1x_2}} ={}& \frac{1}{6}
        %\min_{k\in\{1,2\}} \left(\frac{1}{N_k^{grad}} \right) 
        \frac{1}{N_{x_1x_2}^{grad}}
        \sum_i^2 \sum_j^3 \left( \frac{\partial }{\partial x_i} V_j^{HR} - \frac{\partial }{\partial x_i} V_j^{SR} \right)^2 \\
        L_G^{grad_{x_3}} ={}& \frac{1}{3N_{x3}^{grad}} \sum_i^3 \left( \frac{\partial }{\partial x_3} V_i^{HR} - \frac{\partial }{\partial x_3} V_i^{SR} \right)^2
    \end{align}
    where $\frac{\partial }{\partial x_1}$ and $\frac{\partial }{\partial x_2}$ are the derivatives in horizontal direction and $\frac{\partial }{\partial x_3}$ the derivative in vertical direction. At the boundaries, the one-sided derivatives are used.
    $N^{grad_{x_1 x_2}}$ and $N^{grad_{x_3}}$ are normalization factors that are calculated from the maximum value of all partial high-resolution and super-resolution derivatives $\frac{\partial}{\partial x} V$  involved in each equation through
    %\begin{align}
    %    N_k^{grad} ={}& \max \Biggl( \max_{j\in\{1,2,3\}} \left( \left| \frac{\partial}{\partial x_k} V_j^{HR} \right| \right), \frac{1}{100} \max_{j\in\{1,2,3\}} \left( \left| \frac{\partial}{\partial x_k} V_j^{SR} \right| \right) \Biggr)
    %\end{align}
    \begin{align}
        N_{x_1x_2}^{grad} ={}& \max \Biggl(
        \max_{j\in\{1,2\},k\in\{1,2,3\}} \left( \left| \frac{\partial}{\partial x_k} V_j^{HR} \right| \right),
        \frac{1}{100} \max_{j\in\{1,2,3\},k\in\{1,2\}} \left( \left| \frac{\partial}{\partial x_k} V_j^{SR} \right| \right) 
        \Biggr) \\
        N_{x_3}^{grad} ={}& \max \Biggl(
        \max_{k\in\{1,2,3\}} \left( \frac{\partial}{\partial x_k} V_3^{HR} \right),
        \frac{1}{100} \max_{k\in\{1,2\}} \left( \frac{\partial}{\partial x_k} V_3^{SR}\right) 
        \Biggr)
    \end{align}
    
    Similarly to the gradient, the divergence in two and three dimensions is included in the loss through
    \begin{align}
        L_G^{div} ={}&  
        %\min_{k\in\{1,2,3\}} \left( \frac{1}{N_k^{div}} \right) 
        \frac{1}{N^{div}}
        \left( \sum_i^3 \frac{\partial }{\partial x_i} V_i^{HR}  - \frac{\partial }{\partial x_i} V_i^{SR} \right)^2 \\
        L_G^{div_{x_1 x_2}} ={}&  
        %\min_{i\in\{1,2\}} \left(\frac{1}{N_i^{div}} \right) 
        \frac{1}{N_{x1x2}^{div}}
        \left( \sum_i^2 \frac{\partial }{\partial x_i} V_i^{HR}  - \frac{\partial }{\partial x_i} V_i^{SR} \right)^2
    \end{align}
    Again, the normalization factor is computed over all involved derivatives
    %\begin{align}
    %    N_k^{div} ={}& \max \Biggl( \max \left( \left| \frac{\partial}{\partial x_k} V_k^{HR} \right| \right), 
    %    \frac{1}{100} \max \left( \left| \frac{\partial}{\partial x_k} V_k^{SR} \right| \right) \Biggr)
    %\end{align}
    \begin{align}
        N^{div} ={}& \max \Biggl(
        \max \left( \left| \sum_{k=1}^3 \frac{\partial}{\partial x_k} V_k^{HR} \right| \right), 
        \frac{1}{100} \max \left( \left| \sum_{k=1}^3 \frac{\partial}{\partial x_k} V_k^{SR} \right| \right) \Biggr)\\
        N_{x_1x_2}^{div} ={}& \max \Biggl( \max \left( \left| \sum_{k=1}^2 \frac{\partial}{\partial x_k} V_k^{HR} \right| \right), 
        \frac{1}{100} \max \left( \left| \sum_{k=1}^2 \frac{\partial}{\partial x_k} V_k^{SR} \right| \right) \Biggr)
    \end{align}
    For each of the above loss matrices, the mean is then computed over all matrix components.
    In addition to the aforementioned components, the adversarial generator loss $L_G^{adv}$ is calculated based on the relativistic average discriminator as shown in Equation~\ref{eq:RaGAN_G_loss}. The full loss $L_G$ of the generator can then be written as
    \begin{align}
        \label{eq:g_loss_total}
        L_G ={}&  \eta_1 L_G^{pix} + \eta_2 L_G^{grad_{x_1x_2}} + \eta_3 L_G^{grad_{x_3}} + \eta_4 L_G^{div} + \eta_5 L_G^{div_{x_1x_2}}+\eta_6 L_G^{adv}
    \end{align}
    where $\eta_1$ to $\eta_6$ are the weights of each component with respect to the total loss and can be treated as additional hyperparameters.
    In contrast to the generator loss, the discriminator loss is calculated purely from the relativistic average adversarial loss in Equation~\ref{eq:RaGAN_D_loss}.

    \subsection{Quantitative evaluation metrics}
    \label{subsec:EvaluationMetrics}
    Throughout this work, the peak signal-to-noise ratio (PSNR), the absolute pixel error (short: pix), the pixel-vector error, and the pixel-vector relative error are employed.
    The peak signal-to-noise ratio (PSNR) is commonly used to measure the reconstruction quality of lossy transformation (e.g. image inpainting, image compression), and a higher value is preferable. In the current context, it is defined via the mean squared error ($MSE$) computed from the original wind field and its noisy (trilinear interpolation or super-resolution) approximations. It is computed as
    \begin{equation}\label{eq:psnr}
        PSNR=20 \cdot log_{10}(MAX_u)-10\cdot log_{10}(MSE)
    \end{equation}
    where $MAX_u$ is the maximum wind magnitude.

    The pixel error takes the difference between each "pixel" in the reconstructed and real images in each dimension separately. Applied to 3D vector fields, it is defined as
    \begin{equation}
        pix = \frac{1}{3}\sum_{i=1}^3 MAE(u_i)
    \end{equation}
    where $u_i$ are the components of a vector, and $MAE$ is the mean absolute error over all vectors in the vector field.

    The pixel-vector error is instead calculated from the absolute value of the difference between the real and super-resolved wind vectors $u^{HR}$ and $u^{SR}$ with
    \begin{equation}
        pix^{vector} =  MAE(|u^{HR}-u^{SR}|)
    \end{equation}
    and the pixel-vector relative error is calculated from the average value of $pix^{vector}$ divided by the average wind speed of the sample, i.e. 
    \begin{equation}
        pix^{vector}_{relative} =  \frac{pix^{vector}}{Mean(|u|)}
    \end{equation}
    where $Mean$ averages over the sample. 

\section{Method and set-up}
\label{sec:methodandsetup}

    \subsection{Data generation and investigation}
\begin{figure}
        \centering
        \includegraphics[width=\linewidth]{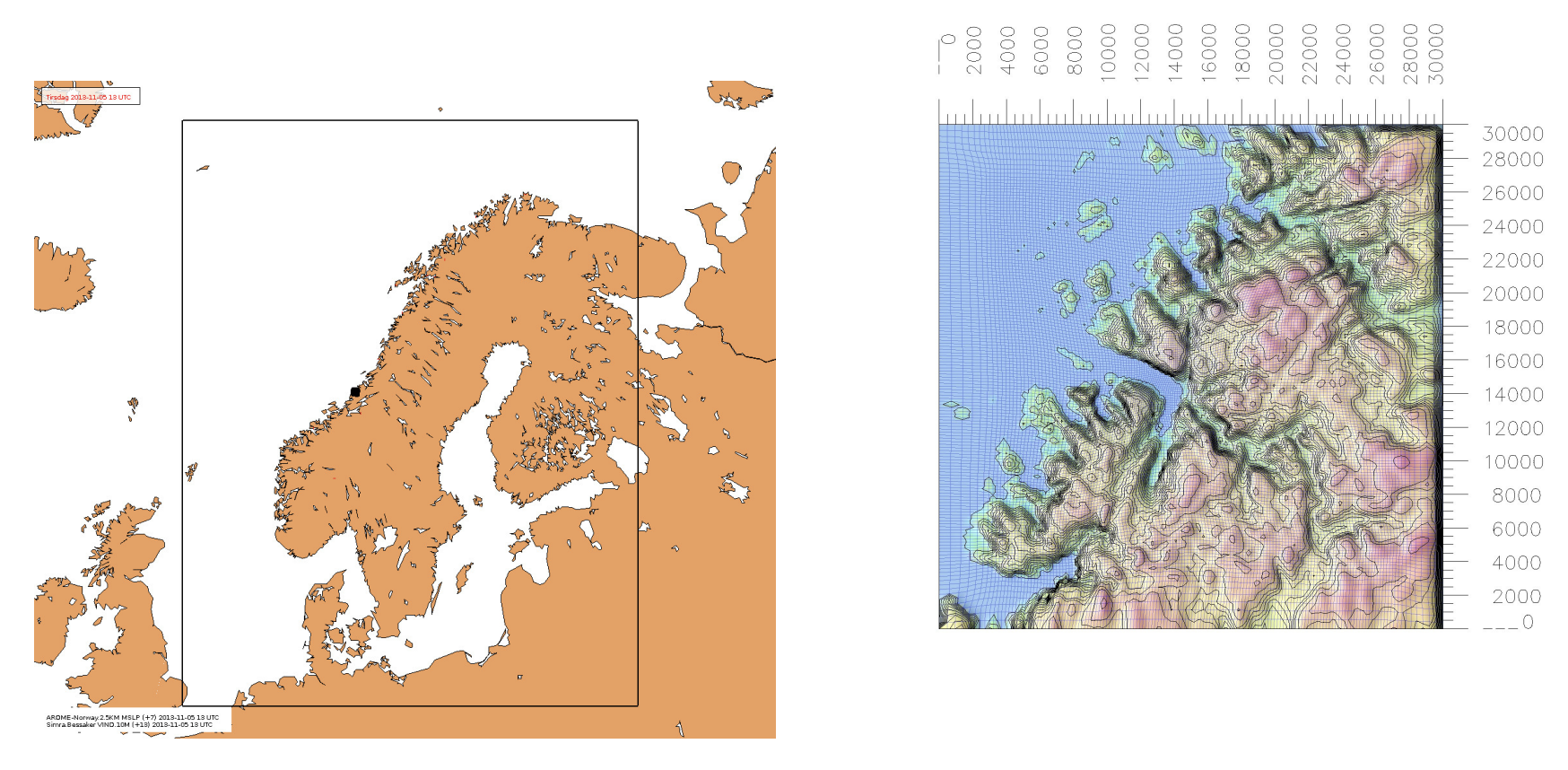}
        \caption[HARMONIE-SIMRA domain]{Area covered by the HARMONIE-SIMRA model, HARMONIE to the left, and SIMRA to the right, as it has been previously presented by Rasheed et. al. \cite{Rasheed2014amw}}
        \label{fig:HARMONIE-SIMRA}
    \end{figure}

\begin{figure}
        \centering
        \includegraphics[width=\linewidth]{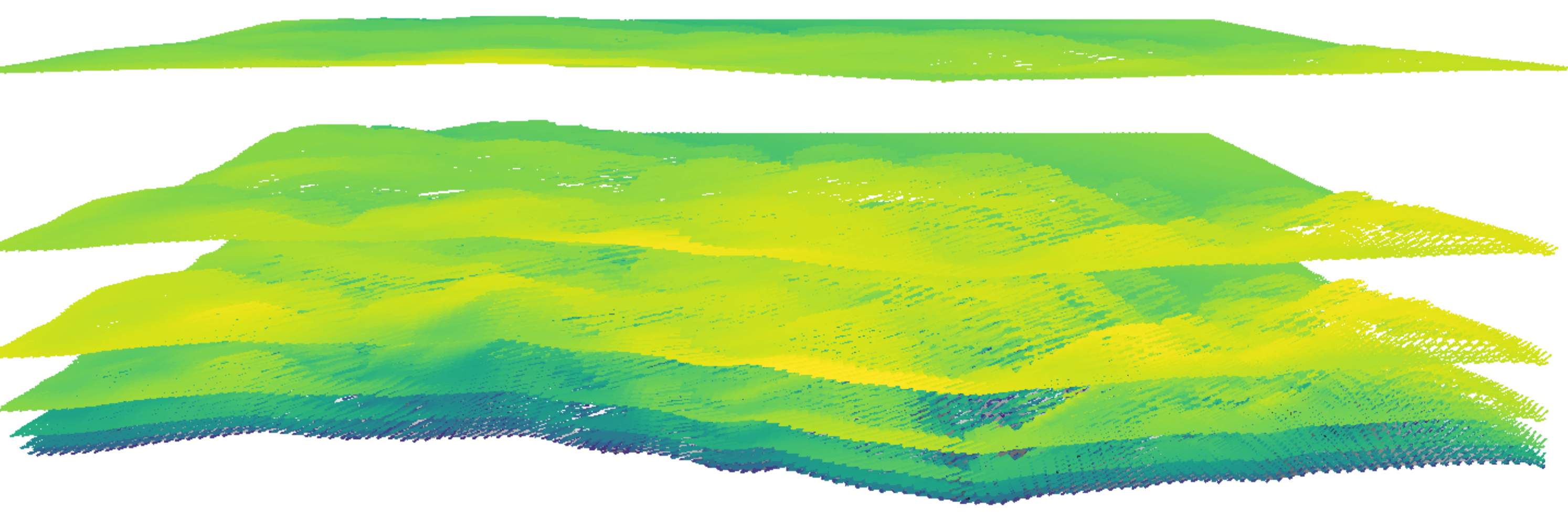}
        \caption[Shape of the Dataset]{Shape of the dataset. Showing every fifth layer of data in the vertical coordinate. The $z$-coordinate has been multiplied by five to highlight the shape. Yellow marks high wind speed, and blue marks low wind speed.}
        \label{fig:entire_area}
    \end{figure}

    The data used in this work are generated by the multiscale numerical simulation model explained in (Section~\ref{subsec:microscale}). 
    The $750 \times 960 \times 65$ number of grid points comprising the mesh of the HARMONIE model covers a $1875~\mathrm{km} \times 2400~\mathrm{km} \times 26~\mathrm{km}$ area, giving it a resolution of $2.5~\mathrm{km} \times 2.5~\mathrm{km}$ in the horizontal domain. The simulated HARMONIE data is interpolated and used as input for the SIMRA model, which simulates the wind field on a  domain size of $30~\mathrm{km} \times 30~\mathrm{km} \times 2.5~\mathrm{km}$ with $150 \times 150 \times 41$ mesh elements. The SIMRA area is chosen around the Bessakerfjellet wind farm located at (64\degree~13'~N,~10\degree~23'~E) on the Norwegian coast. The area features ocean, coast, fjords, cliffs, mountains, and other interesting features that result in highly complex terrain. The resolution of the nested SIMRA model increases slightly towards the center. The actual geographical domains of HARMONIE and SIMRA, as well as the horizontal SIMRA mesh can be seen in Figure~\ref{fig:HARMONIE-SIMRA}.  It should be noted that the SIMRA mesh resolution adapts based on the terrain variation to better resolve their effects on the wind field. Furthermore, the SIMRA mesh has an uneven vertical spacing, with horizontal layers closely following the terrain shape near the ground and gradually turning flatter higher up. This pattern can be seen clearly in Figure~\ref{fig:entire_area}.
    The hourly generated SIMRA data on a horizontal mesh with roughly $200m\times200m$ resolution depending on the distance from the center of the mesh is available at \url{https://thredds.met.no/thredds/catalog/opwind/catalog.html}. This is the dataset that will be used as the original high-resolution wind field in this work.
    
\begin{figure}
        \centering
        \includegraphics[width=\linewidth]{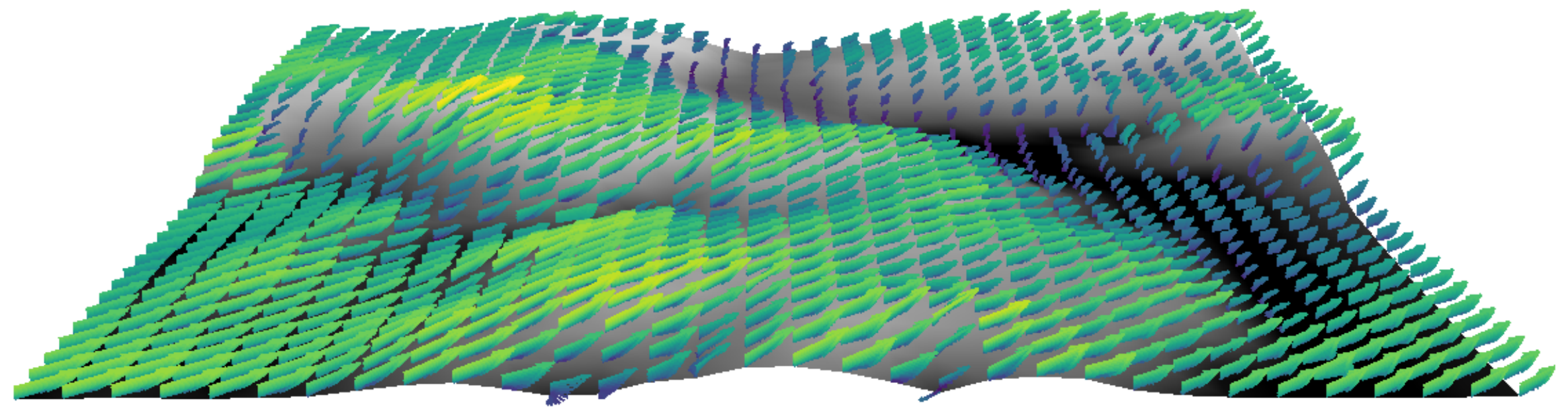}
        \caption[Zoomed In Unscaled Wind Field]{Zoomed in unscaled wind field. Showing the 10 bottommost layers in a 32x32 slice of the data area. Yellow marks high wind speed, and blue marks low wind speed.}
        \label{fig:zoom1}
    \end{figure}

    Zooming in on a slice of the data set with $32 \times 32 \times 10$ points or 10 height layers with size $6.4~\mathrm{km} \times 6.4~\mathrm{km}$, and not scaling the vertical axis, the wind field is shown in more detail in Figure~\ref{fig:zoom1}. It can be seen how the terrain affects the wind, causing higher wind speed uphill, and less, even reversed, wind speed downhill. Figure~\ref{fig:zoom1} also demonstrates how dense the horizontal layers are close to the ground, a normal property of simulated data, to help model the complexities of near-surface wind flow.
    For the purpose of this work, the unevenness in the vertical spacing is an advantage and a challenge at the same time. Higher information density in areas that are hard to approximate is beneficial for accurate modeling, and when designing wind farms one is interested in the wind speed relatively close to the ground, where turbines can harvest the wind energy. At the same time, it poses a challenge for applying CNNs, as they, in applying the same convolutional filter across the entire spatial dimension, assume that the same relations hold between neighboring points in one part of the data as others. If this is not the case, the CNNs have to compromise between features in different parts of space. To address this challenge, different approaches to handle the vertical spacing are tested and compared in \textit{Experiment~1}, which is further explained in Section~\ref{subsec:cases}.
    
\begin{figure}[h]
    \centering
    \begin{subfigure}{\linewidth}
    \includegraphics[width=\linewidth]{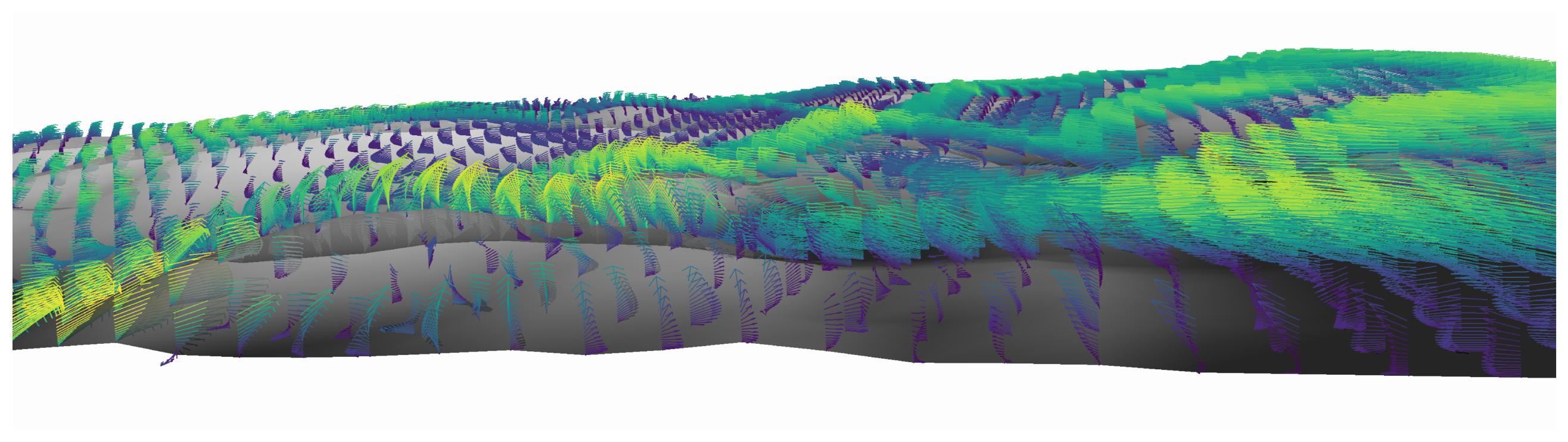}
    \caption{High-resolution wind field}
    \label{subfig:swirlsHR}
    \end{subfigure}
    \begin{subfigure}{\linewidth}
    \includegraphics[width=\linewidth]{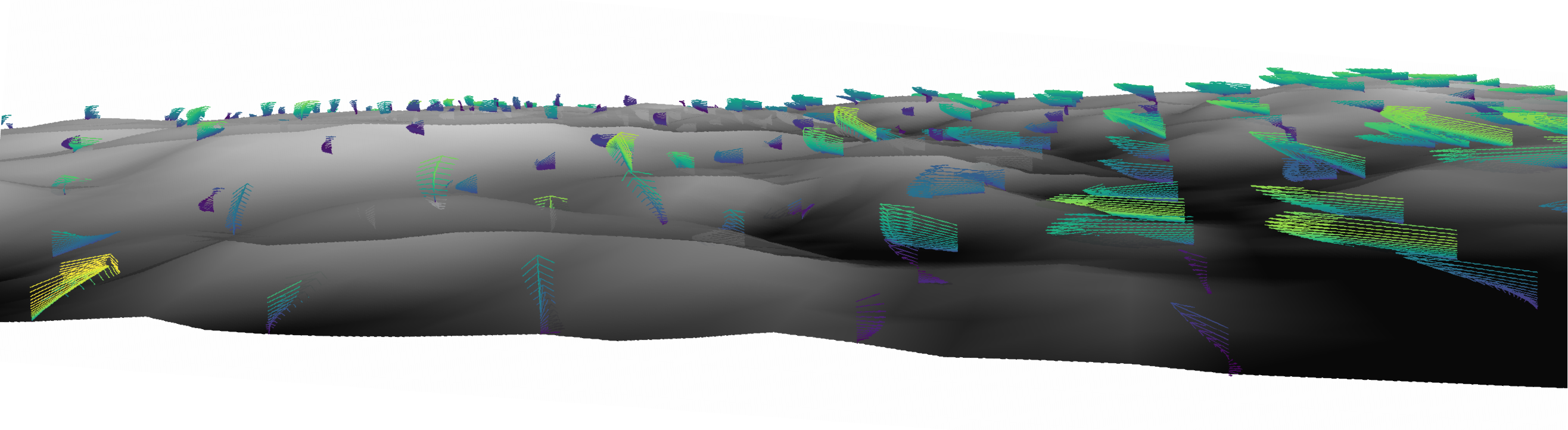}
    \caption{Low-resolution wind field}
    \label{subfig:swirlsLR}
    \end{subfigure} 
    \caption[Turbulent Terrain Effects]{Turbulent Terrain effects. Showing the 16 bottommost layers in a chaotic region of the wind field.}
    \label{fig:swirls}
    \end{figure}

    Zooming further, in Figure \ref{fig:swirls}, some of the flow characteristics can be seen that make it hard to model near-surface wind fields in complex terrain. Dips and bumps in the landscape induce swirls and chaotic wind patterns. The figure shows the difficulty of super-resolving the wind field, as a perfect generator would have to infer the full swirling pattern of the high-resolution wind field in Figure~\ref{subfig:swirlsHR} from the low-resolution wind field in Figure~\ref{subfig:swirlsLR}. Comparing such areas in the generated and real data, therefore, provides a key indicator of how well the model has learned to model the physics of atmospheric wind flow. 
    These patterns also demonstrate an advantage of working with full 3D data rather than 2D slices, as these effects are hard to pick up from one 2D slice.    

    \subsection{Data preprocessing}
    \label{subsec:data_processing}
    The SIMRA model is evaluated on a mesh with size $150\times150\times41$.
    This corresponds to a mesh resolution of $200m\times200m$ horizontally. The vertical spacing increases towards larger heights. A table with the spacing is available in the appendix~\ref{tab:simra_vertical_spacing}. The simulation tends towards low accuracy close to the edges due to relatively coarse resolution; therefore, the outermost points are discarded so that the horizontal dimensions are reduced to $128\times128$ points. Furthermore, the wind field becomes more complex closer to the terrain. Due to limited computational resources available, this work focuses on 10 lower layers where the wind flow prediction is most challenging (layers 2-11, the first layer is dropped as it represents the wind speed at height zero over the terrain which is always zero). The highest layer is therefore around 35 meters above the ground. Note that this restriction is motivated purely by available computational resources and does not limit the significance of the study since it can be assumed that flows in higher layers are equal or easier to interpolate due to the relatively lesser influence of terrain on them. The resulting wind field can be formulated as an array with shape $128 \times 128 \times 10$.
    
\begin{figure}
        \centering
        \includegraphics[width=\linewidth]{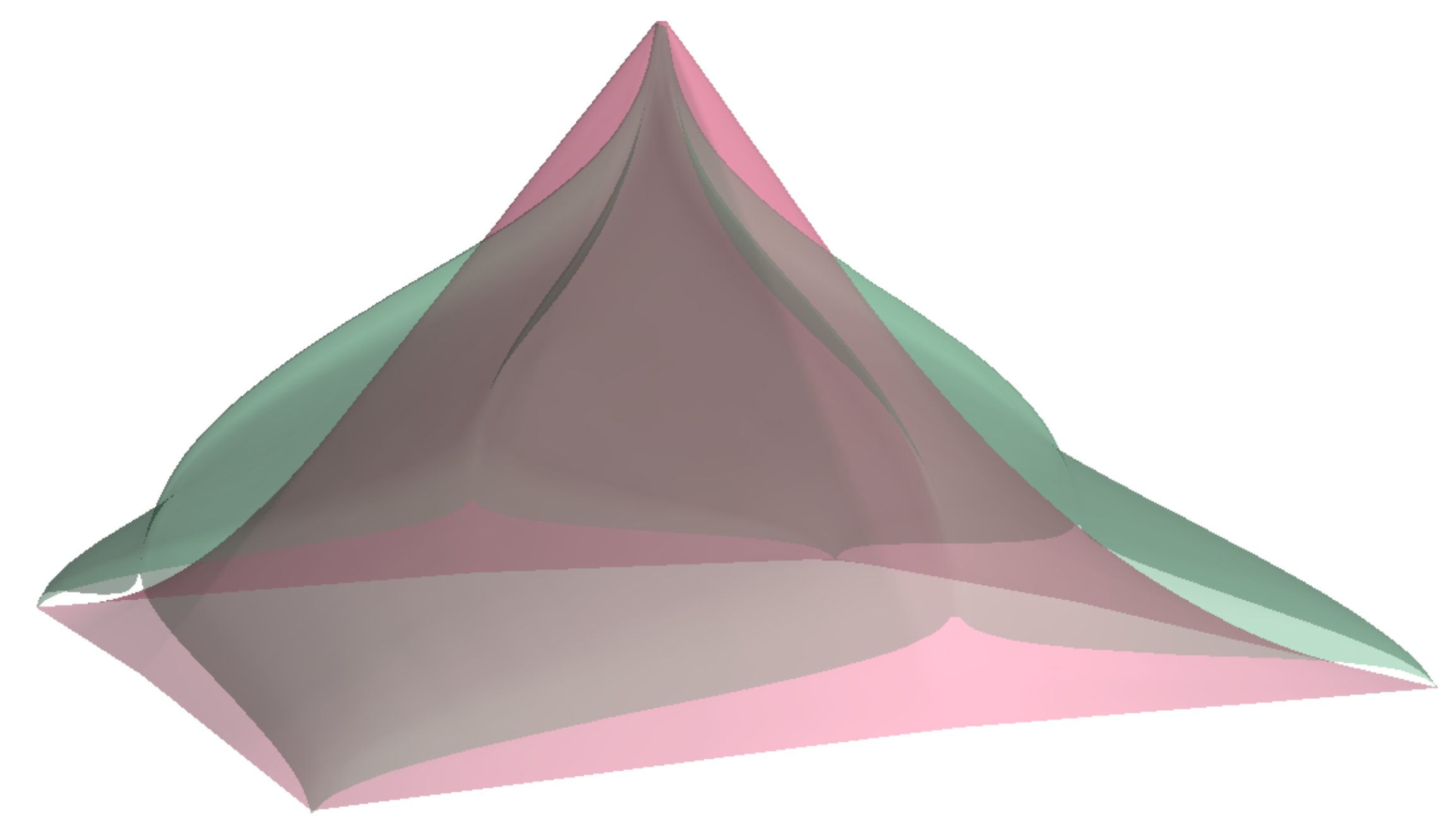}
        \caption{Probability for pixels being included in a subsample using uniform and beta (0.25,0.25) distribution for subsample selection. The beta distribution shifts the probability towards the edges and is used here.}
        \label{fig:beta}
    \end{figure}

    While many distinct wind fields are available for training the generator and discriminator, the wind fields have all been calculated for the same terrain, which poses the risk of the NN to memorize the terrain. To reduce the potential for such overtraining, a smaller slice of the wind field is used for training, so that each wind field slice has a slightly different underlying terrain. The size of the slices is chosen to $64\times64\times10$ of the entire terrain $128\times128\times10$. This results in $65\times65$ possibilities for taking a slice. 
    If sampled uniformly, the values at the edges are less likely to be included in the data set. To alleviate this, a beta distribution with $(0.25,0.25)$ is used to draw samples. The probability of a pixel being included in the data set is shown in Figure~\ref{fig:beta}. As the translation invariance of convolutional networks limits the effect of this strategy, the wind field is also horizontally mirrored and rotated in $90\degree$ steps, resulting in $65\times65\times2\times4=33800$ potentially correlated but not identical terrains from which to draw wind fields.
    
    Next, the wind field is downsampled in horizontal directions to a lower resolution using the nearest-neighbor method. Another viable option could have involved running the multiscale model at a resolution of $800m\times800m$, instead of resorting to the nearest neighbor approximation. However, we firmly believe that this approach not only conserves the time required for an 800-meter resolution simulation, but also introduces a beneficial element of noise into the input data. This noise has been shown to improve the robustness of GANs during training. The downsampling factor in this work is 4 unless explicitly stated otherwise. The result is a high-resolution wind field (referred to as HR) with $64\times64\times10$ wind vectors as the generator target and a low-resolution wind field (referred to as LR) with $16\times16\times10$ wind vectors as the generator input.

    While overtraining of the terrain in the feature maps should be avoided, the generator should still have a chance to learn from the terrain corresponding to the input wind field. Therefore, the low-resolution input may also include as additional features the low-resolution height above the mean sea level, low-resolution terrain height, and/or height over the terrain in low resolution. Additionally, the low-resolution pressure can potentially contribute information and can be included as a feature in the low-resolution input. Each of these additional contributions can be toggled with a new hyperparameter, and the effect of each contribution is investigated as part of this work as explained in \textit{Experiment~1} in Section~\ref{subsec:cases}.

\begin{figure}
        \centering
        \begin{subfigure}{0.9\linewidth}
            \includegraphics[width=\linewidth]{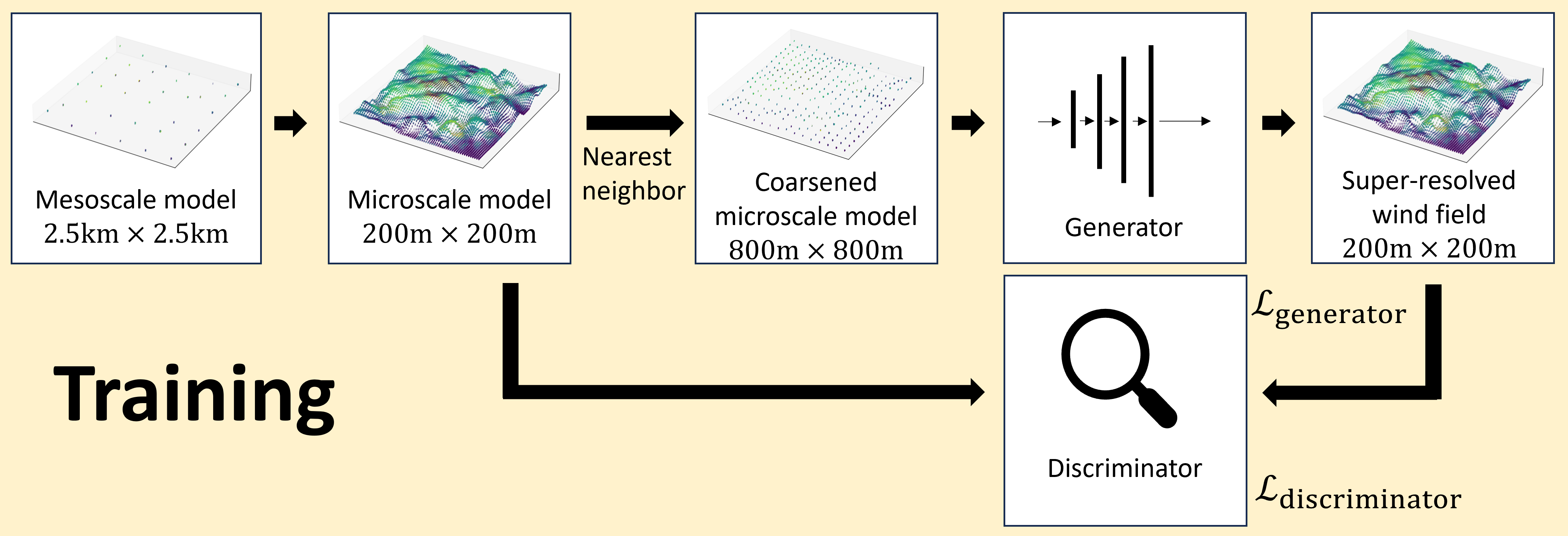}
            \caption{The training and optimization of the model.}
            \label{fig:pipeline_train}
        \end{subfigure}
        \begin{subfigure}{0.9\linewidth}
            \includegraphics[width=\linewidth]{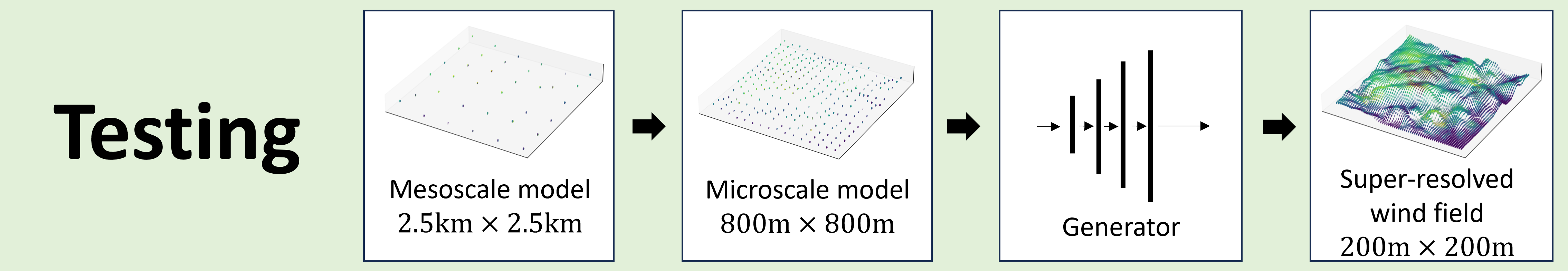}
            \caption{The testing of the model.}
            \label{fig:pipeline_application}
        \end{subfigure}
        \caption{The pipeline for training, optimization, and testing of the model. The training data is generated by nesting the microscale simulation into the mesoscale simulation. The generator uses coarsened microscale simulations as input and creates a super-resolved wind field. In the GAN setup, the discriminator tries to distinguish the super-resolved wind field and the microscale simulation, and the result is used to improve the generator. Once the model is optimized, it needs to be tested before it can be applied to the mesoscale model.}
        \label{fig:pipeline}
    \end{figure}

    It is well known that normalizing data tends to ease the training of NNs, and so here too all values are normalized to $[0,1]$, except for vector components, which are normalized to $[-1,1]$. One additional hyperparameter is added that toggles whether all input values are interpolated in the vertical direction to be evenly spaced with respect to the ground. The preprocessed data is then used as input (LR) and target (HR) of the generator and as ground truth (HR) for the discriminator. Note that the 3D vector field is kept in a cartesian coordinate system so that the wind vectors are parametrized in $u_x$, $u_y$, and $u_z$. The data is split into training, validation, and test data set with ratios [0.8,0.1,0.1] respectively. 
    Moreover, during the testing phase, we conducted additional $800\times800m$ simulations to serve as input data for the trained model. The full training and testing pipeline is depicted in Figure \ref{fig:pipeline}.

    \subsection{Model structure}
    \label{subsec:modelstructure}   
\begin{figure}[htb!]
    \centering
    \includegraphics[width=\linewidth]{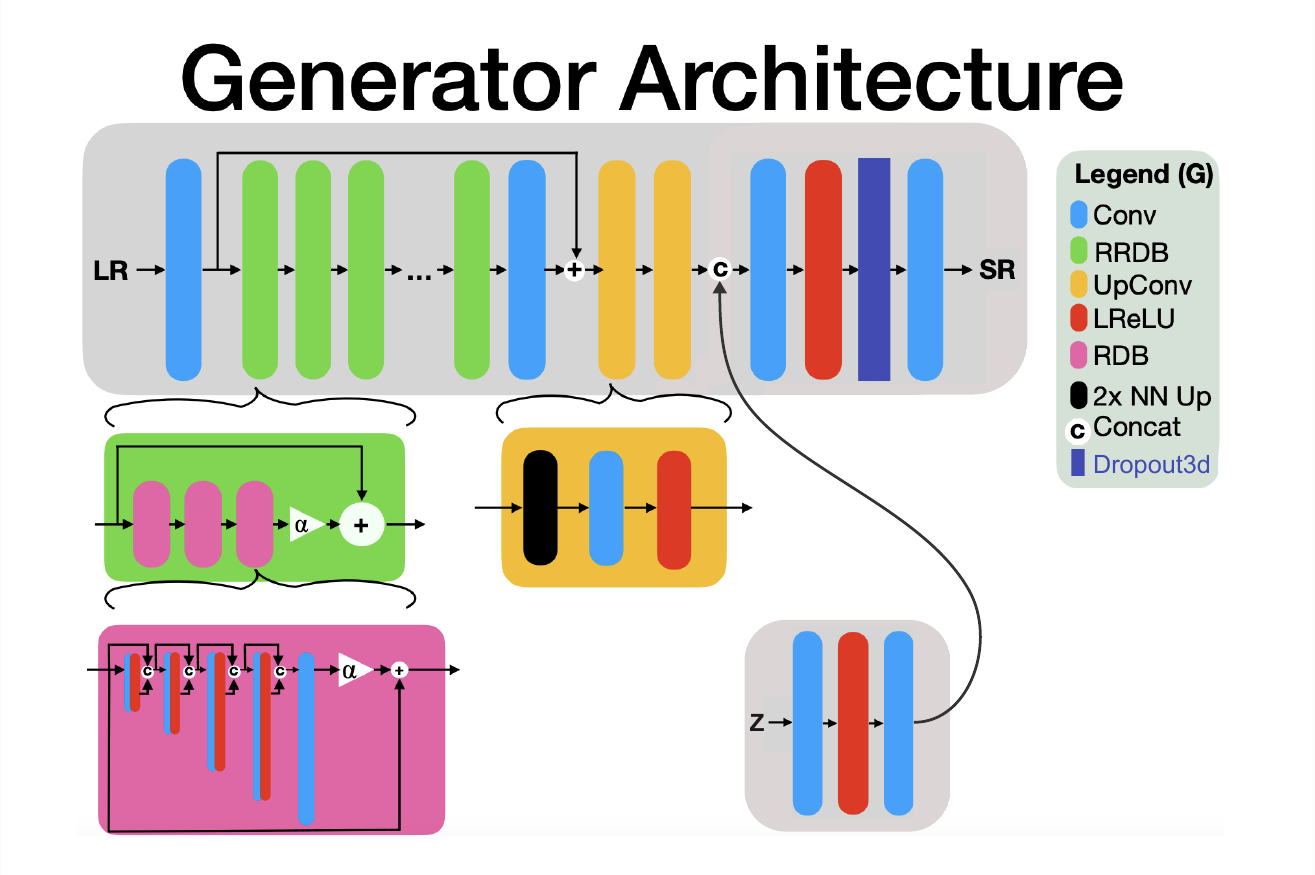}
    \includegraphics[width=\linewidth]{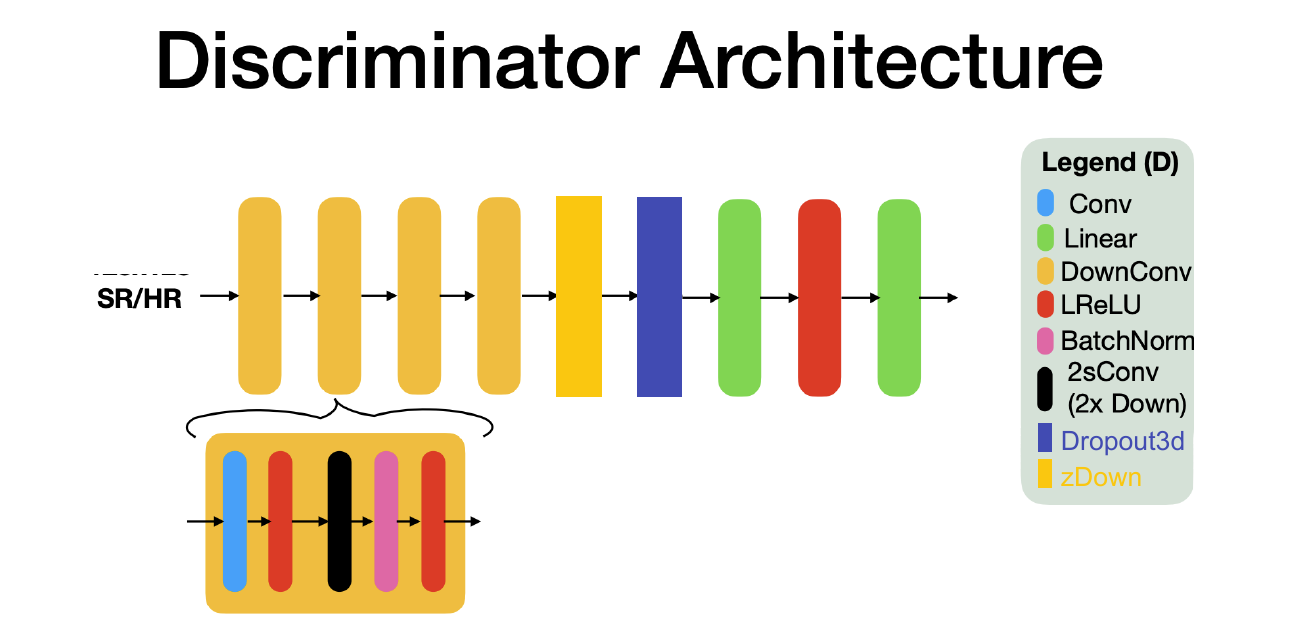}
    \caption{Generator and discriminator structure used in this work. A modified version of ERSDGAN \cite{Vesterkjaer2019e}. Noteworthy changes are the increase from 2D to 3D, a new terrain feature extractor that takes high-resolution terrain information as input, additional feature dropout layers in the generator and discriminator, and the inclusion of a DownConv block with stride width 2 in vertical and 2 or 1 in horizontal direction depending on whether slice data augmentation is enabled or not.}
    \label{fig:newESRGAN}
    \end{figure}

    The Generator and Discriminator architectures in this work are based on the ESRDGAN implementation \cite{Vesterkjaer2019e}, which itself is based on the ESRGAN model by Wang\cite{Wang2019ees}.
    The architectures of both networks are shown in Figure~\ref{fig:newESRGAN}. 
    
    \paragraph{Generator} The generator uses the 3D low-resolution data (LR) and the high-resolution terrain height (Z) as input.
    The LR data is first processed by a 3D convolutional layer, whereafter multiple residual-in-residual dense blocks (RRDB) are stacked. An RRDB uses multiple residual dense blocks (RDB) which are implemented with residual connections, whose output is weighed with $\alpha=0.2$ before being added to the input. In the RDB, 3D convolutional layers with leaky rectified linear units (LReLU) activations are densely connected (after each output all previous inputs are summed up as input for the next layer). The RDB block finishes with a final convolutional network, whose output is weighted with $\alpha=0.2$ before being added to the input of the block.
    After the RRDB blocks, an additional convolutional layer is added, and a residual connection bypassing all RRDB is added. It follows upsampling convolutional blocks (UpConv) which are built from a nearest-neighbor upsampling (2x NN Up) and a convolutional layer with LReLU activation.
    Next, the output is concatenated with the terrain extractor, which takes the high-resolution terrain as input and extracts features through two convolutional layers where the first has LReLU activation.
    The concatenated output is processed by two final convolutional layers, where the former one has LReLU activation and a 3D feature dropout layer (Dropout3d). The output fo the final layer is then the super-resolved wind vector field (SR).
    
    \paragraph{Discriminator}
    The discriminator uses downsampling convolutional blocks (DownConv), where each block starts with a convolutional layer with LReLU activation, followed by a depthwise separable convolutional layer for downsampling by a factor 2, batch normalization, and another LReLU activation. A modified DownConv block is added with separate stride widths for horizontal and vertical directions to account for unequal dimensions in the input data. The block is followed by a 3D dropout layer and two linear layers where the first uses LReLU activation.
    
    \subsection{Hyperparameters}
    \label{subsec:hyperparameters}
    Several hyperparameters have been changed from the original ESRGAN  \cite{Vesterkjaer2019e}, and from the earlier work on wind field interpolations \cite{Tran2020ges,Larsen2020ota}. An exhaustive list of all default hyperparameters used in this work can be found in the appendix in Table~\ref{tab:full-fyperparameters}. The hyperparameters are altered throughout the work only where mentioned explicitly. Specifically, in addition to the wind field, pressure and the $z$-coordinates of the data are included as low-resolution input channels to the generator. Data is augmented by slicing, mirroring, and rotating as specified in Section~\ref{subsec:data_processing}, and the generator and the discriminator are trained by alternating between them for 90k iterations. Instance noise is added to the input samples in the early training phase to confuse the discriminator, and one-sided label smoothing is included to prevent overconfidence inside the discriminator. A set of new hyperparameters that deserve special attention is connected to the choice of the loss function and is addressed in Section~\ref{subsec:loss}.

    \subsection{Cases}
    \label{subsec:cases}
    As mentioned earlier, additional features and additional hyperparameters are introduced. The hyperparameter consists of weighting coefficients for the different components of the loss function. To tune these new hyperparameters and test the impact of features the following three experiments are conducted.
    
    \subsubsection{Experiment 1: Evaluating the impact of input features}
    First, the impact of various additional input features is tested. As touched upon earlier, the terrain shape and altitude significantly affect wind flow, but the irregular spacing of the vertical coordinates poses a problem for applying CNNs. This experiment aims to find the best way to address this, either by giving the generator input that contains information about the terrain, altitude, and vertical spacing or by interpolating the data vertically to get a more CNN-suited grid. To this end, the GANs are trained and evaluated with several combinations of pressure, absolute height, height over terrain, terrain height, and vertical interpolation as low-resolution input, while other parameters are kept as specified in Section~\ref{subsec:hyperparameters}. Two different seeds are used to estimate whether the results are reliable and can be reproduced.

\newcolumntype{R}{>{\raggedleft\arraybackslash}X}
\newcolumntype{L}{>{\raggedright\arraybackslash}X}
\renewcommand\tabularxcolumn[1]{m{#1}}
\newcommand{\cmark}{\ding{51}} % checkmark
\newcommand{\grayline}{\arrayrulecolor{gray}\hline\arrayrulecolor{black}} % gray lines

\begin{table*}
    \centering
        \caption[Experiment 1 Combinations]{Combinations tested in \textit{Experiment~1}. The column \textbf{name} is what the combination will be referred to, column \textbf{interpolate} indicates whether the data is interpolated vertically, while columns \bm{$z$}, \bm{$p$}, \bm{$z_{alt}$}, and \bm{$z_t$} respectively refer to height above sea level, pressure, altitude, and terrain height respectively.}
    \label{tab:exp1}
    \begin{tabularx}{\linewidth}{L|XRRRR}
        \textbf{name} & \textbf{interpolate} & \bm{$z$} & \bm{$p$} & \bm{$z_{alt}$} & \bm{$z_t$} \\ \hline
        only\_wind &  &  &  &  &  \\ \grayline
        $z$\_channel &  & \cmark & &  &  \\ \grayline
        $p$\_channel &  &  & \cmark &  &  \\ \grayline
        $p\_z$\_channels &  & \cmark & \cmark &  &  \\ \grayline
        $z_{ground}$\_channels &  &  &  & \cmark & \cmark \\ \grayline
        $p\_z_{ground}$\_channels &  &  & \cmark & \cmark & \cmark \\ \grayline
        only\_wind\_interp & \cmark &  &  & &  \\ \grayline
        $z$\_channel\_interp & \cmark & \cmark &  &  & \\ \grayline
        $p$\_channel\_interp & \cmark &  & \cmark &  &  \\ \grayline
        $z\_p$\_channels\_interp & \cmark & \cmark & \cmark &  &  \\ \hline
    \end{tabularx}
\end{table*}

    \subsubsection{Experiment 2: Evaluating the impact of the loss components}
\label{subsec:methods_exp2} Next, the second set of hyperparameters, the weights of the components that make up the loss functions, is investigated. In \textit{Experiment~2}, the importance of the components is investigated by setting some weights to 0 while scaling up others, so that certain components of the loss function have higher contributions while others are deactivated.
In addition to the loss function, another slightly modified configuration was made in response to the results of \textit{Experiment~1}. The training ratio between the discriminator and generator (\textit{D\_G\_train\_ratio}) is changed to 2 after 60k iterations and \textit{niter} to 100k so that the generator is trained for 60k+20k iterations and the discriminator is trained for 60k+40k iterations in total.
    Otherwise, the setup remains unchanged, i.e. only the pressure and height above mean sea level are added to the low-resolution input of the generator. In \textit{Experiment~2} the combinations of loss component weights $\eta_i$ in Table~\ref{tab:exp2} are tested with two different seeds. The pixel loss and adversarial loss are kept constant at $\eta_1=0.15$ and $\eta_6=0.005$. 

\begin{table}
    \centering
     \caption[Experiment 2 Setup]{Combinations tested in \textit{Experiment~2}. The column \textbf{name} is what the combination will be referenced to as, and the other columns \bm{$\eta_i$}s are the cost weighing constants defined in Equation \eqref{eq:g_loss_total}}
    \label{tab:exp2}
    \begin{tabularx}{\linewidth}{l|RRRR}
    
        \textbf{name} &$\bm{\eta_2}$ & $\bm{\eta_3}$ & $\bm{\eta_4}$ & $\bm{\eta_5}$ \\ \hline
        \textit{only\_pix\_cost} & 0.0 & 0.0 & 0.0 & 0.0 \\ \grayline
        \textit{grad\_cost} & 1.0 & 0.2 & 0.0 & 0.0 \\ \grayline
        \textit{div\_cost} & 0.0 & 0.0 & 0.25 & 0.25 \\ \grayline
        \textit{$xy$\_cost} & 1.0 & 0.0 & 0.25 & 0.0 \\ \grayline
        \textit{std\_cost} &  1.0 & 0.2 & 0.25 & 0.25 \\ \grayline
        \textit{large\_grad\_cost} & 5.0 & 1.0 & 0.25 & 0.25 \\ \grayline
        \textit{large\_div\_cost} & 1.0 & 0.2 & 1.25 & 1.25 \\ \grayline
        \textit{large\_$xy$\_cost} & 5.0 & 0.2 & 1.25 & 0.25 \\ \hline
    \end{tabularx}
\end{table}
    
    The \textit{std\_cost} combination is equivalent to the \textit{$p\_z$\_channels} combination in \textit{Experiment~1}, so it will not be evaluated again, but this means that it hasn't been changed as specified above, instead running 90k iterations for both discriminator and generator. The effect will be addressed in Section~\ref{sec:resultsanddiscussions}.
        
    \subsubsection{Experiment 3: Loss weight optimization}
    \label{subsec:methods_exp3}
    
    Following the fixed-value evaluation of the loss components, a hyperparameter search was conducted in the following search space which was chosen based on results of \textit{Experiment~2}
    \begin{align}
        0.0 < & \eta_1 < 1.0 \\
        0.5 < & \eta_2 < 32.0 \\
        0.25 < & \eta_3 < 16.0 \\
        0.25 < & \eta_4 < 16.0 \\
        0.25 < & \eta_5 < 16.0, 
    \end{align}
    The search space is sampled logarithmically with base 2, except for $\eta_1$ which is sampled uniformly. The hyperparameter search is implemented using the ray tune framework \cite{Liaw2018tar}, with the Optuna \cite{Akiba2019oana} optimizing framework and an Asynchronous Successive Halving Algorithm (ASHA) \cite{Li2020asf} scheduler. The Optuna search algorithm implements a Bayesian tree search algorithm, it incorporates the results of previous samples when choosing configurations to test. The ASHA scheduler asynchronously stops poorly performing configurations, allowing tests of more combinations with the same computational resources. Note that this biases the search towards parameter combinations that result in good performance in the early training stages, and therefore assumes that strong initial performance indicates strong performance in later stages. Ray tune is a flexible framework for implementing the parallel processing of the search. The maximum training iterations is set to 35k and the minimum iterations before stopping a run is set to 1.2k. The search is initialized with promising combinations based on results from \textit{Experiment~2}.
    
    \subsection{Software and hardware framework}
    All the data used in this project was available in a NetCDF (Network Common Data Form) file format through an OpenDap server hosted by the Norwegian Meteorological Institute. For processing the data NetCDF library was utilized. All the codes for the GANs framework were developed in Python 3.9.13 using the PyTorch 2.0.1 library, which is an open-source software library developed by Facebook's AI group with a focus on the implementation of various neural network architectures. 
    
    The HARMONIE-SIMRA codes were run on the supercomputing facility ``Vilje", which is an SGI Altix ICE X distributed memory system that consists of $1440$ nodes interconnected with a high-bandwidth low-latency switch network (FDR Infiniband). Each node has two 8-core Intel Sandy Bridge ($2.6\,\mbox{Ghz}$) and $32\,\mbox{GB}$ memory, yielding the total number of cores to $23040$. The system is applicable and intended for large-scale parallel MPI (Message Passing Interface) applications. The results are converted into NetCDF and realized through an OPeNDAP server. The use of OPeNDAP (Open-source Project for a Network Data Access Protocol) \cite{OPeNDAPh} precludes the redundant copying of the results files on numerous machines for post-processing. A set of Python routines are implemented to read and post-process the hosted files on the fly. A brief overview of the computational set-up is given in Table~\ref{table:compmodels}. The HARMONIE model runs on $1840$ cores and to complete a $48$ hours forecast it takes approximately $87$ minutes. SIMRA on the other hand, running on $48$ cores, takes 13 minutes to finish one hourly averaged simulation each for the next $12$ hours. 
    
    The listed experiments and training of the models were done with NVIDIA A100m40 GPUs on the IDUN cluster at the Norwegian University of Science and Technology (NTNU). With one GPU, training the model for 100k iterations takes slightly less than two days. The GPU memory usage strongly depends on the input data size and the batch size. With batch size 32 and $16\times16\times10$ to $64\times64\times10$ interpolation, training the model requires approximately 22GB GPU memory.  The full code and its history can be found on GitHub at \url{https://github.com/jacobwulffwold/GAN_SR_wind_field}. The results are reproducible using the repository. The code is modified from Larsen's \cite{Larsen2020ota} code, which again is an adaptation of Vesterkjær's \cite{Vesterkjaer2019e} code that is accessible at \url{https://github.com/eirikeve/esrdgan}. All parts of the code have been modified, but Vesterkjær's basic structure remains.

\section{Results and discussions}
    \label{sec:resultsanddiscussions}
        In this section, the results of \textit{Experiments 1-3} are presented and discussed. The best combination of hyperparameters is identified, the model is evaluated with those hyperparameters, and the resulting wind fields are discussed in detail. Finally, the effects of overfitting the terrain and increasing the resolution enhancement factor are investigated.
        
        \subsection{Experiment 1: Evaluating the impact of input features}
        \begin{figure}[h]
        \centering
        \includegraphics[width=\linewidth]{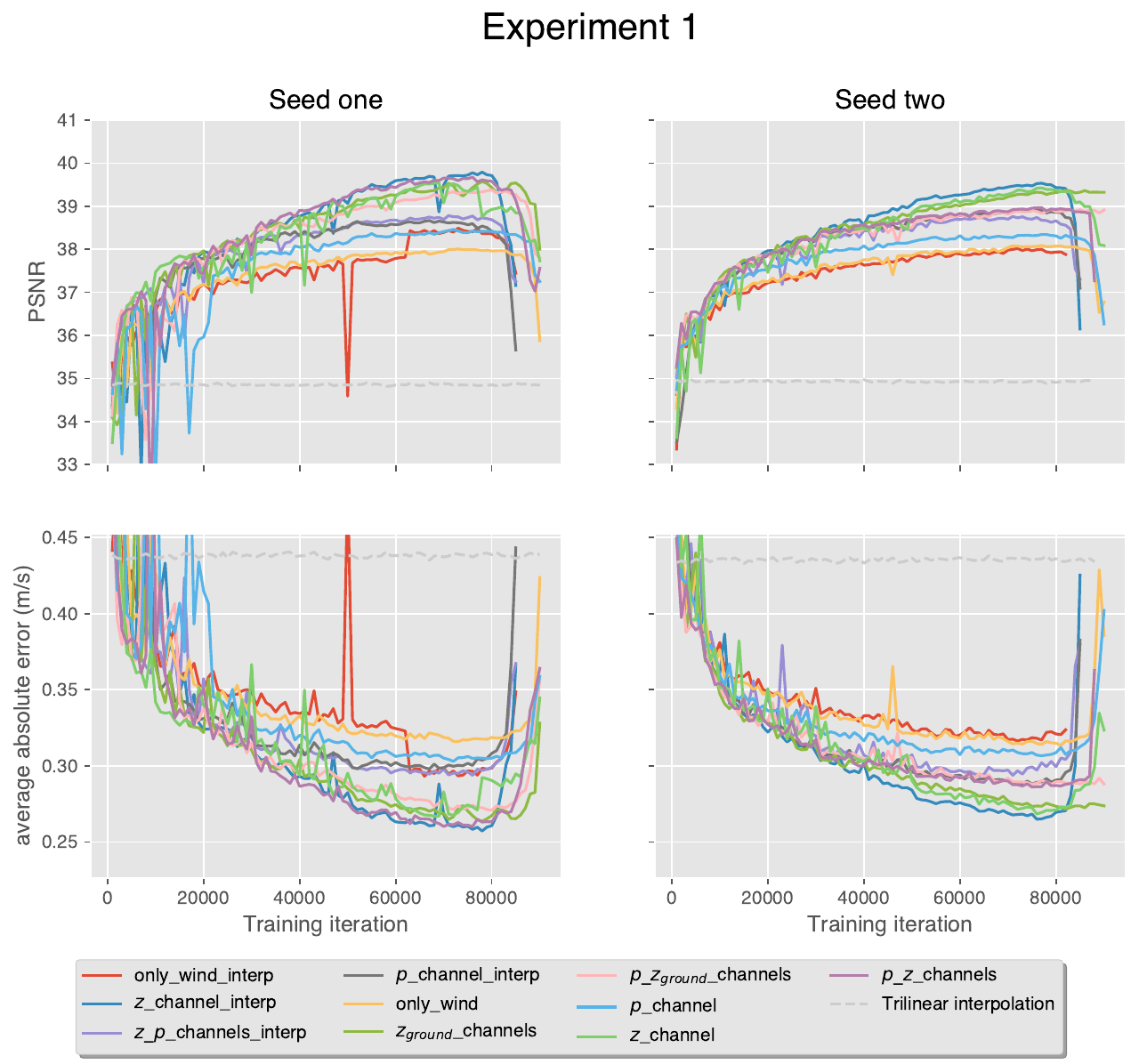}
        \caption[Experiment 1 Training Results]{Experiment 1 validation results during training for two different seeds, colored according to the combinations specified in Table~\ref{tab:exp1}. The top row shows the validation peak signal-to-noise ratio, and the bottom row shows the average absolute error of a wind component of the generated wind field. A dashed line for trilinear interpolation performance is included for comparison.}
        \label{fig:Exp1}
    \end{figure}
        The first experiment targets the additional input of the generator on two seeds as explained in Section~\ref{subsec:cases}. The peak-signal-to-noise ratio and average absolute error of each seed over the number of training iterations are shown in Figure~\ref{fig:Exp1}. It can be immediately seen that all configurations perform significantly better than the trilinear interpolation and that the performances of all GANs follow a similar pattern during training. However, the performance significantly drops after around 80k training iterations (80k epochs), which was verified to be caused by the adversarial loss component. The reason for this phenomenon is that the contributions from all loss components except adversarial loss have become so small that adversarial loss dominates the error and explodes the network. This finding is the reason for the change in the training strategy in \textit{Experiment~2} and \textit{Experiment~3} as explained in Section~\ref{subsec:cases}.
        
        The performance of the GANs saved at 80k training iterations is listed in Table~\ref{tab:results-exp1}. Note that in some cases, the degradation of performance already starts before reaching 80k epochs. For this reason, Figure~\ref{fig:Exp1} is used for the evaluation of the experiment. It can be seen that including pressure and/or any information on height improves the prediction over the \textit{only\_wind} setup. Interpolation of the input alone does not provide reproducible benefits. Adding only pressure results in minor benefits for both seeds. However, augmenting the input with altitude or height above mean sea level brings significant improvements. In both seeds, the five best-performing input constellations include some form of information on the height of the wind vectors either relative to terrain or to mean sea level.
        In both seeds, the best results are achieved by including the relative height above mean sea level and using interpolation, i.e. by setup \textit{$z$\_channel\_interp}.

        Limited computational resources prevented the evaluation of more seeds, which makes it impossible to draw more nuanced conclusions between input constellations. However, for this work, it is sufficient to identify the best performing model, to show that the generator based on convolutional layers performs well despite the irregular coordinates, and to prove that the quality of the generator can be improved by including information on the height of each wind vector in the input.

    \subsection{Experiment 2: Evaluating the impact of the loss components}
    \begin{figure}[h]
        \centering
        \includegraphics[width=\linewidth]{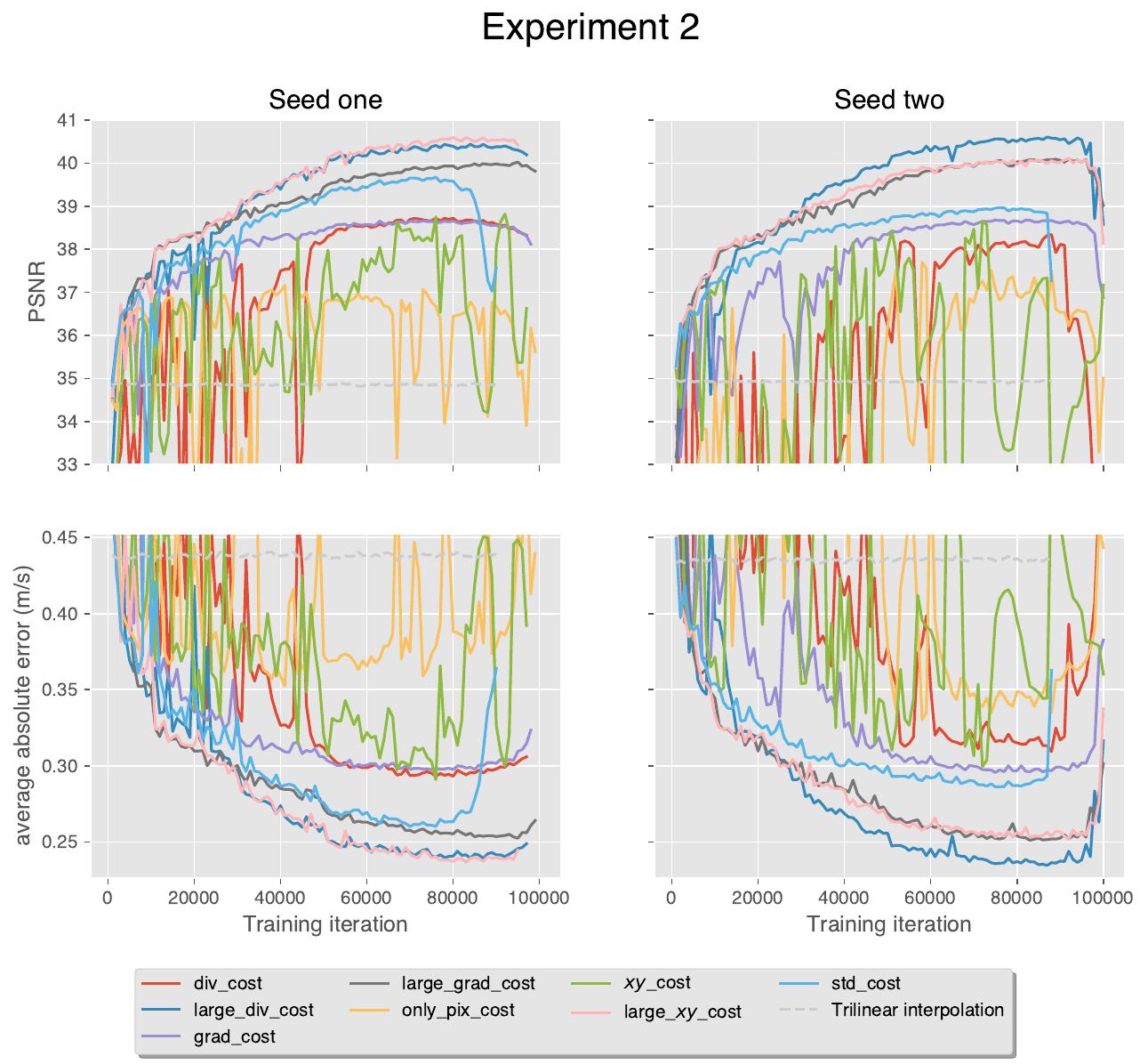}
        \caption[Experiment 2 Training Results]{\textit{Experiment~2} validation results during training for two different seeds, colored according to the combinations specified in Table~\ref{tab:exp2}. The top row shows the validation peak signal-to-noise ratio, and the bottom row shows the average absolute error of a wind component of the generated wind field. A dashed line for trilinear interpolation performance is included for comparison.}
        \label{fig:Exp2}
    \end{figure}
    The objective of \textit{Experiment~2} is to identify the most important components of the loss by adjusting the weights as explained in Section~\ref{subsec:cases}. Similar to Experiment 1, the peak-signal-to-noise ratio and the average absolute error are shown in Figure~\ref{fig:Exp2}. The results of using the generator saved on the training iteration 80k for \textit{std\_cost} and 90k for the rest of the combinations in the test set are collected in Table~\ref{tab:results-exp2}. Again, Figure~\ref{fig:Exp2} is used for the main evaluation. As a response to the strong network degradation due to the adversarial loss at around 80k epochs in experiment 1, the discriminator was trained more frequently than the generator by only evaluating the generator every second training iteration. It can be seen in Figure~\ref{fig:Exp2} that still a degradation sets in at around 100k training iterations, which corresponds to training the generator for 80k epochs. The exception is the default configuration \textit{std\_cost}, which has not been retrained, and therefore experiences 80k generator epochs and, therefore, training collapse after 80k training iterations. Therefore, this approach did not prevent degradation.
    Furthermore, it can be seen that using only the pixel- and adversarial loss (i.e. \textit{only\_pix\_loss}) results in very poor performance. Neglecting the derivatives in the vertical direction (i.e. combination \textit{$xy$\_cost}) results in bad quality of reconstructed wind fields too. Finally, training the GANs without gradient loss (\textit{div\_cost}) or divergence loss (\textit{grad\_cost}) reduces the quality of the predictions, albeit not as significantly. Large weights for any of the gradient components (\textit{large\_grad\_cost}) divergence components (\textit{large\_div\_cost}) and horizontal derivatives (\textit{large\_$xy$\_cost}) represent the best constellations.

    These observations suggest that most derivative-based loss components are important for optimal training. The exceptions are the vertical components of the gradient and divergence, where it has only been shown that one of them brings an important contribution, but no investigation has been carried out to separate them. While they could be inspected more closely here by introducing an additional constellation, their contributions are instead investigated as part of \textit{Experiment~3}.
    Additionally, constellations with large weights on loss components perform better than the initial values. A learning rate that is smaller than optimal could cause this. Further investigation has shown that increasing the learning rate indeed improves performance.
    
    \subsection{Experiment 3: Loss weight optimization}
\begin{figure}[h]
        \centering
        \begin{subfigure}{\linewidth}
            \includegraphics[width=\linewidth]{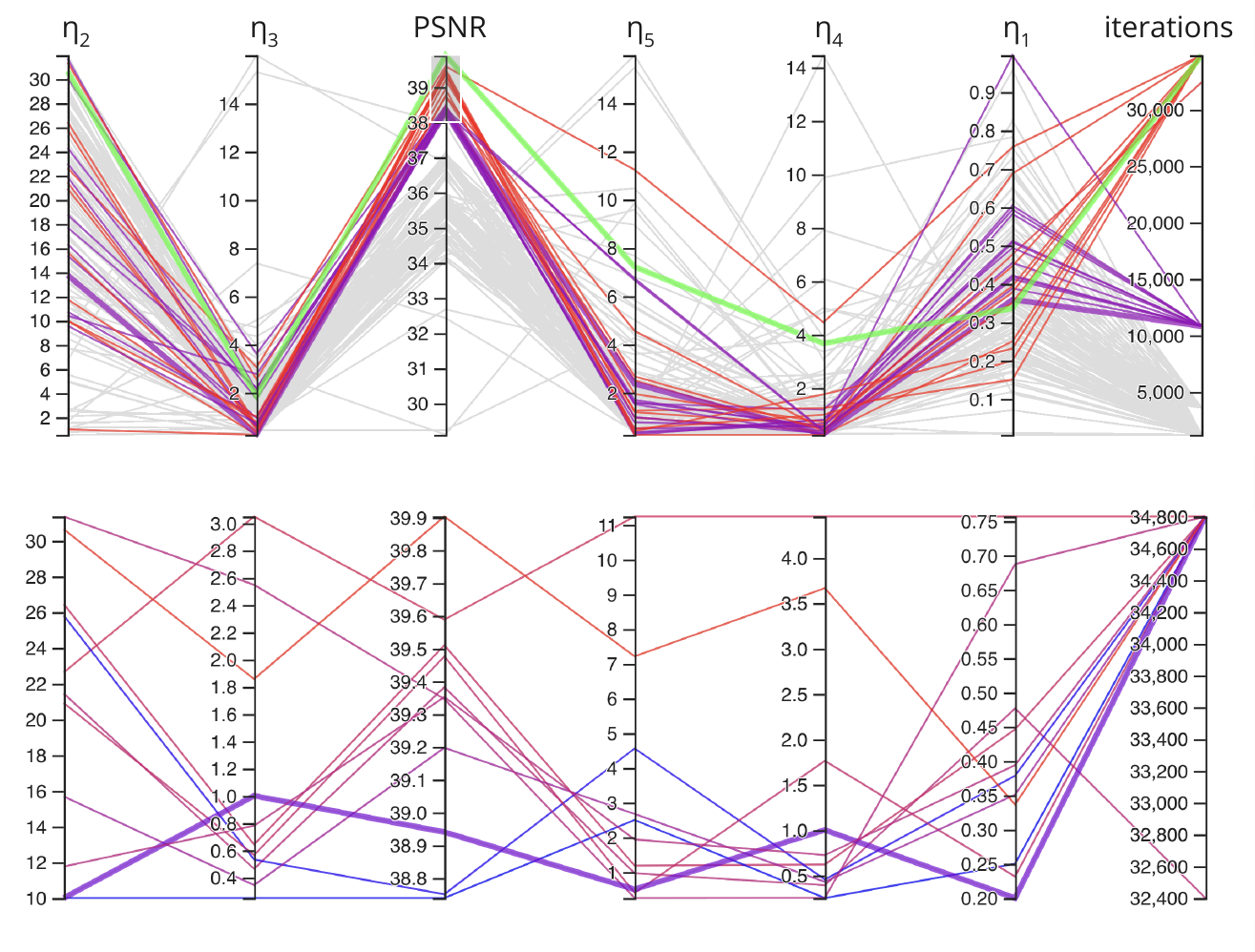}
        \end{subfigure}
        \caption[Experiment 3 Results]{Results of \textit{Experiment~3}. In the upper plot, the scores of all tested combinations and their parameters are displayed, with the best-performing combinations highlighted. The $\eta_i$s are the loss weighing constants defined in \eqref{eq:g_loss_total}, "iterations" is what iteration the run was stopped at, and PSNR is the validation PSNR score after that iteration. In the bottom plot, we see the same table for only the best performers with the parameter axes scaled according to their spanned parameter range. Plot generated with \href{https://docs.ray.io/en/latest/tune/tutorials/tune-output.html}{ray.tune.}}
        \label{fig:param_search}
    \end{figure}

\begin{figure}
        \centering
        \includegraphics[width=\linewidth]{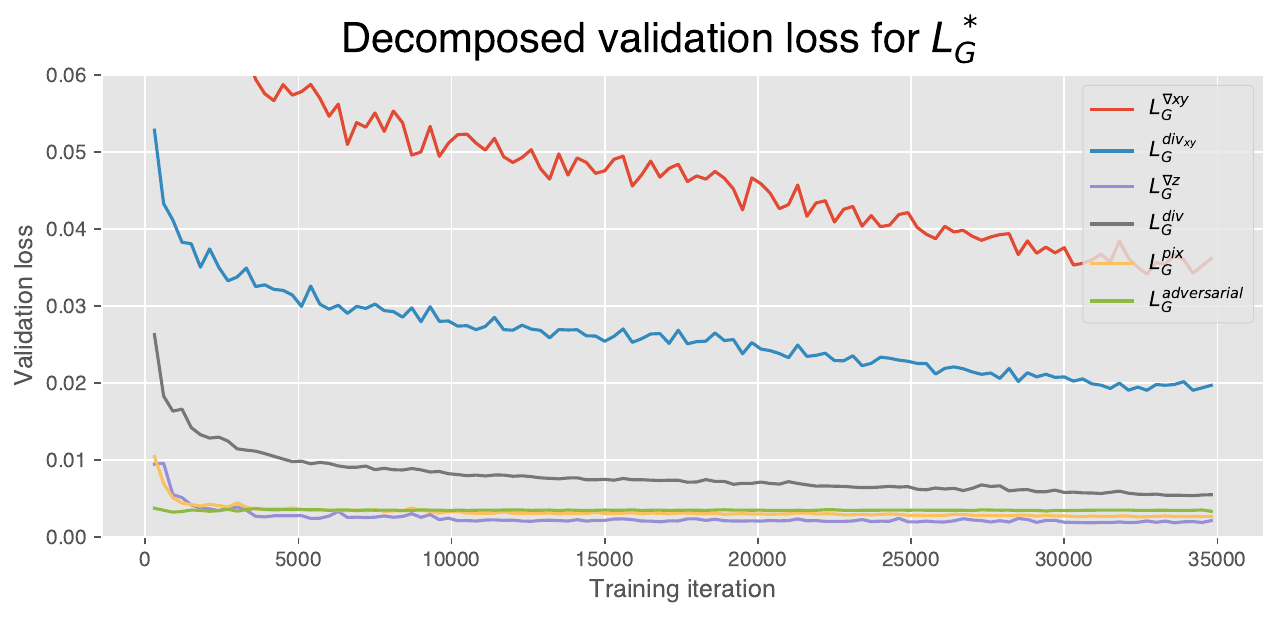}
        \caption[Decomposed Validation Loss]{Decomposed validation loss for the best-performing loss weights in the parameter search. The loss components in the plot are as the ones defined in Equation~\ref{eq:g_loss_total}.}
        \label{fig:Exp25Best}
    \end{figure}

    In \textit{Experiment~3} the weights of each loss component are optimized by a hyperparameter search as explained in Section~\ref{subsec:methods_exp3}. The results of the parameter search can be seen in Figure~\ref{fig:param_search}. In the upper plot, all evaluated combinations are shown, while the lower plot gives a zoom-in on the region covered by the best-performing models according to PSNR. The loss weights of the five best-performing evaluations and their PSNR and pixel loss are shown in Table~\ref{tab:exp25}. Both Figure~\ref{fig:param_search} and Table~\ref{tab:exp25} show that high values for $\eta_2$, i.e. the gradient in the horizontal directions, are preferred over very high values for $\eta_3$ and $\eta_5$. The pixel loss weight $\eta_1$ and the divergence loss weight $\eta_5$ of the best-performing models span almost their whole parameter space, indicating rather low importance. 
    \begin{table*}
        \centering
        \caption[Best Performing Combinations in Experiment 3]{Best performing combinations in \textit{Experiment~3}. The $\bm{\eta_i}$s are the loss weighing constants defined in Equation~\ref{eq:g_loss_total}, and \textbf{PSNR} and \textbf{pix} are the metrics as defined in Section~\ref{subsec:EvaluationMetrics}.}
        \label{tab:exp25}
        \begin{tabularx}{\linewidth}{XXXXX|RR}
        $\bm{\eta_1}$ & $\bm{\eta_2}$ & $\bm{\eta_3}$ & $\bm{\eta_4}$ & $\bm{\eta_5}$ & \textbf{PSNR~(db)} & \textbf{pix~(m/s)} \\ \hline
        $0.336$ & $30.6$ & $1.86$ & $7.21$ & $3.66$ & $39.9$ & $0.259$ \\ \grayline
        $0.757$ & $22.6$ & $3.05$ & $11.2$ & $4.45$ & $39.6$ & $0.265$ \\ \grayline
        $0.446$ & $26.4$ & $0.643$ & $1.18$ & $0.62$ & $39.5$ & $0.271$ \\ \grayline
        $0.231$ & $20.9$ & $0.565$ & $0.398$ & $1.76$ & $39.5$ & $0.277$ \\ \grayline
        $0.477$ & $21.4$ & $0.464$ & $0.965$ & $0.391$ & $39.4$ & $0.271$ \\
        \hline
        \end{tabularx}
    \end{table*}
    Note that the weights just determine the scaling of the loss component. To identify which losses actually contribute most to the optimization, the weighted loss components are evaluated separately during training on the validation data set and shown for the best-performing model in Figure~\ref{fig:Exp25Best}. The loss is dominated by the horizontal gradient component $L^{grad_{xy}}$ and the horizontal divergence $L^{div_{xy}}$ but also experiences contributions from the full divergence $L^{div}$. The pixel loss component $L^{pix}$ and vertical gradient component $L^{grad_z}$ have minor contributions in the early training stage, but quickly decay. The adversarial loss has barely any contributions throughout the 35k iterations and does not change except for minor fluctuations in the first 3k and last 500 epochs. 
    Subsequent testing showed that the steep drop in performance around 80k epochs of generator training is, in fact, caused by the discriminator loss. At 80k epochs, all other loss components are optimized to the point where the adversarial loss becomes the main component of the loss. Further investigation showed that this drop can be prevented by adjusting the hyperparameters of the discriminator, which will be discussed in more detail in Section~\ref{ssec:best}.

    \subsection{Evaluation of the best model}
    \label{ssec:best}
    \subsubsection{Hyperparameter adjustment}
    Based on the results of \textit{Experiment~1}, \textit{Experiment~2}, \textit{Experiment~3} and further testing, a number of changes were made to the hyperparameters of the model.
    \textit{Experiment~1} gives reason to remove the pressure from the generator input, while \textit{Experiment~2} and \textit{Experiment~3} motivate to adjust the learning rate and the loss-component weights. Most significantly, adversarial learning has been dropped. 
    While the degradation in training around 80k epochs reported earlier can be prevented by adjusting the alternation between generator and discriminator training from every epoch to every 50th epoch, changing the interval between the training of discriminator and generator, experimenting with turning on or off instance noise and label smoothing, adjusting the adversarial loss weight, and initiating with a generator pre-trained on content loss gave no significant increase in performance compared to the loss function with purely physically-motivated loss components. Therefore, the best model architecture in this work is the one without adversarial learning. Note that this means that the best super-resolution neural network in this work is a generative convolutional network, but not a generative adversarial neural network. In fact, it is a major result of this work that the physics-based loss components perform so well that including adversarial losses does not further improve performance.
    
    The other loss component weights are set according to the best-performing values from \textit{Experiment~3}, but normalized by a factor of 10, except for the $L^{grad_z}$ component, which had negligible contributions to the total loss, and the pixel loss weight $\eta_1$, which was set dynamically. The generator was first trained for 100k iterations with pixel loss $\eta_1=0.034$ and then trained for 150k iterations with $\eta_1=0.136$. Restarting with four times as high pixel loss not only improved pixel-wise error but also caused lower gradient-based losses. Furthermore, while improving the PSNR and pixel error, larger loss weights with constant learning rate result in larger gradients in the loss, i.e. larger steps per evaluation. However, this collides with the gradient clipping previously employed. In fact, clipping results in constant gradients for almost all evaluations throughout the full training, which causes a significantly slower parameter change than the gradients suggest. Therefore, the learning rate here was increased by a factor of 8, and gradient clipping was dropped. The full set of modified generator parameters is listed in the appendix in Table~\ref{tab:best-setup}. Otherwise, the generator setup is unchanged from the full list of hyperparameters in the appendix in Table~\ref{tab:full-fyperparameters}.
    
    \subsubsection{Metrics and Overall Performance}
\begin{figure}
        \begin{subfigure}[b]{0.475\linewidth}
            \includegraphics[width=\linewidth]{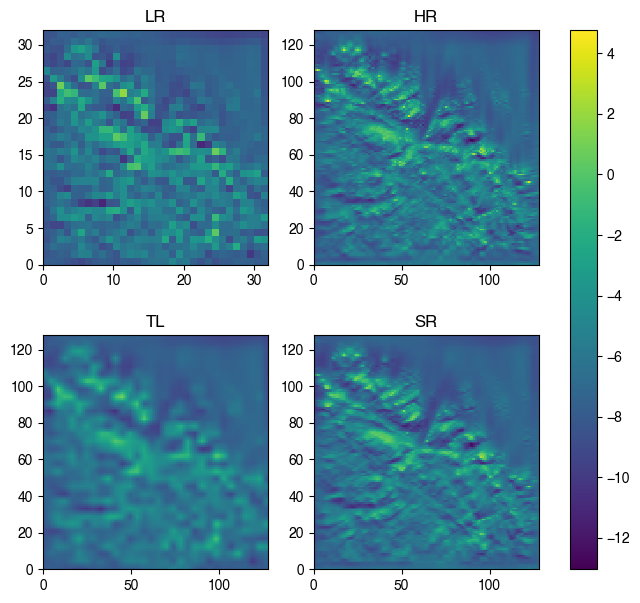}
            \caption{Wind velocity along the $x$-axis}
        \end{subfigure}
        \begin{subfigure}[b]{0.475\linewidth}
            \includegraphics[width=\linewidth]{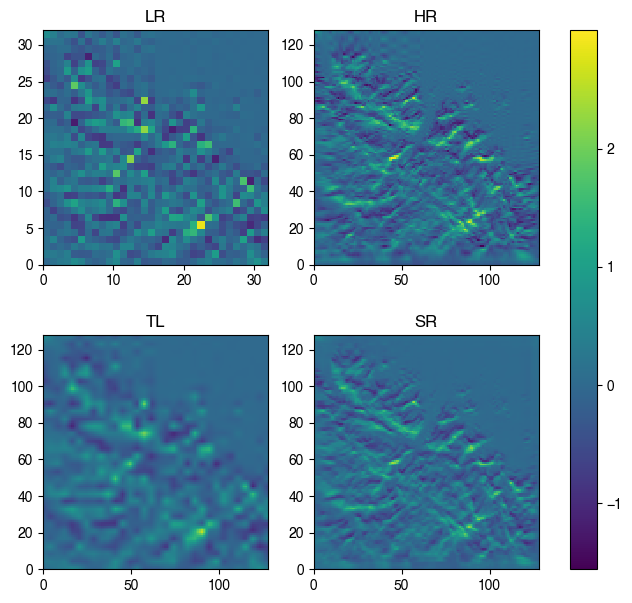}
            \caption{Wind velocity along the $z$-axis}
        \end{subfigure} \hfill
        \begin{subfigure}[b]{0.475\linewidth}
            \includegraphics[width=\linewidth]{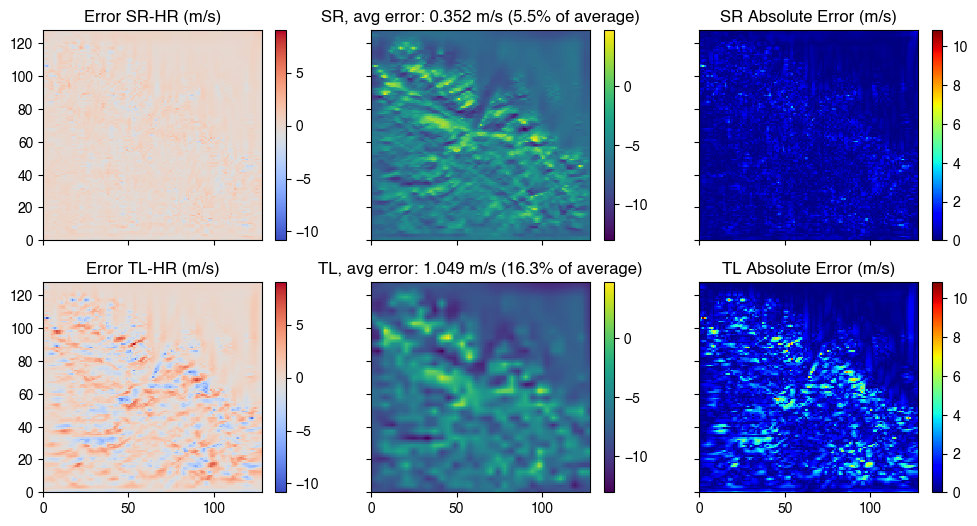}
            \caption{Error for the wind component pointing along the $x$-axis}
        \end{subfigure}
        \begin{subfigure}[b]{0.475\linewidth}
            \includegraphics[width=\linewidth]{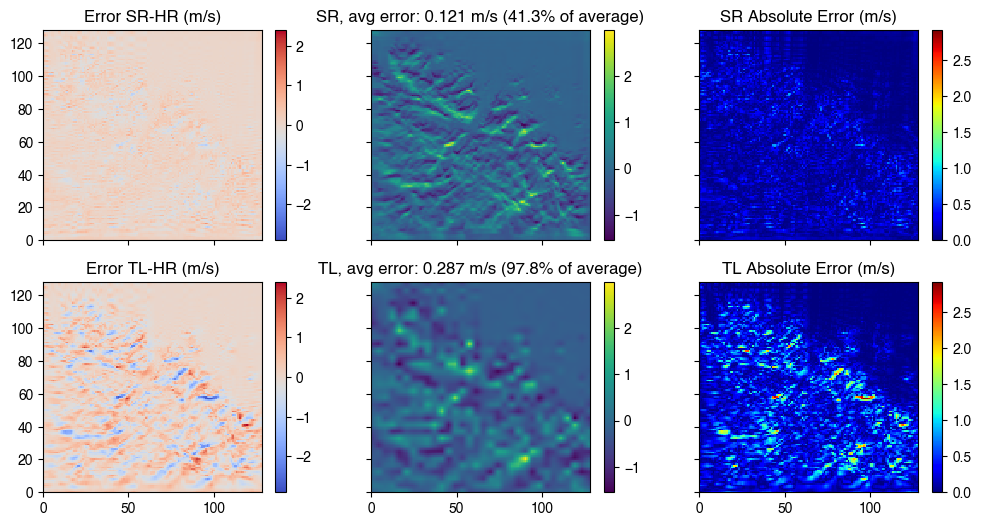}
            \caption{Error for the wind component pointing along the $z$-axis}
        \end{subfigure}
        \caption[2D Best Model Results 1]{Comparison of the second lowest horizontal 2D slice of the wind field for a randomly selected wind field. LR means low resolution, the input of the model, TL means trilinear interpolation of the LR data, SR means the super-resolved wind field generated by the trained network and HR means the true high-resolution wind field. ``Avg error" refers to the average absolute error for the displayed wind component in the displayed slice, the "\% of average" means this value divided by the average value of that wind component in that slice for the HR wind field.}
        \label{fig:u0w0}
    \end{figure}

    The average test scores of the final model are compared against the trilinear interpolation in Table~\ref{tab:results-best_model}. With the adjusted parameters, the generator achieves even higher PSNR and lower pixel error. Additionally, the absolute error of the vector length in 3D, i.e. the "pixel vector" of the vector length, pix-vector, and the pix-vector relative to the maximum wind speed in the sample are shown. The generator achieves significantly better results than the trilinear interpolation on all metrics.
    \begin{table}
        \caption[Test result of the final model]{Test result of the final model. Metrics are defined in Section~\ref{subsec:EvaluationMetrics}.
        Though being interpolated, the generated wind field is not interpolated back and compared to the original data as in Table~\ref{tab:results-exp1}, as this did not affect the results significantly. 
        Trilinear interpolation performance is included for comparison.}
        \label{tab:results-best_model}
        \centering
        \begin{tabularx}{\linewidth}{L|RR}
            \textbf{Interpolation} & Trilinear & Generator  \\ \hline
            \textbf{PSNR (db)} & 36.35 & 47.14 \\ \grayline
            \textbf{pix (m/s)} & 0.385 & 0.116 \\ \grayline
            \textbf{pix-vector (m/s)} & 0.80 & 0.24 \\ \grayline
            \textbf{pix-vector relative} & 18.5\% & 6.12\%\\ \hline
        \end{tabularx}
    \end{table}
    In Figure~\ref{fig:u0w0} the second lowest horizontal layer of the 3D wind field can be seen in low resolution (LR), high resolution (HR), trilinear interpolated (TL), and super-resolved by the generator (SR). The generator adds finer details to the reconstructed image that the trilinear interpolation cannot capture and includes information from the underlying terrain. More importantly, the relative and absolute error of the wind in both horizontal and vertical directions is reconstructed significantly better by the generator. The same figure for the second-highest horizontal layer of the wind field can be found in the appendix as Figure~\ref{fig:u9w9}.

    \subsubsection{Generated 3D Wind Field}
\begin{figure}[h]
        \begin{subfigure}{0.48\linewidth}
            \includegraphics[width=\linewidth]{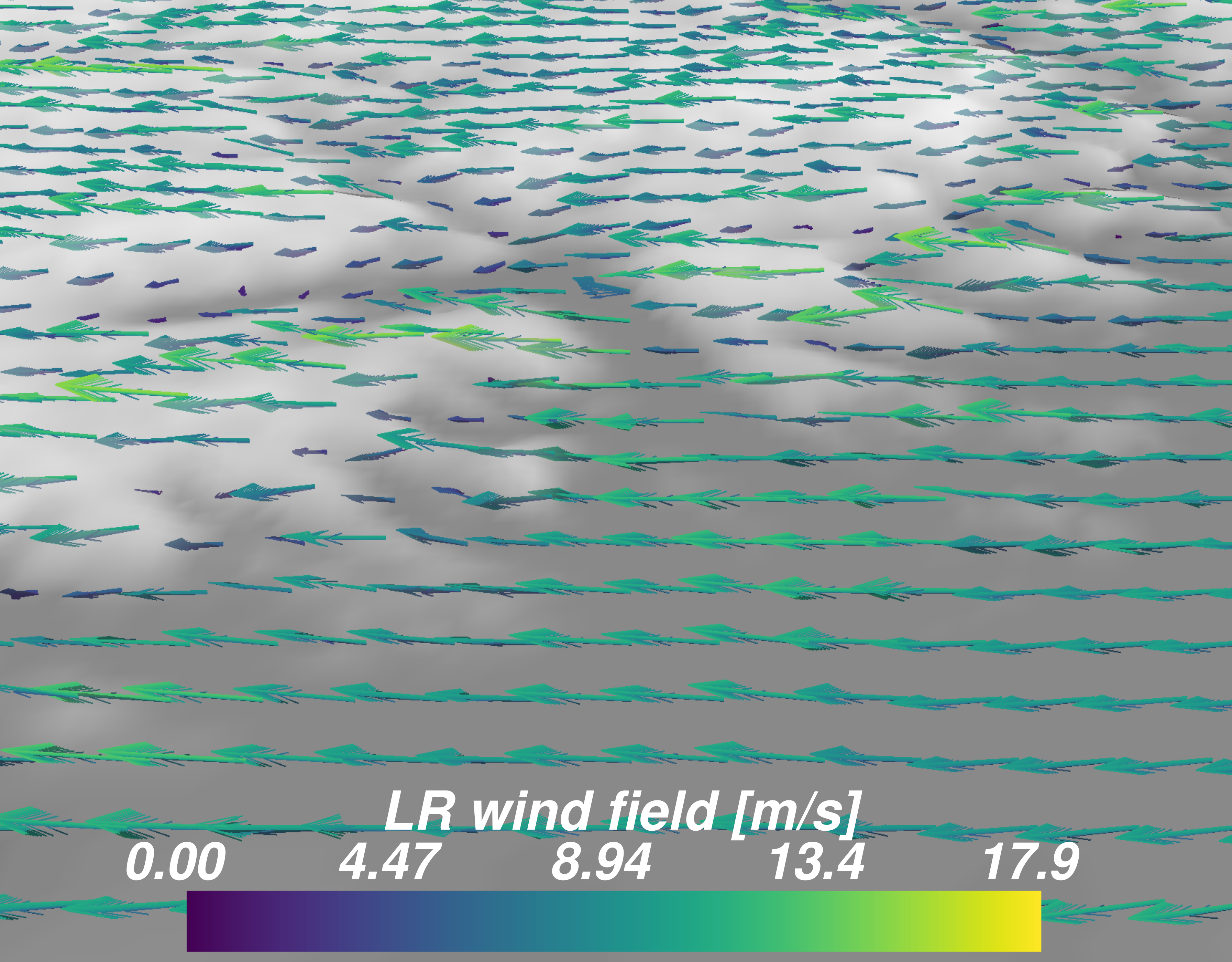}
            \caption{LR wind field}
        \end{subfigure}
        \begin{subfigure}{0.48\linewidth}
            \includegraphics[width=\linewidth]{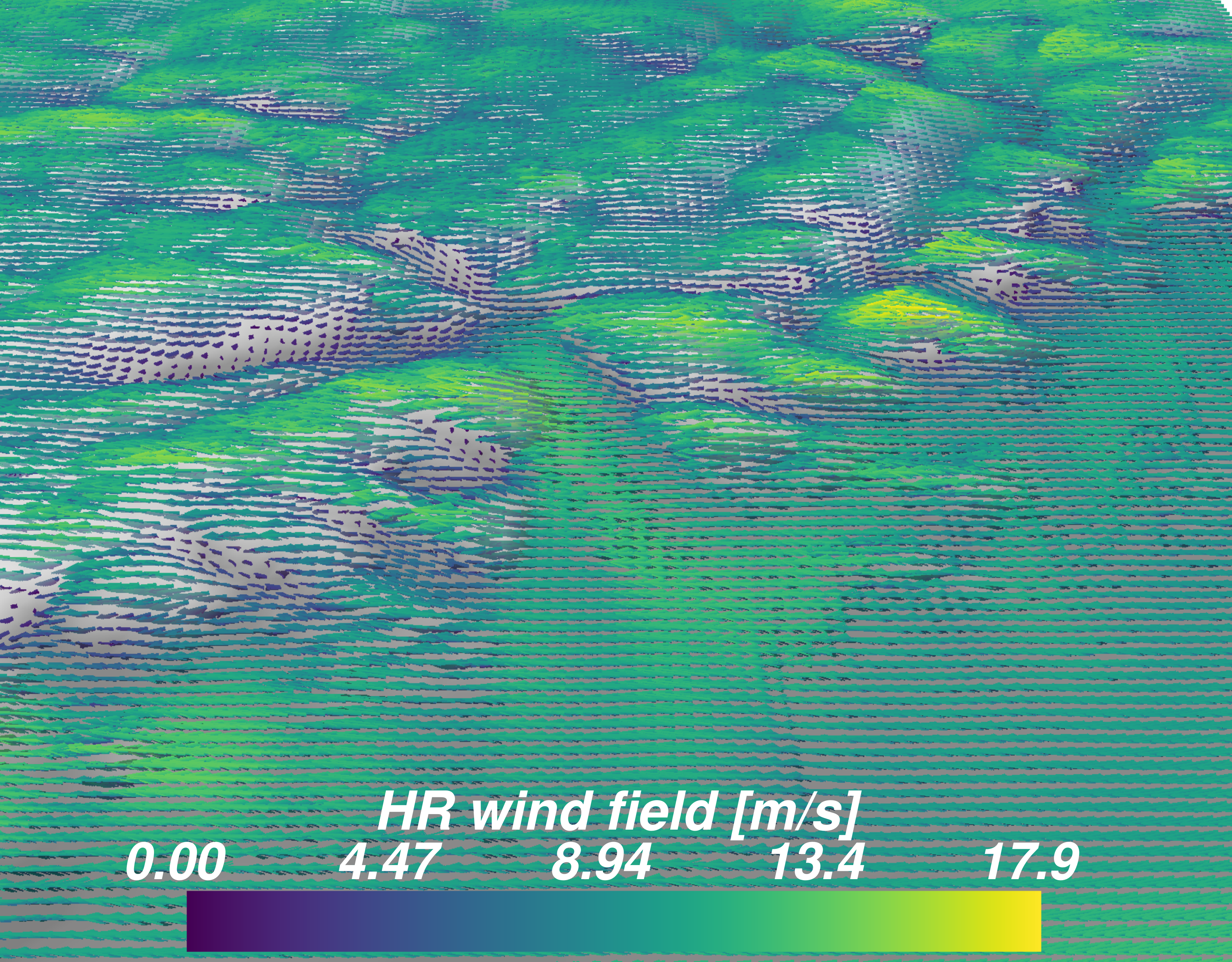}
            \caption{HR wind field}
        \end{subfigure}\\
        \begin{subfigure}{0.48\linewidth}
            \includegraphics[width=\linewidth]{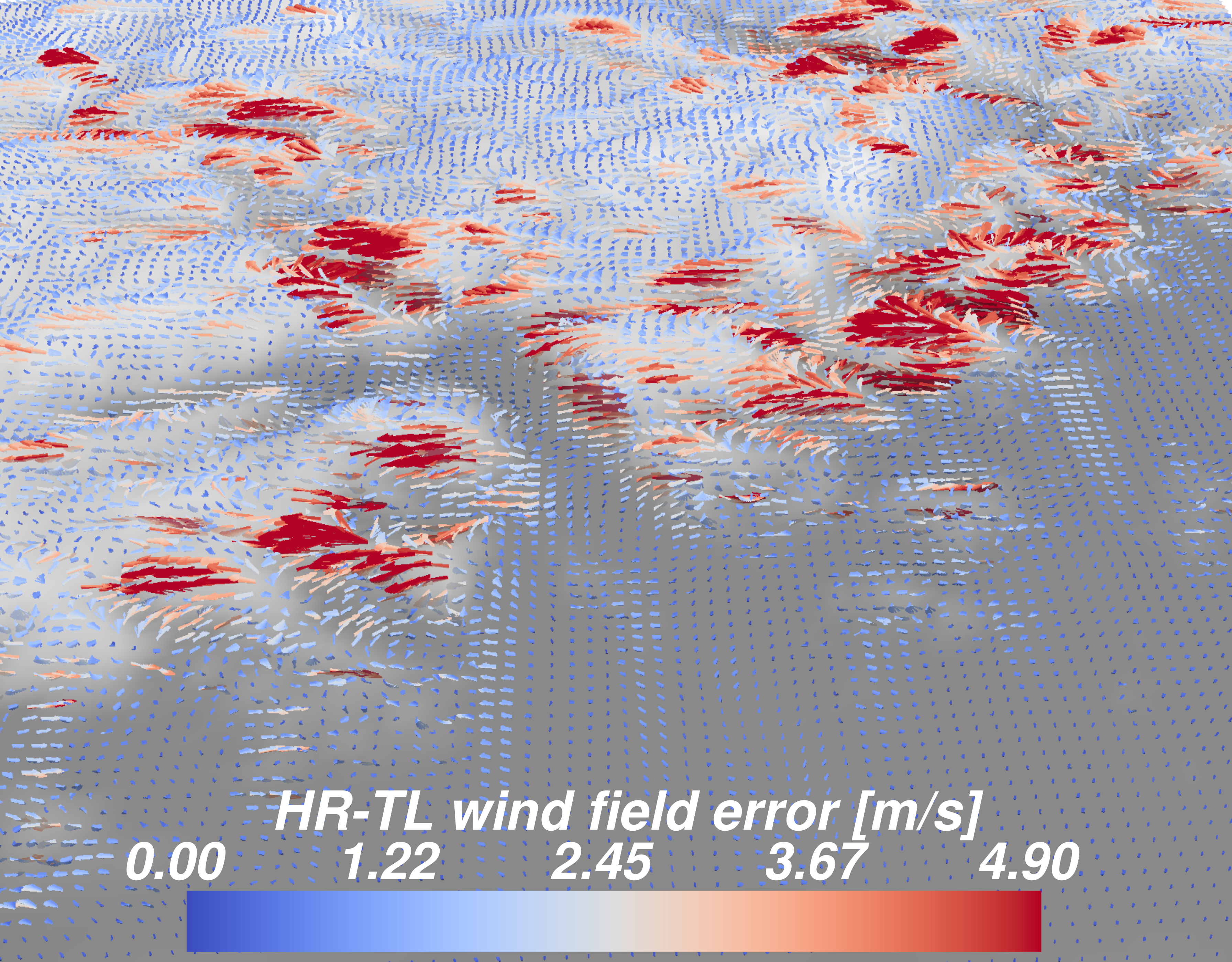}
            \caption{TL error, HR-TL}
            \label{subfig:3Derror_TL}
        \end{subfigure}
        \begin{subfigure}{0.48\linewidth}
            \includegraphics[width=\linewidth]{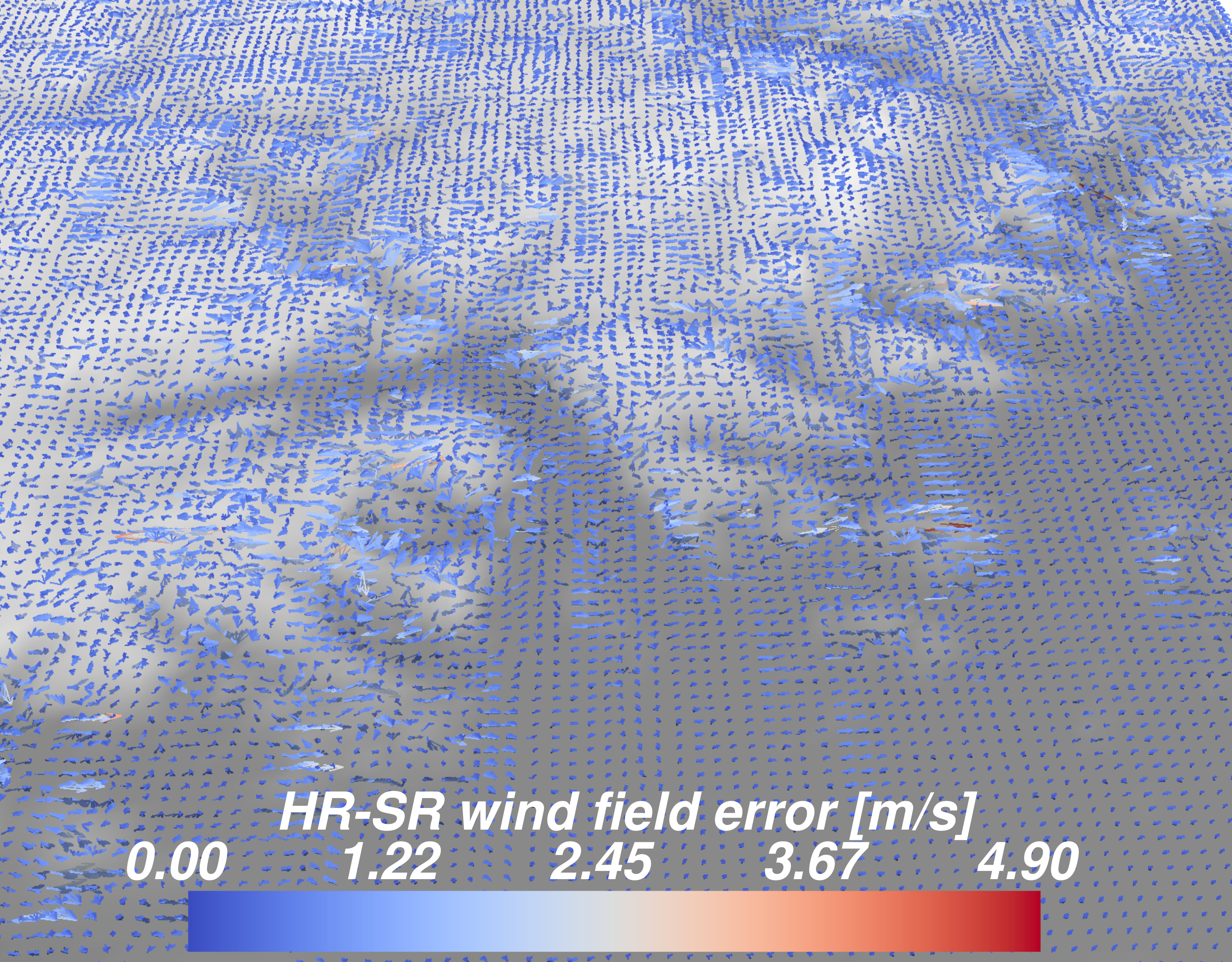}
            \caption{SR error, HR-SR}
            \label{subfig:3Derror_SR}
        \end{subfigure}\\
        \begin{subfigure}{0.48\linewidth}
            \includegraphics[width=\linewidth]{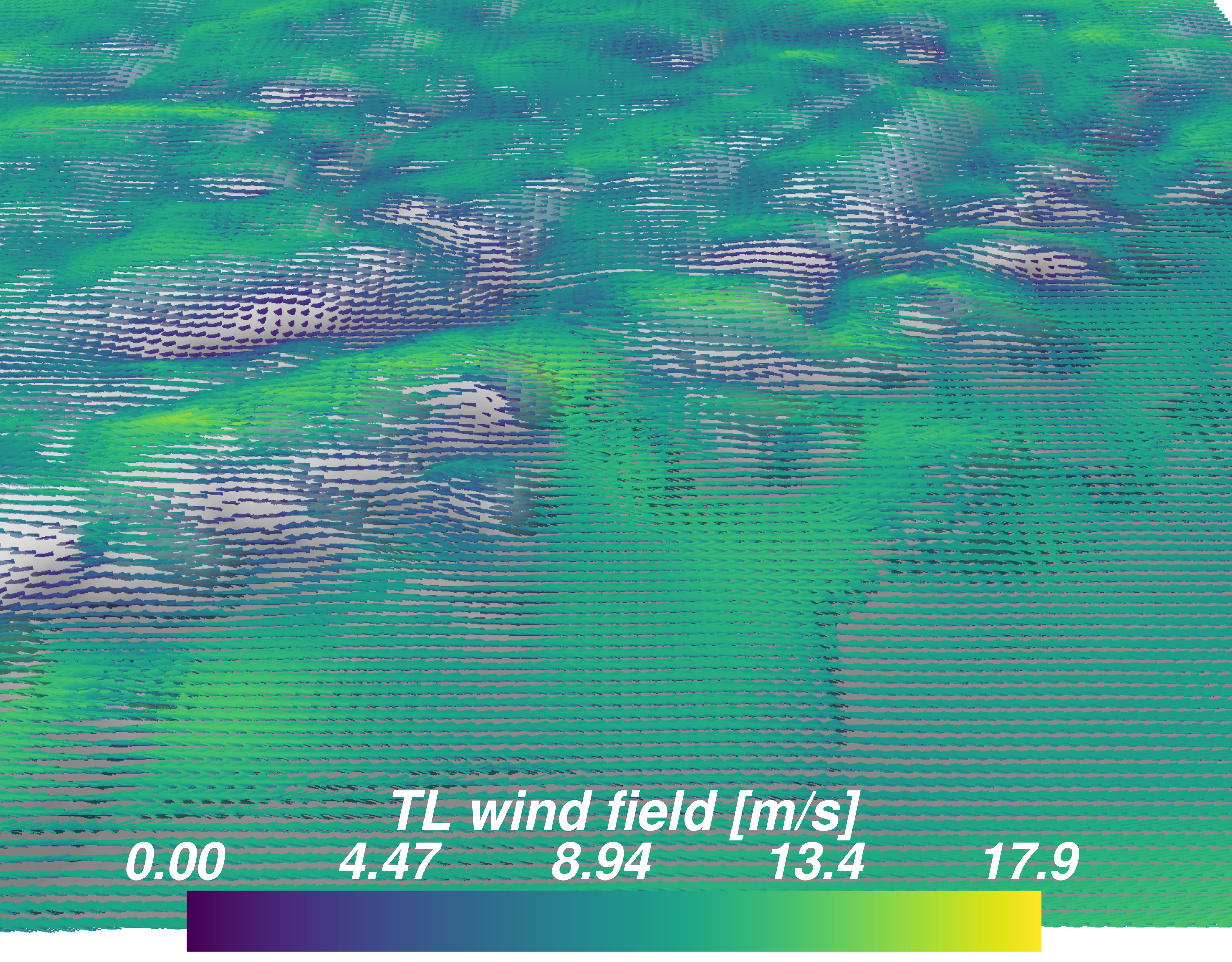}
            \caption{TL wind field}
        \end{subfigure}
        \begin{subfigure}{0.48\linewidth}
            \includegraphics[width=\linewidth]{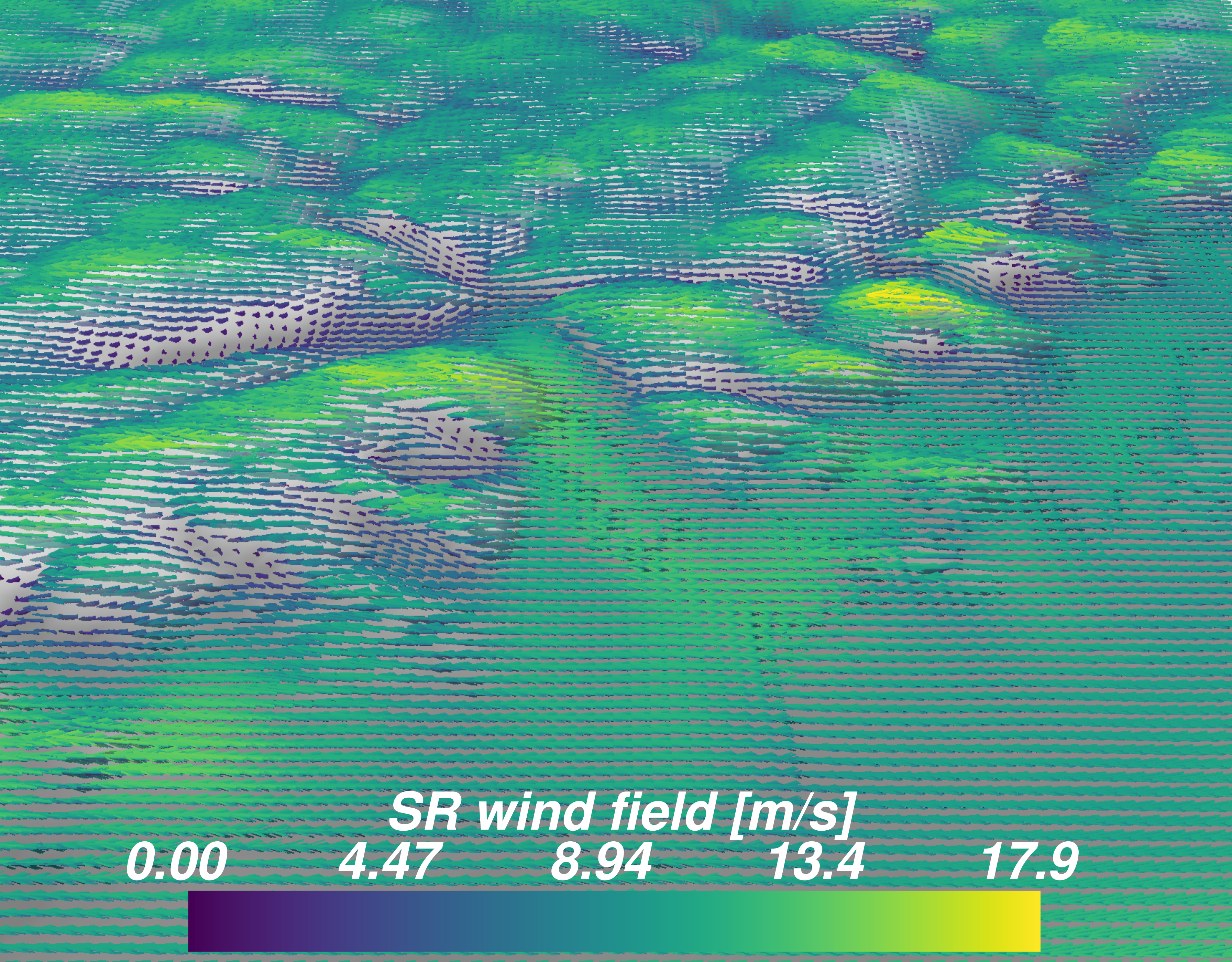}
            \caption{SR wind field}
        \end{subfigure} 
        \caption[3D Generated Wind Fields From Best Model 1]{Comparison of LR, HR, TL, and SR wind fields from a randomly selected sample with high average wind speed. LR means low resolution, the input of the model, TL means trilinear interpolation of the LR data, SR means the super-resolved wind field generated by the trained network and HR means the true high-resolution wind field.}
        \label{fig:overview3D}
    \end{figure}

    Next, the entire wind field is compared in 3D.
    In Figure~\ref{fig:overview3D} a sample with a mean wind speed at 8.2 m/s is investigated. For this wind field, the average length of the error wind vector of the SR field is 0.41 m/s (5\% of average) and the average absolute error of the TL field is 1.34 m/s (16\% of average). In figure Figure~\ref{fig:overview3D} it can be seen how the generated wind fields adapt to the terrain, comparing LR, HR, TL, and SR data, as before. 
    
\begin{figure}[h]
      %  \centering
        \begin{subfigure}[b]{0.49\linewidth}
            \includegraphics[width=\linewidth]{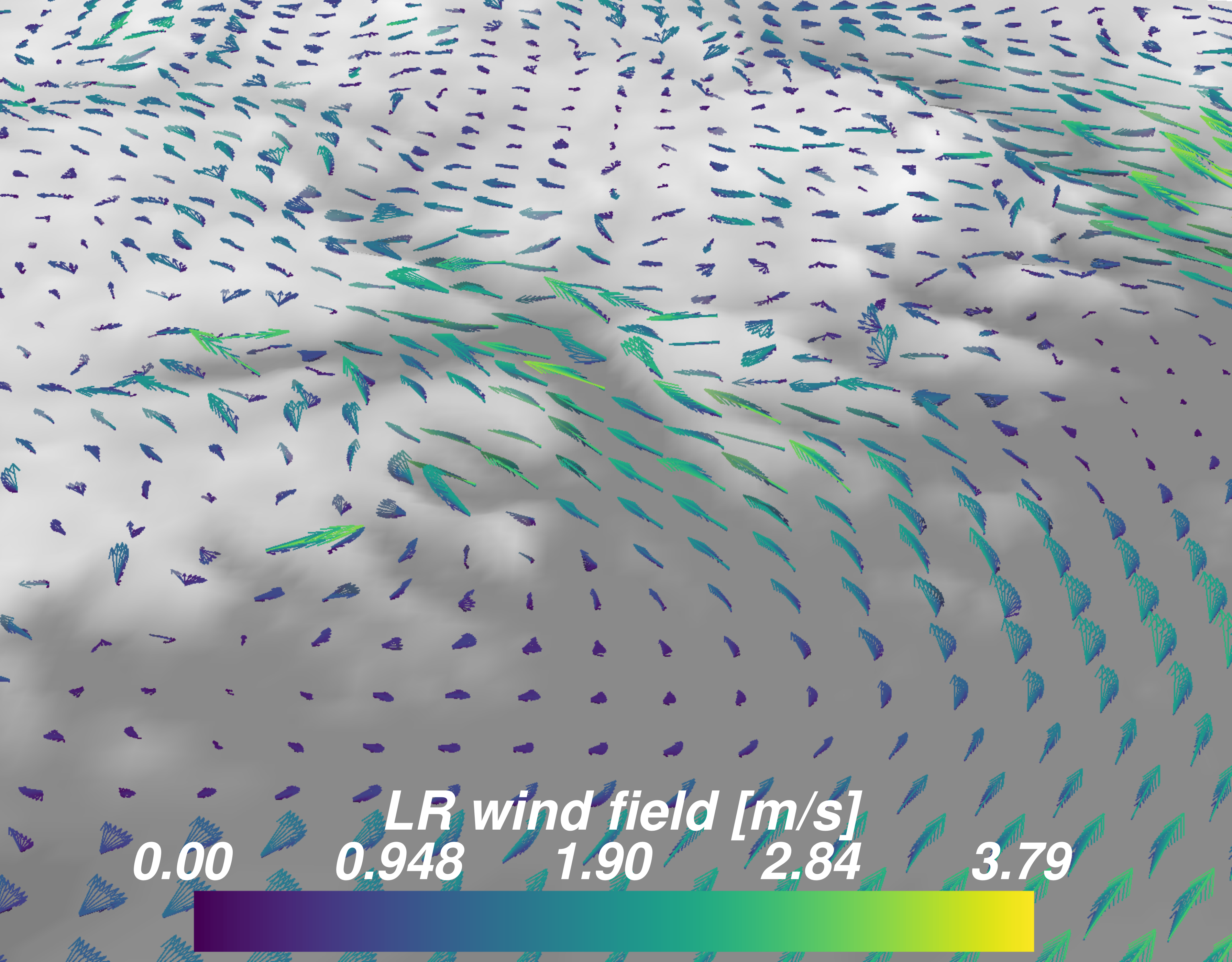}
            \caption{LR wind field}
        \end{subfigure}
        \begin{subfigure}[b]{0.49\linewidth}
            \includegraphics[width=\linewidth]{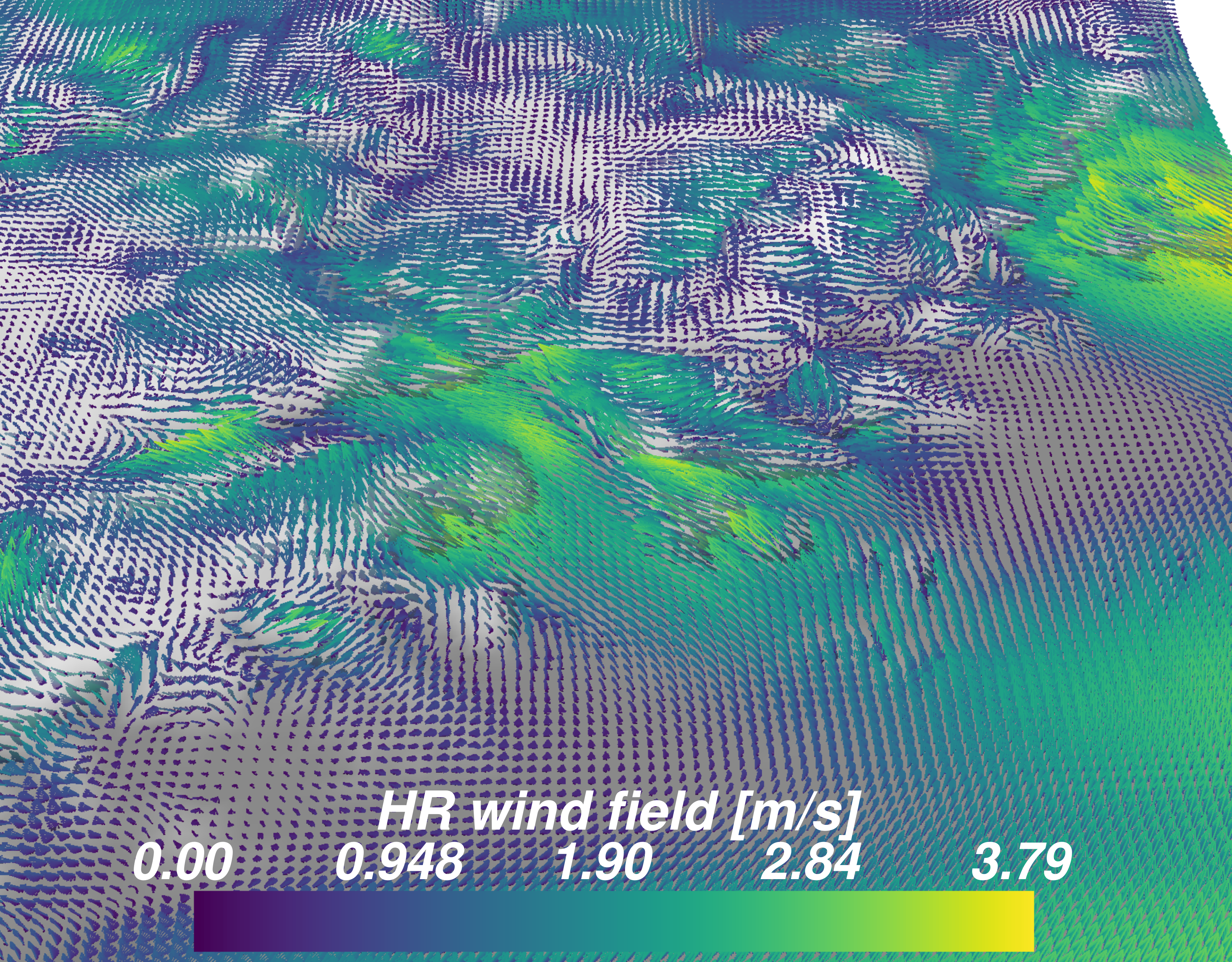}
            \caption{HR wind field}
        \end{subfigure}\\
        \begin{subfigure}[b]{0.49\linewidth}
            \includegraphics[width=\linewidth]{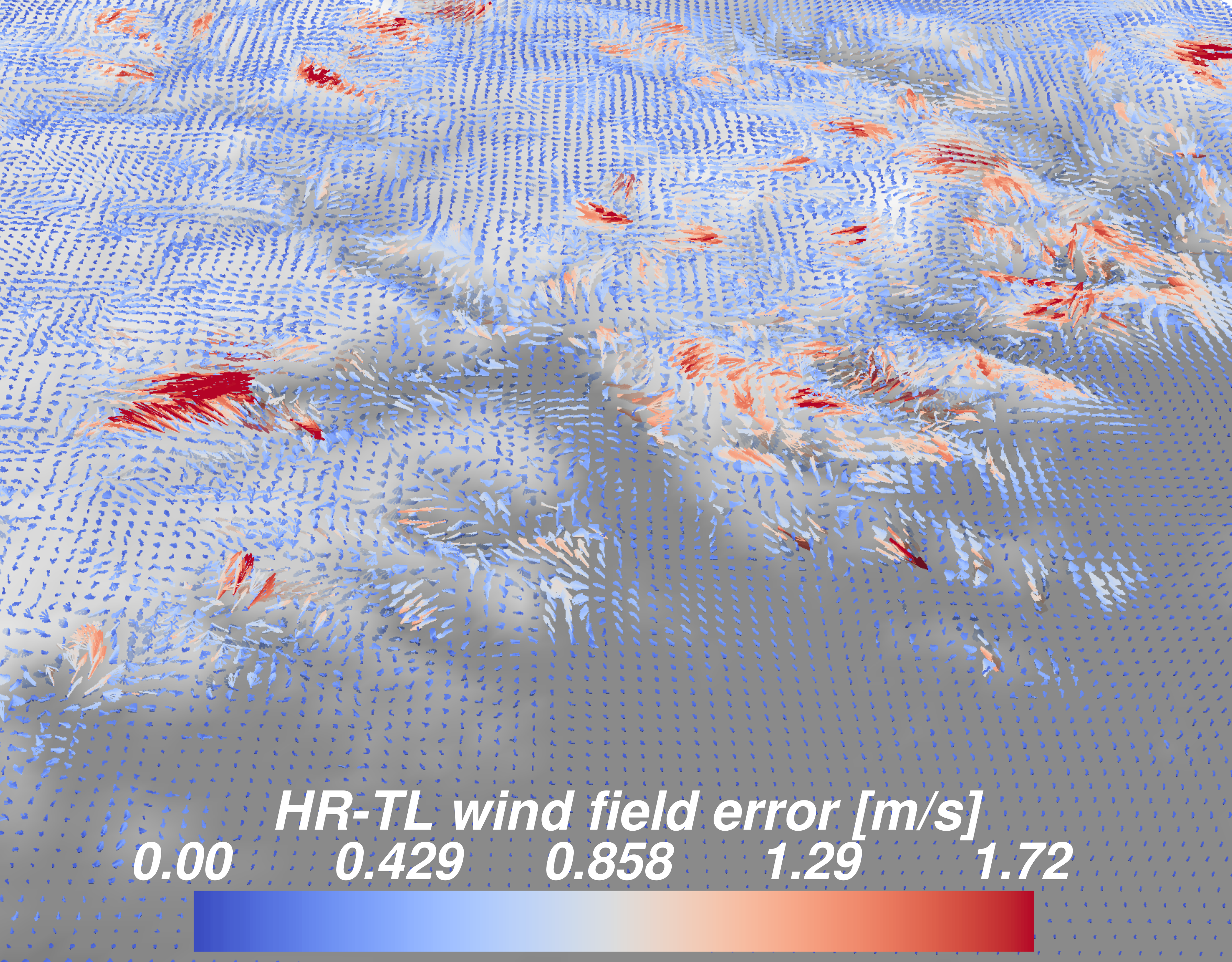}
            \caption{TL error, HR-TL}
        \end{subfigure}
        \begin{subfigure}[b]{0.49\linewidth}
            \includegraphics[width=\linewidth]{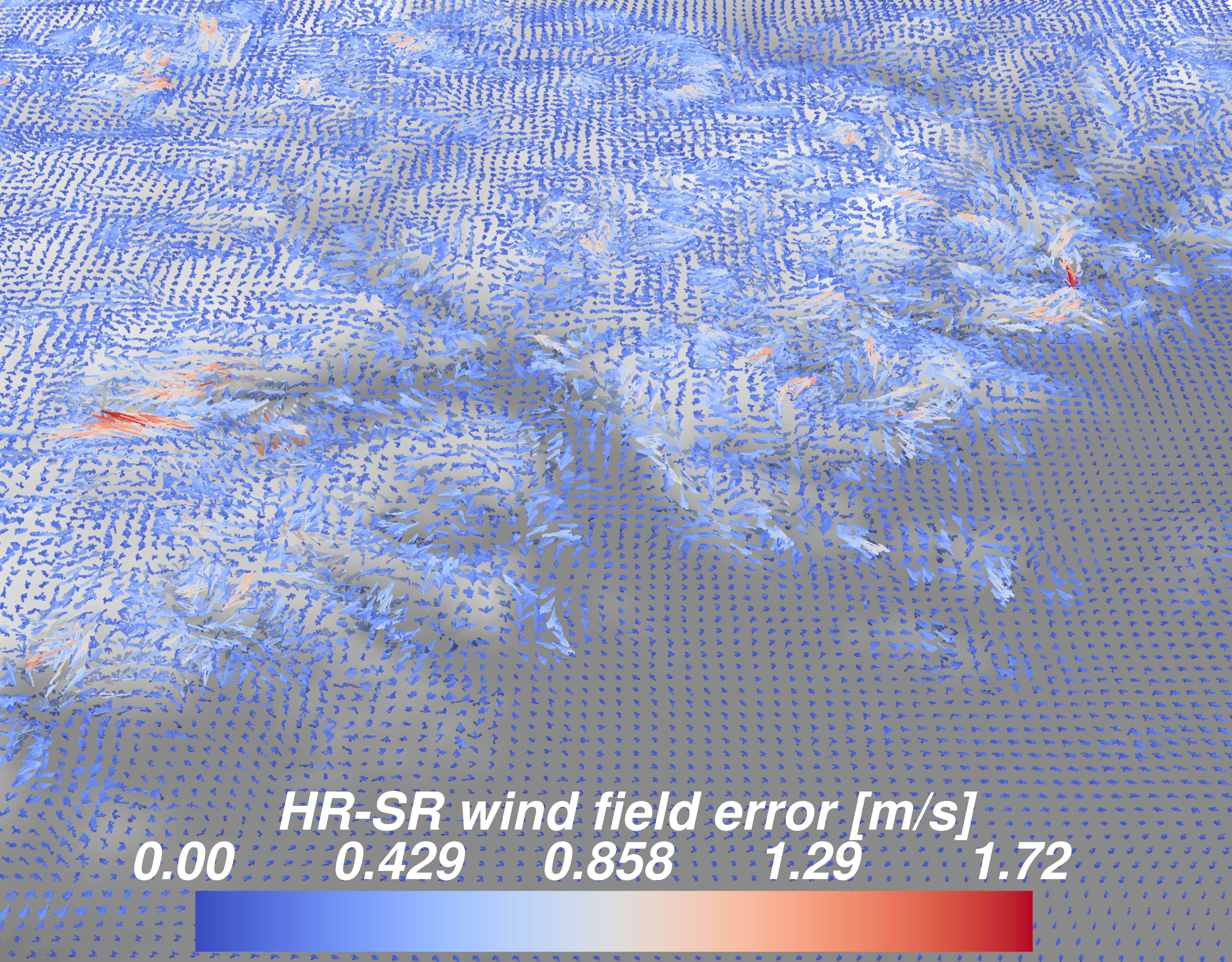}
            \caption{SR error, HR-SR}
            \label{subfig:slow3DErr-SR}
        \end{subfigure}\\
        \begin{subfigure}[b]{0.49\linewidth}
            \includegraphics[width=\linewidth]{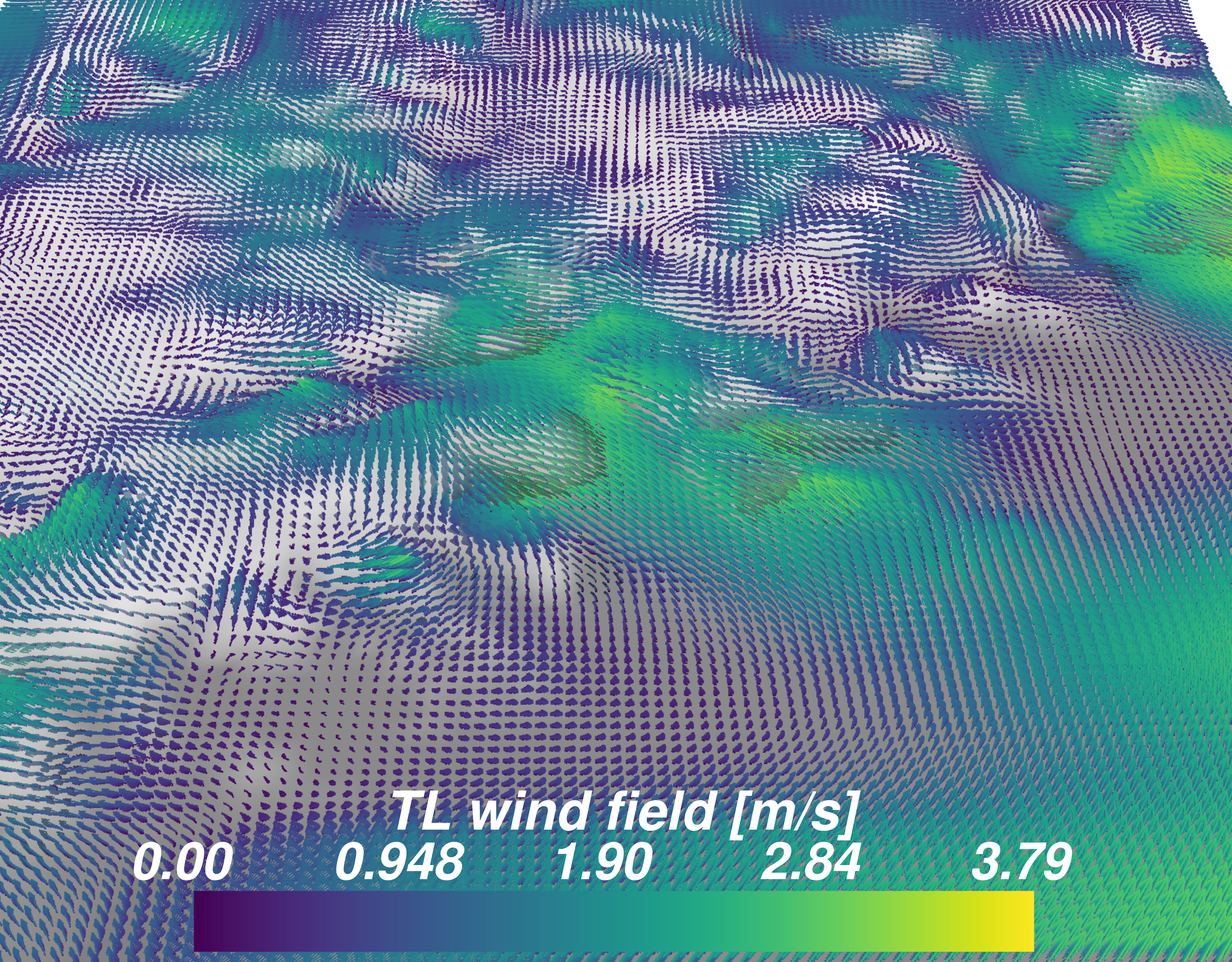}
            \caption{TL wind field}
            \label{subfig:slow3DErr-TL}
        \end{subfigure}
        \begin{subfigure}[b]{0.49\linewidth}
            \includegraphics[width=\linewidth]{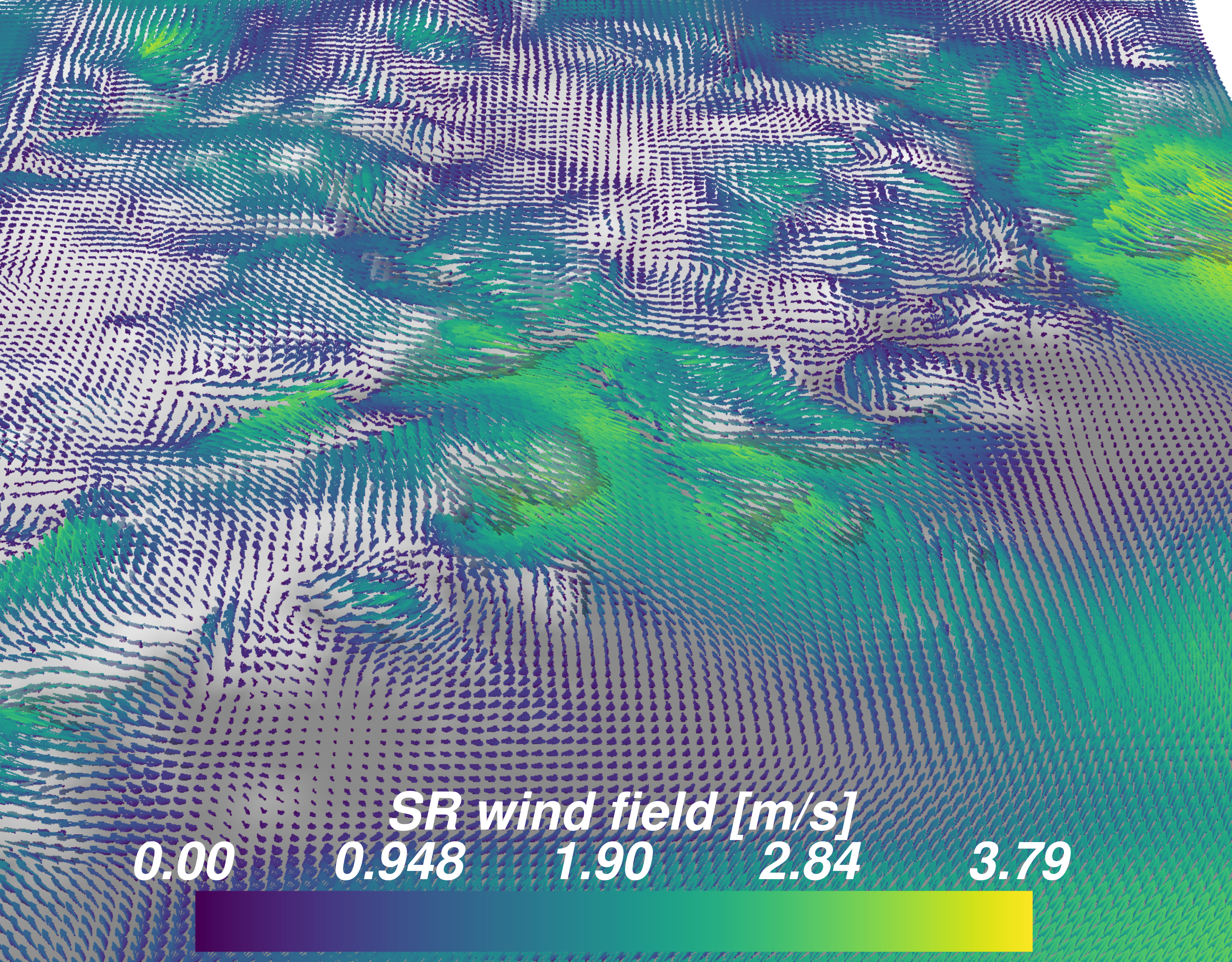}
            \caption{SR wind field}
        \end{subfigure} 
        \caption[3D Generated Wind Fields From Best Model 2]{Comparison of LR, HR, TL, and SR wind fields from a randomly selected sample with low average wind speed. LR means low resolution, the input of the model, TL means trilinear interpolation of the LR data, SR means the super-resolved wind field generated by the trained network and HR means the true high-resolution wind field.}
        \label{fig:overview_slow_3D}
    \end{figure}

    Figure~\ref{fig:overview3D} demonstrates that the super-resolved wind field adapts much more crisply to the terrain than the interpolated data. The trilinear interpolation error plot in Figure~\ref{subfig:3Derror_TL} follows regular terrain patterns, with the interpolation systematically misjudging regions of rapidly changing terrain. Meanwhile, Figure~\ref{subfig:3Derror_SR} captures even fast-changing terrain patterns well, and the smaller error is more randomly distributed, indicating that the generator performs well regardless of terrain features. The model, however, is trained to minimize absolute error in gradients and component-wise wind speed, which means that for any batch during training, the larger wind fields dominate the loss function. Therefore, Figure~\ref{fig:overview_slow_3D} shows a sample with an average wind speed of 1.04 m/s. 
    For this area, the average absolute error is 0.19 m/s (18\% of average speed) for the generated super-resolution win field and 0.31 m/s (30\% of average speed) for the interpolated wind field, i.e. a lower absolute error but a higher relative error for trilinear interpolation and generated super-resolution wind field.
    Comparing Figure~\ref{subfig:slow3DErr-SR} and Figure~\ref{subfig:slow3DErr-TL} more closely, it can be seen that the super-resolution field is performing worse than trilinear interpolation in quite large areas of low wind speed but still, the maximum and average error is smaller than that of triangular interpolation.
    It can be concluded that the model's emphasis on absolute error makes it perform worse on very weak wind fields. However, this is desirable as the absolute error is generally more important in the context of wind fields.
    
    Figure~\ref{fig:overview_slow_3D} also demonstrates a problem with the vertical cutoff of the training data. In the figure, there are large areas in which the wind seems to stop completely. Here the wind is pushed upwards by the terrain. With terrain-following coordinates only up to 40m above ground level and weak winds, the data doesn't cover the wind continuing higher up, instead making it seem like the wind only blows at the top of hills. Given that this work tries to push the model towards learning traits like mass conservation, this is not ideal. Extending the input data to higher altitudes above ground may therefore improve the performance.
    
\begin{figure}[h]
    %    \centering
        \begin{subfigure}[b]{0.48\linewidth}
            \includegraphics[width=\linewidth]{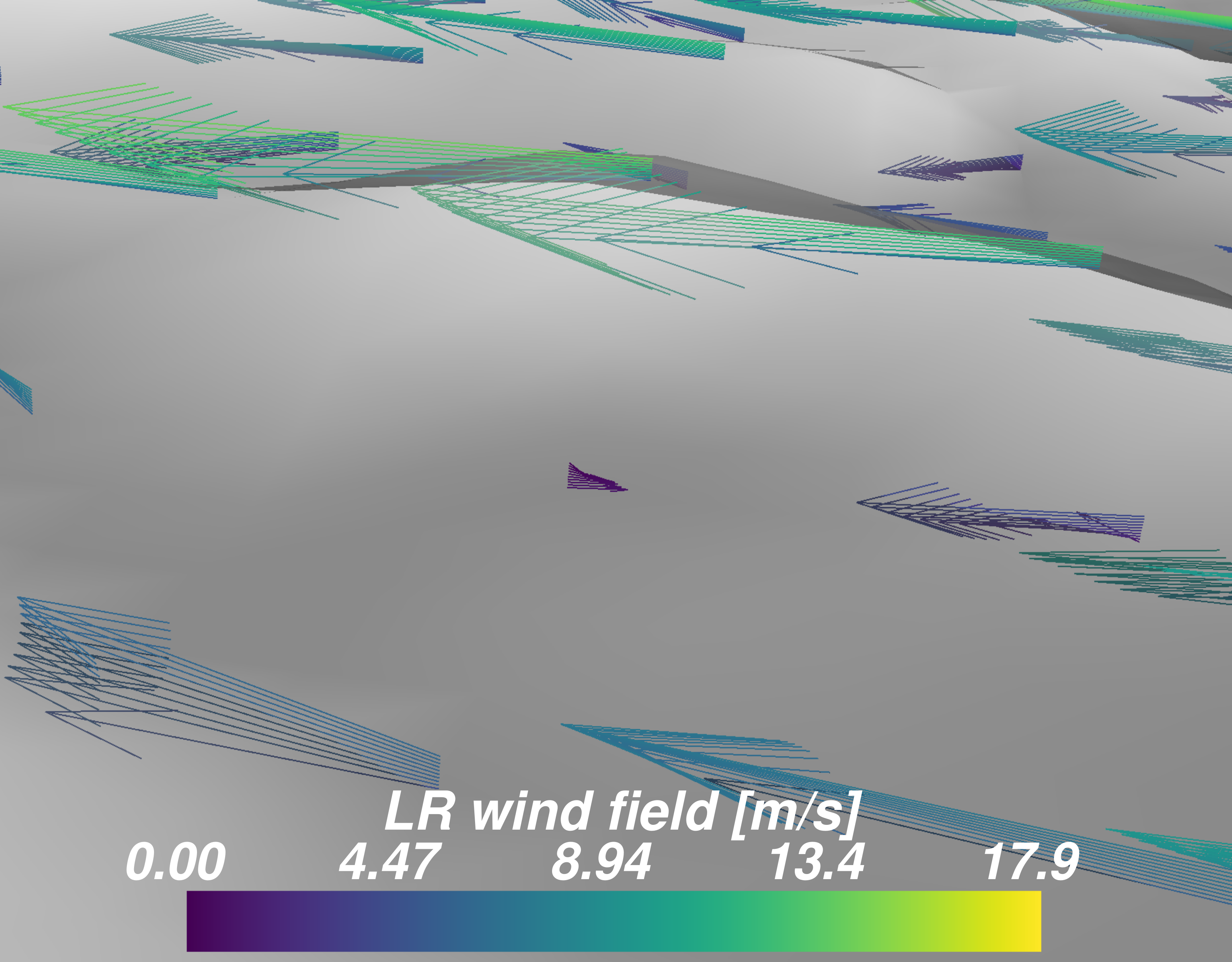}
            \caption{LR wind field}
            \label{subfig:turb_LR}
        \end{subfigure}
        \begin{subfigure}[b]{0.48\linewidth}
            \includegraphics[width=\linewidth]{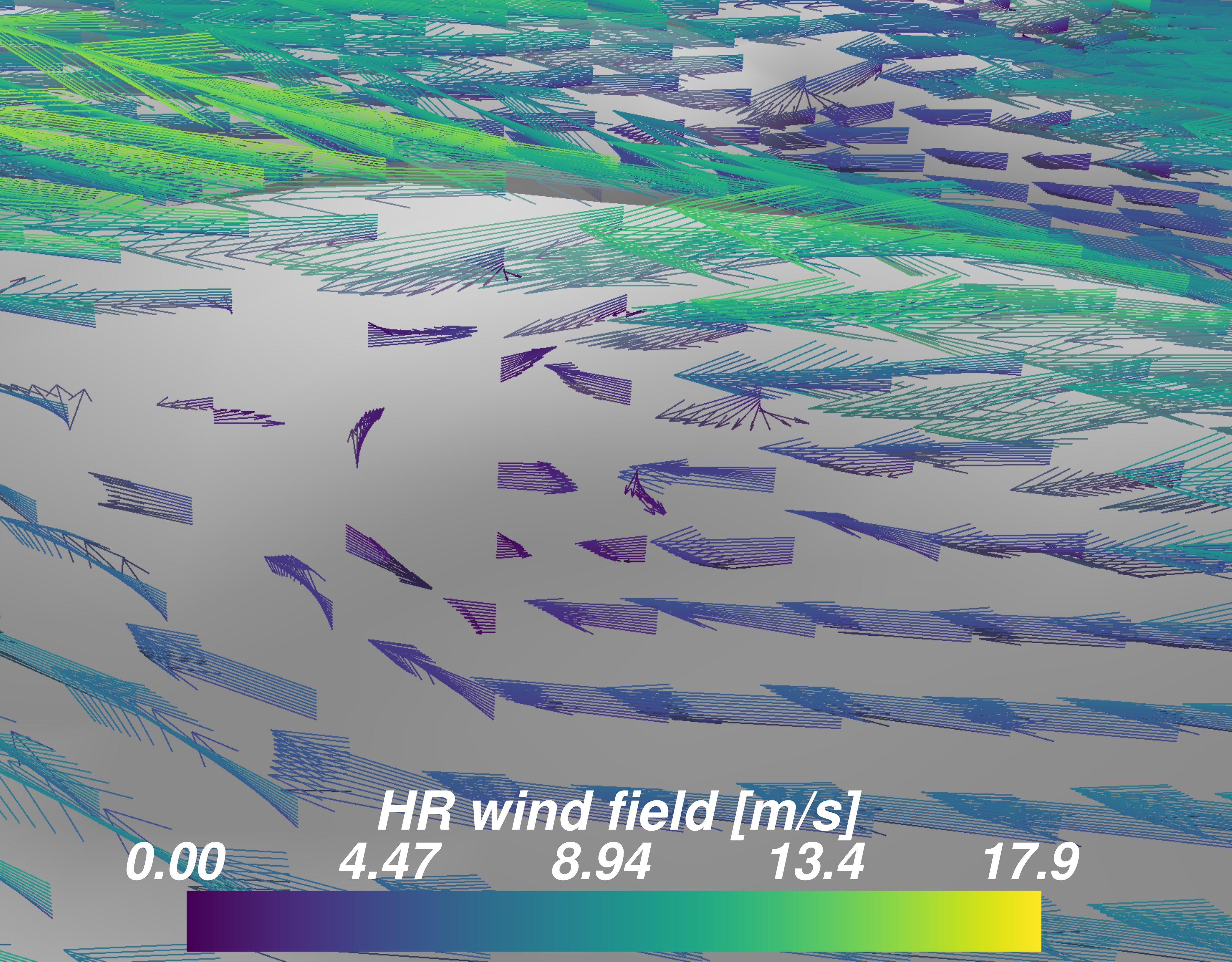}
            \caption{HR wind field}
            \label{subfig:turb_HR}
        \end{subfigure} \\
        \begin{subfigure}[b]{0.48\linewidth}
            \includegraphics[width=\linewidth]{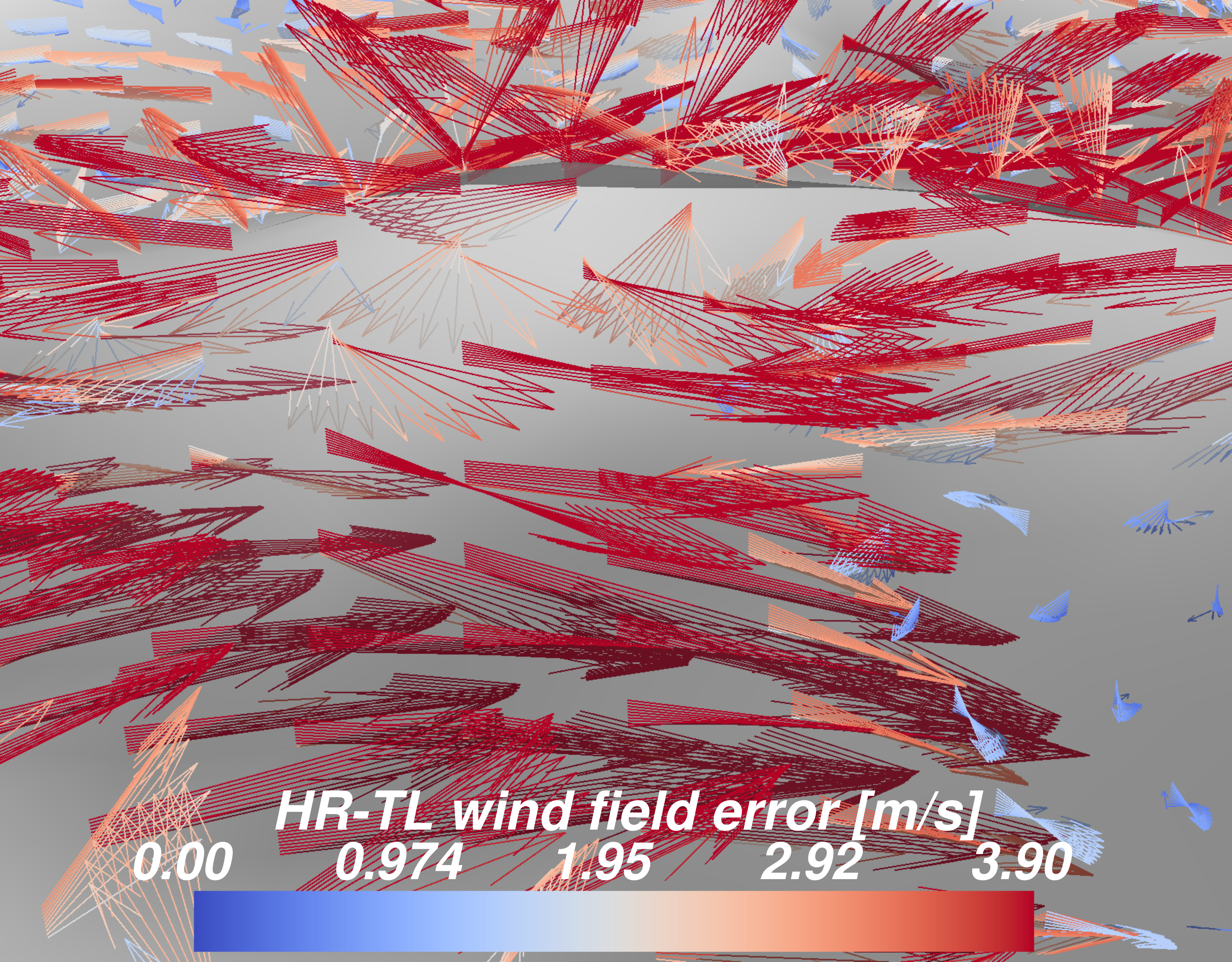}
            \caption{TL error, HR-TL}
            \label{subfig:turb_error_TL}
        \end{subfigure}
        \begin{subfigure}[b]{0.48\linewidth}
            \includegraphics[width=\linewidth]{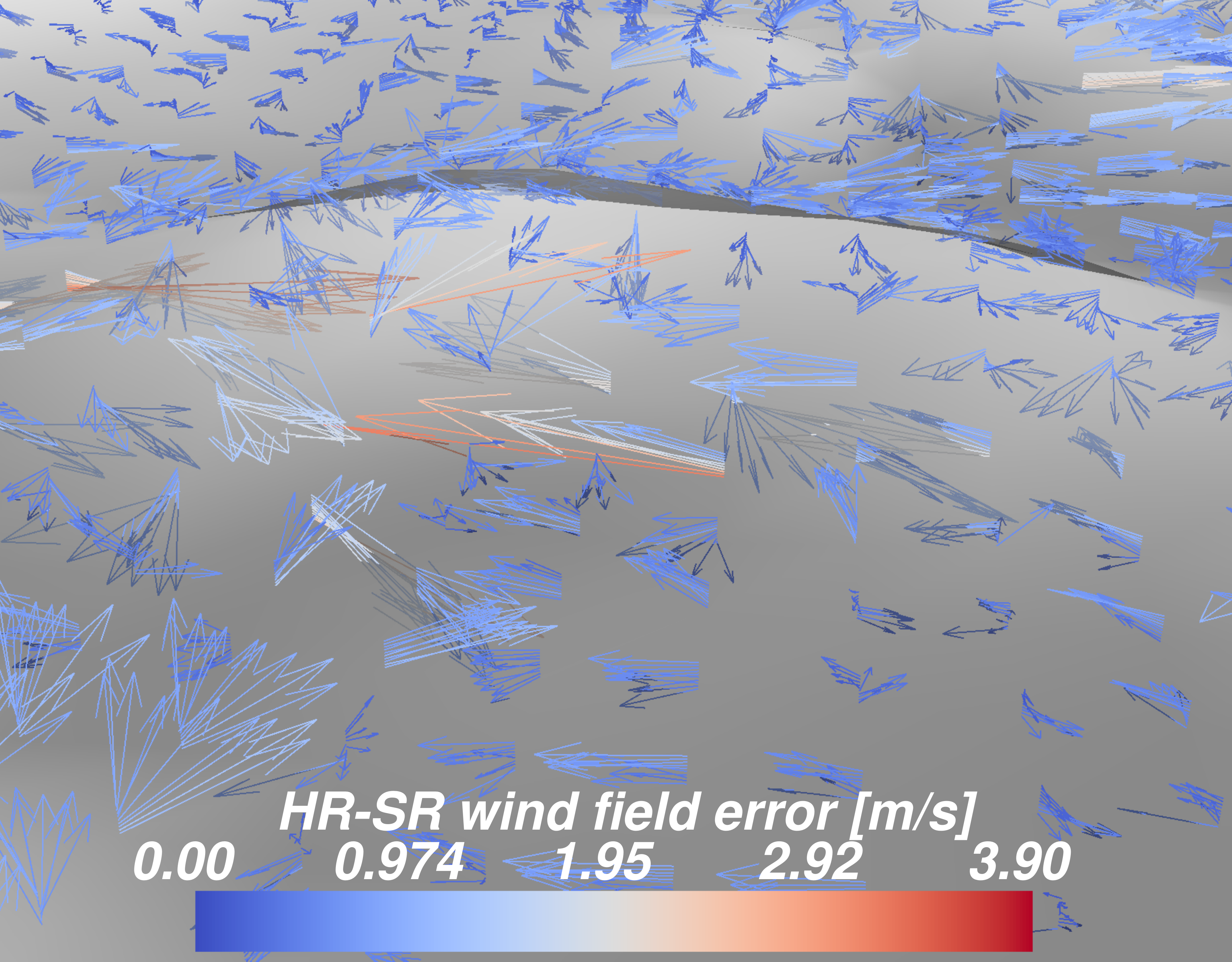}
            \caption{SR error, HR-SR}
            \label{subfig:turb_error_SR}
        \end{subfigure}\\ 
        \begin{subfigure}[b]{0.48\linewidth}
            \includegraphics[width=\linewidth]{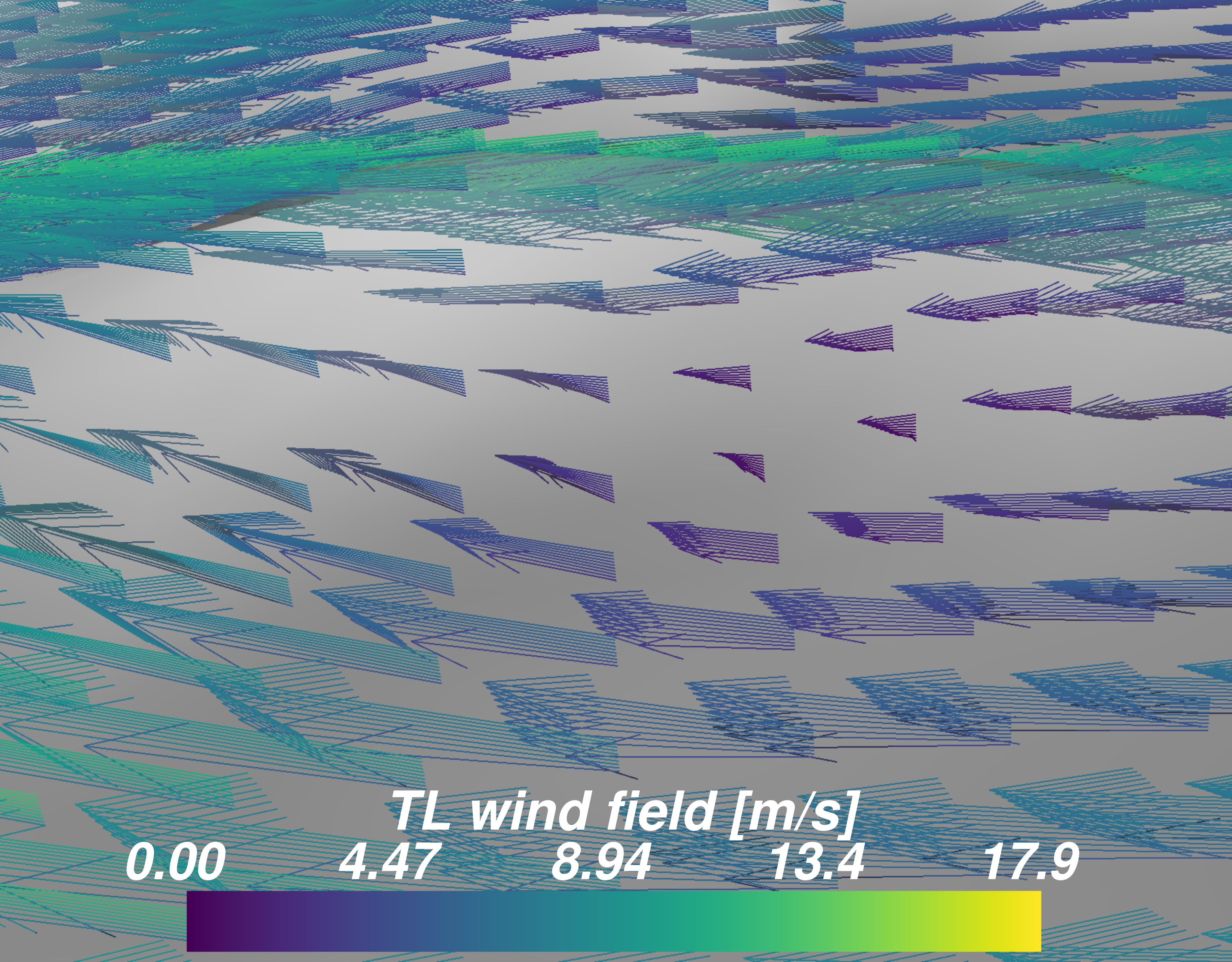}
            \caption{TL wind field}
        \end{subfigure}
        \begin{subfigure}[b]{0.48\linewidth}
            \includegraphics[width=\linewidth]{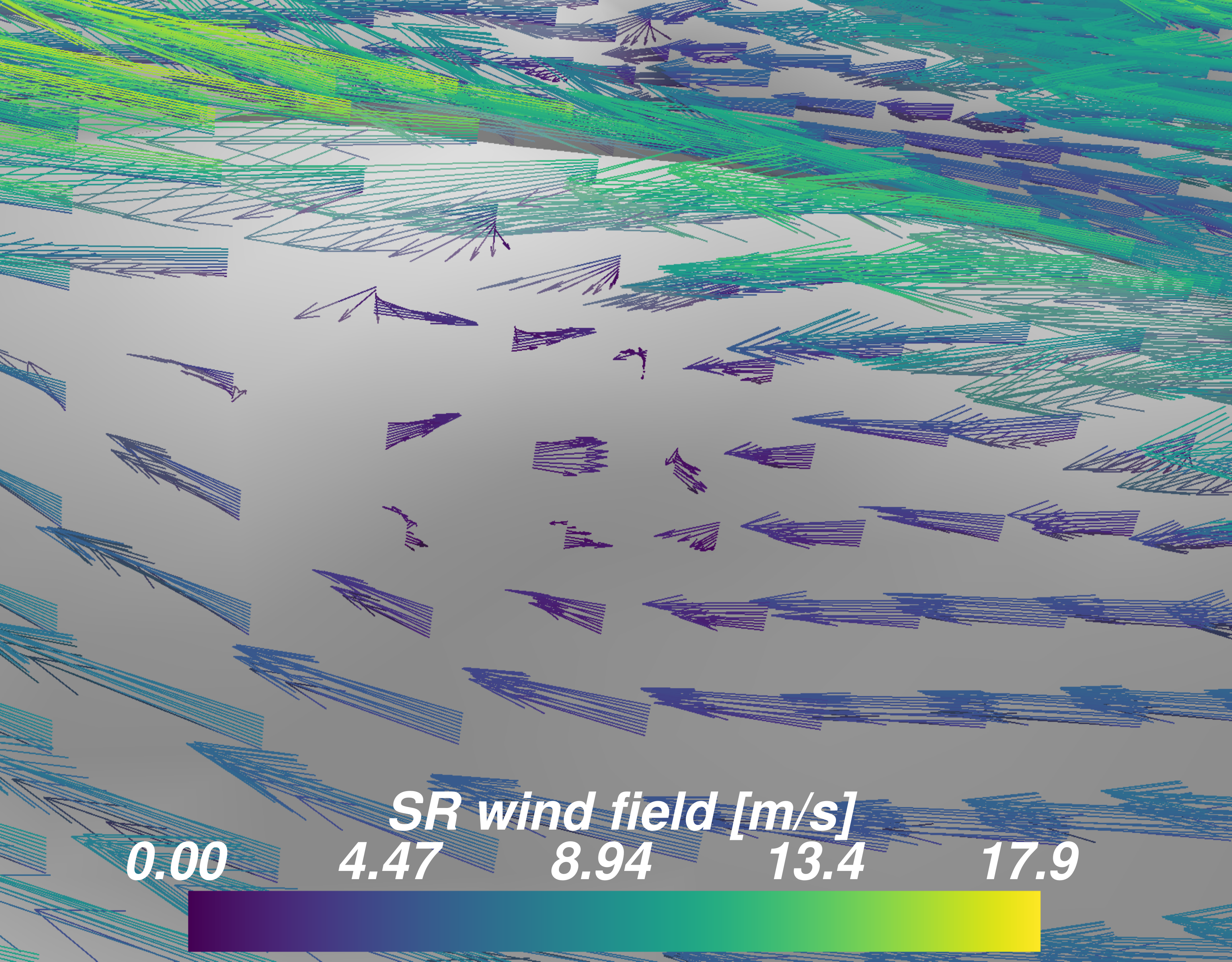}
            \caption{SR wind field}
        \end{subfigure} 
        \caption[3D Generated Turbulence From Best Model]{Comparison of LR, HR, TL and SR wind fields in a turbulent region. LR means low resolution, the input of the model, TL means trilinear interpolation of the LR data, SR means the super-resolved wind field generated by the trained network and HR means the true high resolution wind field.}
        \label{fig:best_turb}
    \end{figure}

    Turbulent effects in the wind field are investigated in Figure~\ref{fig:best_turb}, which zooms in on a turbulent region of the high-resolution (HR), low-resolution (LR), trilinear interpolated (TL), and super-resolved (SR) wind fields. 
    While interpolation almost completely smoothes out the turbulence area, the super-resolved wind field is remarkably accurate. With only the datapoints in Figure~\ref{subfig:turb_LR} it recreates the swirling pattern of Figure~\ref{subfig:turb_HR}, albeit slightly more regular than the actual data. 
    %This might be a reason why it was hard to improve performance with adversarial learning. For chaotic regions, the best guess for the model is probably a smoothed out swirling pattern like we see above. However when the HR data is sufficiently chaotic, the discriminator gets an easy way to tell whether the data is real or fake by targeting the chaotic areas. In order to fool the discriminator the generator might have to generate more chaotic patterns, but the more random the pattern, the more likely it is to be wrong, so instead the content loss ensures that the generator stays in line.
    
    \subsection{Influence on super-resolution achieved} \label{subsec:scale-slice}
    Two more tests are carried out with the final configuration. First, higher upsampling factors are targeted, and second, the data slicing, i.e. predicting only a subset of the wind full wind field per evaluation, is turned off.
    Table~\ref{tab:results-upscale} shows the results of super-resolving with a 4x4,  8x8, and 16x16 resolution increase with and without data slicing. 
    It can be clearly seen that the model performs better when data-slicing is disabled, which likely results in one single terrain being overfitted. 
    While this is not desirable when learning underlying equations, it can be used in digital twins to be trained with a pre-computed data set, and upsample low-resolution data to wind flow with high accuracy in real-time, which is impossible with numerical codes. 

\begin{figure*}[h]
        \centering
        \begin{subfigure}[b]{0.475\linewidth}
            \includegraphics[width=\linewidth]{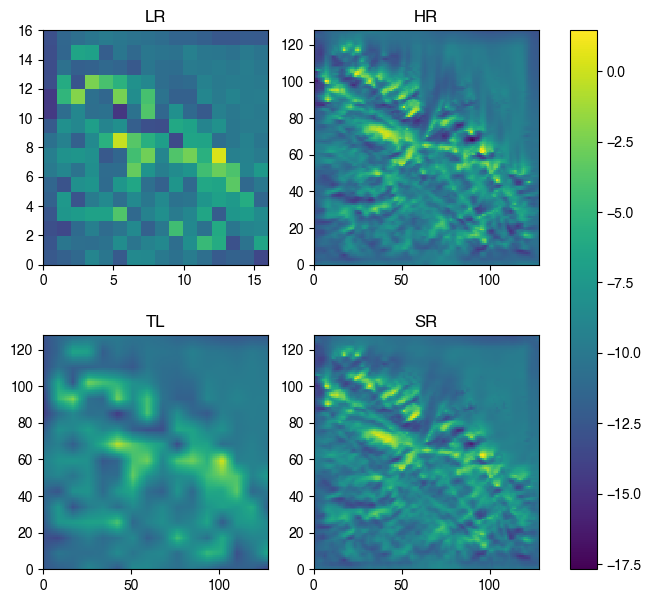}
            \caption{Wind velocity for 8x8 resolution increase}
        \end{subfigure}
        \begin{subfigure}[b]{0.475\linewidth}
            \includegraphics[width=\linewidth]{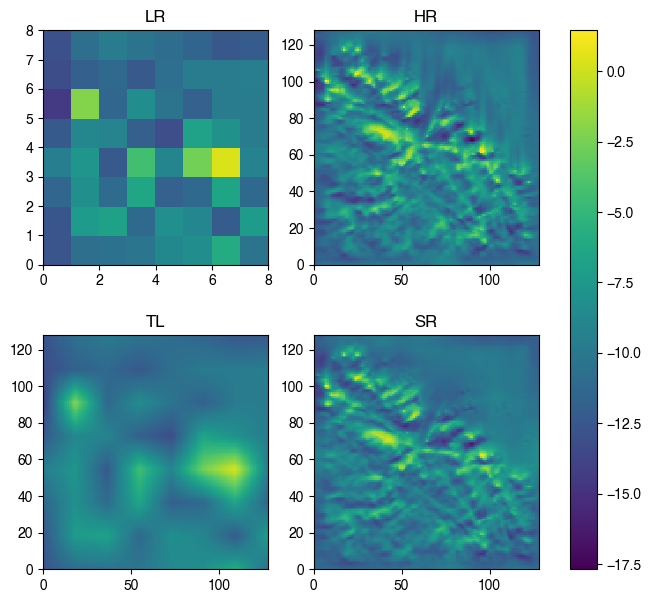}
            \caption{Wind velocity for 16x16 resolution increase}
        \end{subfigure} \hfill
        \begin{subfigure}[b]{0.475\linewidth}
            \includegraphics[width=\linewidth]{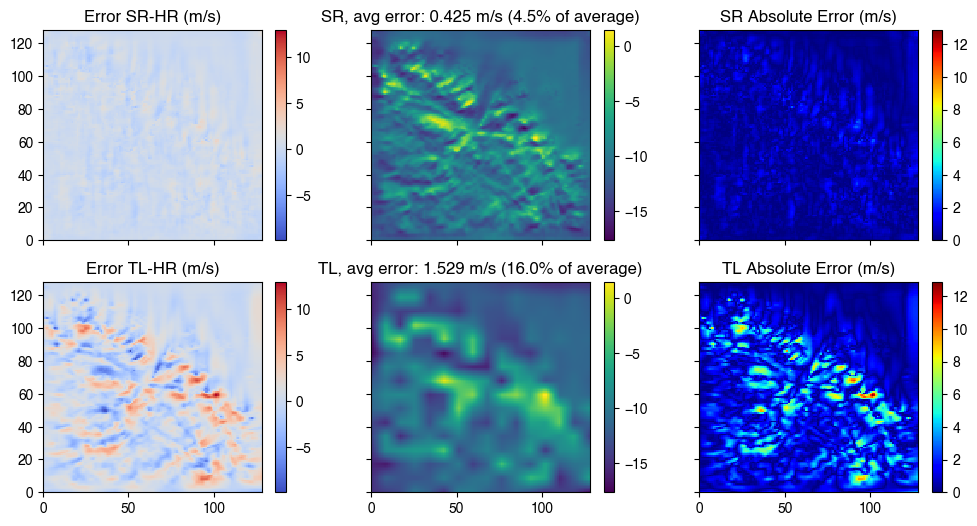}
            \caption{Error for 8x8 resolution increase}
        \end{subfigure}
        \begin{subfigure}[b]{0.475\linewidth}
            \includegraphics[width=\linewidth]{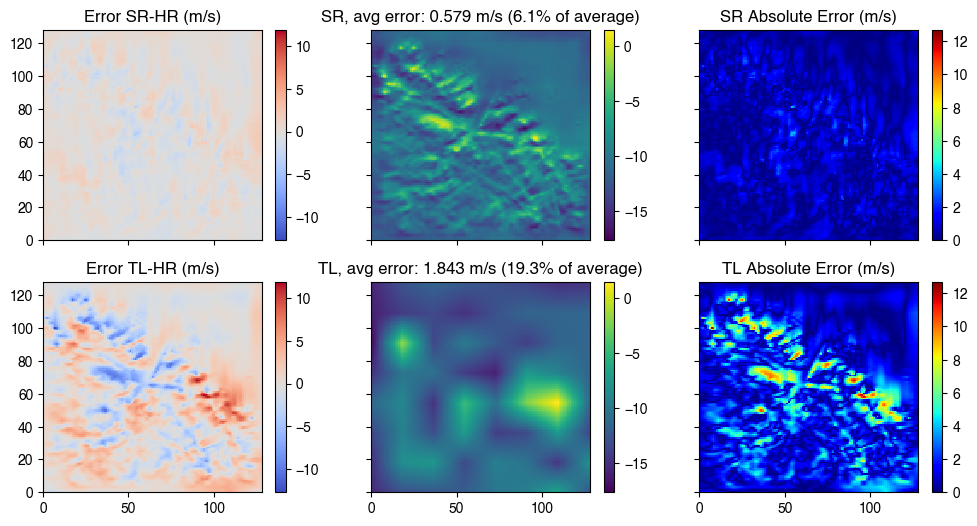}
            \caption{Error for 16x16 resolution increase}
        \end{subfigure}
        \caption[2D Best Model Results 1]{Comparison of 8x8 and 16x16 resolution increase for models trained without slicing on wind velocity along the $x$-axis for the second highest horizontal 2D slice of the wind field for the same wind field as in Figure~\ref{fig:u9w9}. LR means low resolution, the input of the model, TL means trilinear interpolation of the LR data, SR means the super-resolved wind field generated by the trained network and HR means the true high-resolution wind field. "Avg error" refers to the average absolute error for the displayed wind component in the displayed slice, the "\% of average" means this value divided by the average value of that wind component in that slice for the HR wind field.}
        \label{fig:u8x16}
    \end{figure*}

    Furthermore, the results show that big resolution increases can be achieved while still producing reliable super-resolved wind fields. 
    Especially when overtraining a single terrain, the generator achieves better performance on a 16x16 resolution increase than the trilinear interpolation for a 4x4 resolution increase.
    However, while this shows the benefits of overfitting the generator, it also raises the question of to what degree the data-slicing is able to reduce said overfitting. While several data augmentation measures have been taken to avoid this, the training set is still biased towards one terrain, and evaluation on an unseen terrain would be needed to investigate the potential overtraining further.
    \begin{table}
        \caption[Results of changing scaling and slicing]{Test results depending on varying resolution increase and whether data slicing is used in training. 
        Though being interpolated, the generated wind field is not interpolated back and compared to the original data as in Table~\ref{tab:results-exp1}, as this did not affect the results significantly. 
        Trilinear interpolation performance is included for comparison.}
        \centering
           \begin{tabularx}{\linewidth}{L|RRRR}
        \textbf{name} & \textbf{PSNR (db)} & \textbf{pix (m/s)} & \textbf{pix-vector (m/s)} & \textbf{pix-vector relative}  \\ \hline
        $4\times4$ SR & 47.14 & 0.116 & 0.24 & 6.1\% \\
        $4\times4$ SR no slicing &  49.27  & 0.090  & 0.19 & 5.0\% \\
        $4\times4$ TL & 36.35 & 0.385 & 0.8 &  18.5\%\\ \grayline
        $8\times8$ SR & 41.95 & 0.204 & 0.43 & 11.3\% \\
        $8\times8$ SR no slicing & 44.25 & 0.159 & 0.33 & 9.1\% \\
        $8\times8$ TL & 33.86 & 0.528 & 1.12 &  25.8\% \\ \grayline
        $16\times16$ SR & 34.2 & 0.502 & 1.11 & 26.68\% \\
        $16\times16$ SR no slicing & 41.57 & 0.216 & 0.46 & 12.8\% \\
        $16\times16$ TL & 32.77 & 0.609 & 1.29 & 30.6\% \\
            \hline
        \end{tabularx}
        \label{tab:results-upscale}
    \end{table}
    In Figure~\ref{fig:u8x16} we compare the results of $8\times8$ and $16\times16$ resolution increase when not slicing the wind field, for the same horizontal layer as in Figure~\ref{fig:u9w9}. Even with $8\times8$, many details in the wind field can be recovered through the high-resolution terrain.
    
\section{Conclusion and future work}
    \label{sec:conclusionandfuturework}
    %\todo[inline]{Felt a little tentative to call it GANS - because "adversarial" loss term is 0 i.e. turned off. Better to use term like neural network or generative model.}\label{sec:conclusionandfuturework}
This paper presented a neural-network-based 3D wind flow super-resolution algorithm which was successfully applied to reconstruct a simulated 3D wind field of a Norwegian coastal area with a complex topography. The super-resolving neural network was based on the ESRGAN \cite{Wang2019ees}, however, several modifications were made to the architecture and hyperparameters. The GAN architecture was improved to process unevenly spaced 3D vector fields as input, and a terrain feature extractor was incorporated to provide the GAN with high-resolution terrain information. In addition, a feature dropout layer was added to the generator, and the value of complementary input information in the form of height and pressure was explored. Moreover, major changes were made to the loss function. The main findings can be enumerated as follows:

    \begin{enumerate}
    \item We have made 2TB of high-resolution data spanning three years available for use by researchers interested in wind energy. To simplify data access and processing, we have stored the data in the netCDF file format on an OPeNDAP server. This eliminates the need for local storage for the entire dataset. If someone wishes to utilize these data, they can stream them on-the-fly according to their specific requirements. This approach is particularly crucial for the kind of research presented here.  
    \item We demonstrate that microscale 3D atmospheric wind flow can be accurately approximated using Convolutional Neural Networks (CNNs) with low-resolution wind data and high-resolution terrain data, even when the data are irregularly spaced vertically. The final generator was evaluated on a test set of wind fields, achieving a peak signal-to-noise ratio of 47.14~dB, an average wind vector length error of 0.24~m/s, and a relative wind speed error of 6.12\%. Notably, our generator significantly outperformed trilinear interpolation, which achieved scores of 36.42~dB, 0.80~m/s, and 18.5\%, respectively. This superiority extended to higher resolution enhancements, such as $8\times8$ and $16\times16$, especially when applied to the same terrain on which it was trained. In light of our results, it is worth noting that we can accurately simulate a $200m\times200m$ wind field using an $800m\times800m$ mesh, leading to a potential speedup of up to 100 times.
    \item In place of the perceptual loss component in ESRGAN, we introduced gradient- and divergence-based loss components rooted in physical principles. This shift proved to be highly effective in enhancing the performance of the super-resolution generator to the extent that eliminating adversarial loss during training did not lead to a reduction in performance. This observation led us to opt for a generative model over a generative adversarial model for the final model configuration. The proposed fully convolutional 3D generator, incorporating a wind gradient-based loss function, excelled in super-resolving the near-surface 3D wind field, demonstrating impressive results in both quantitative error and visual fidelity.
    \end{enumerate}

    Although these results are encouraging, there is room for further improvement in the following ways:
    
\begin{enumerate} 
    \item The high-resolution terrain feature extractor was implemented, but its architecture was not fully optimized. The ESRGAN in this work could be compared to an encoder-decoder structure such as that employed in the U-net \cite{Ronneberger2015unc}.
    The U-net starts and finishes with the same spatial dimensions, which would allow to include the high-resolution terrain in the same layer as a trilinear interpolated low-resolution wind field. On one hand, this allows the high-resolution terrain to be injected close to the low-resolution wind field, but on the other hand, information might degrade faster, and the interpolated wind field reduces the information density, especially for large zoom factors.

    \item The results of this work discourage the use of adversarial approaches, as opposed to informed content loss. Working with adversarial training is more computationally demanding and potentially less stable (e.g. due to mode collapse and reproducibility). Nonetheless, carefully fine-tuned adversarial learning could bring benefits in the late stages of the training when physically motivated losses yield no further improvement. Additionally, a loss based on discriminator feature maps could be beneficial if the discriminator performs reasonably well. Using the feature extractor of a pre-trained discriminator or of the discriminator saved at some interval during training to add a feature loss might, for instance, give better granularity to the adversarial feedback. 

    \item In this work, an estimate for the real high-resolution wind conditions is being predicted, and a large training set is being used to represent various wind conditions. However, the exact accuracy of the model may vary between predictions. A probabilistic approach could give indications of the uncertainty of individual predictions.
    Two approaches are Bayesian neural networks and mixture density neural networks, which have been used for example in \cite{Sun2020pcb} and \cite{Maulik2020pnn} respectively for fluid flow modeling. However, Bayesian neural networks require many evaluations to map the posterior, which can affect runtime, and application to convolutional super-resolution neural networks requires a novel Bayesian convolutional upsampling layer. Mixture density models require no modification of the underlying neural network layers but require more output parameters and larger neural network architectures. Nonetheless, probabilistic methods can be considered for future research.
    \end{enumerate}

\begin{figure*}[h]
        \centering
        \includegraphics[width=\textwidth]{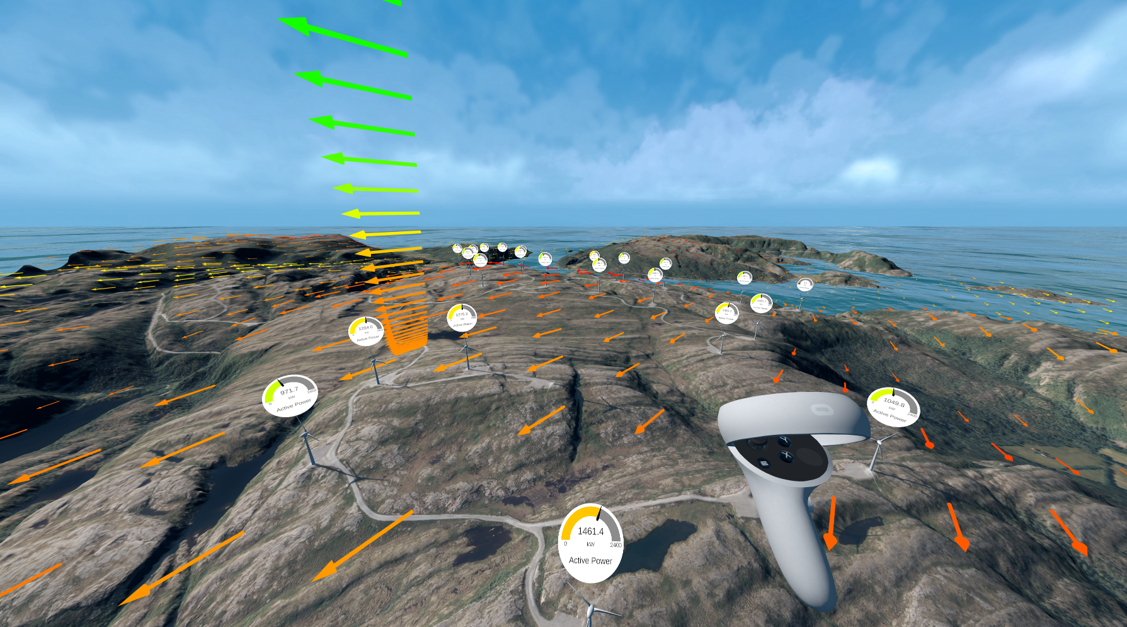}
        \caption{A high-resolution wind field implemented into a digital twin of the Bessakerfjellet wind farm. Here the digital twin is being accessed through a virtual reality headset.}
        \label{fig:bess}
    \end{figure*}

    Once the network has been validated for interpolation using the output of real low-resolution models as input, its fast evaluation allows it to be applied in real-time inside digital twins. The implementation of the high-resolution wind field into a digital twin accessed through a virtual reality interface is demonstrated in \ref{fig:bess}.
    Inside the digital twin, the prediction can be used directly for power forecasting, as has been demonstrated in~\cite{Stadtmann2023doa}, or can be used as input to models for wakes and structural responses.

    Despite the potential for further research, the proposed model should be considered as a proof-of-concept for CNN-driven superresolution of 3D microscale atmospheric wind flow.

\section*{Declaration of competing interest}
    The authors declare that they have no known competing financial interests or personal relationships that could have appeared to influence the work reported in this paper.

\section*{Acknowledgements}
    This publication has been prepared as part of NorthWind (Norwegian Research Centre on Wind Energy)\cite{FMENorthWindnrc} which is co-financed by the Research Council of Norway (project code 321954), industry, and research partners.

\bibliographystyle{AR} 
\bibliography{references}

\begin{thebibliography}{10}
\expandafter\ifx\csname url\endcsname\relax
  \def\url#1{\texttt{#1}}\fi
\expandafter\ifx\csname urlprefix\endcsname\relax\def\urlprefix{URL }\fi
\expandafter\ifx\csname href\endcsname\relax
  \def\href#1#2{#2} \def\path#1{#1}\fi

\bibitem{Gibon2022cni}
T.~Gibon, A.~Hahn~Menacho, M.~Guiton,
  \href{https://unece.org/sed/documents/2021/10/reports/life-cycle-assessment-electricity-generation-options}{Carbon
  {Neutrality} in the {UNECE} {Region}: {Integrated} {Life}-cycle {Assessment}
  of {Electricity} {Sources}}, Tech. rep., United Nations Economic Commission
  for Europe, Palais des Nations, CH-1211 Geneva 10, Switzerland (2022).

\bibitem{2018ida}
E.~Commission,
  \href{https://knowledge4policy.ec.europa.eu/publication/depth-analysis-support-com2018-773-clean-planet-all-european-strategic-long-term-vision_en}{In-depth
  {Analysis} in {Support} of the {Commission} {Communication} {COM}(2018) 773},
  Tech. rep., European Commission, Brussles (2018).

\bibitem{Lee2021gwr}
J.~Lee, Z.~Feng, \href{https://gwec.net/global-wind-report-2021/}{Global {Wind}
  {Report} 2021}, Tech. rep., Global Wind Energy Council, Rue de Commerce 31
  – 1000 Brussels (2021).

\bibitem{Rasheed2020dtv}
A.~Rasheed, O.~San, T.~Kvamsdal,
  \href{https://ieeexplore.ieee.org/document/8972429/}{Digital {Twin}:
  {Values}, {Challenges} and {Enablers} {From} a {Modeling} {Perspective}},
  {\em IEEE Access}, 8:21980--22012 (2020).
\newblock

\bibitem{Stadtmann2023dti}
F.~Stadtmann, A.~Rasheed, T.~Kvamsdal, K.~A. Johannessen, O.~San, K.~Kölle,
  J.~O.~G. Tande, I.~Barstad, A.~Benhamou, T.~Brathaug, T.~Christiansen, A.-L.
  Firle, A.~Fjeldly, L.~Frøyd, A.~Gleim, A.~Høiberget, C.~Meissner,
  G.~Nygård, J.~Olsen, H.~Paulshus, T.~Rasmussen, E.~Rishoff, J.~O. Skogås,
  \href{https://arxiv.org/abs/2304.11405}{Digital {Twins} in {Wind} {Energy}:
  {Emerging} {Technologies} and {Industry}-{Informed} {Future} {Directions}},
  {\em arXiv preprint arXiv:2304.11405} (2023), Publisher: arXiv Version
  Number: 1.
\newblock

\bibitem{Skajaa2015ito}
A.~Skajaa, K.~Edlund, J.~M. Morales,
  \href{http://ieeexplore.ieee.org/document/7005520/}{Intraday {Trading} of
  {Wind} {Energy}}, {\em IEEE Transactions on Power Systems}, 30:3181--3189
  (2015).
\newblock

\bibitem{Knudsen2015sow}
T.~Knudsen, T.~Bak, M.~Svenstrup,
  \href{https://onlinelibrary.wiley.com/doi/10.1002/we.1760}{Survey of wind
  farm control—power and fatigue optimization}, {\em Wind Energy},
  18:1333--1351 (2015).
\newblock

\bibitem{VanDijk2017wfm}
M.~T. Van~Dijk, J.-W. Van~Wingerden, T.~Ashuri, Y.~Li,
  \href{https://linkinghub.elsevier.com/retrieve/pii/S0360544217300518}{Wind
  farm multi-objective wake redirection for optimizing power production and
  loads}, {\em Energy}, 121:561--569 (2017).
\newblock

\bibitem{Stanley2020wfl}
A.~P.~J. Stanley, J.~King, A.~Ning,
  \href{https://iopscience.iop.org/article/10.1088/1742-6596/1452/1/012072}{Wind
  {Farm} {Layout} {Optimization} with {Loads} {Considerations}}, {\em Journal
  of Physics: Conference Series}, 1452:012072 (2020).
\newblock

\bibitem{Ernst2012ios}
B.~Ernst, J.~R. Seume,
  \href{http://www.mdpi.com/1996-1073/5/10/3835}{Investigation of
  {Site}-{Specific} {Wind} {Field} {Parameters} and {Their} {Effect} on {Loads}
  of {Offshore} {Wind} {Turbines}}, {\em Energies}, 5:3835--3855 (2012).
\newblock

\bibitem{Midjiyawa2023ncf}
Z.~Midjiyawa, J.~V. Venås, T.~Kvamsdal, A.~M. Kvarving, K.~H. Midtbø,
  A.~Rasheed,
  \href{https://www.sciencedirect.com/science/article/pii/S0167610523002003}{Nested
  computational fluid dynamic modeling of mean turbulent quantities estimation
  in complex topography using {AROME}-{SIMRA}}, {\em Journal of Wind
  Engineering and Industrial Aerodynamics}, 240:105497 (2023).
\newblock

\bibitem{Rasheed2017wfm}
A.~Rasheed, M.~Tabib, J.~Kristiansen,
  \href{https://asmedigitalcollection.asme.org/OMAE/proceedings/OMAE2017/57786/Trondheim,%20Norway/282088}{Wind
  {Farm} {Modeling} in a {Realistic} {Environment} {Using} a {Multiscale}
  {Approach}}, in {\em Volume 10: {Ocean} {Renewable} {Energy}}, American
  Society of Mechanical Engineers, American Society of Mechanical Engineers,
  Trondheim, Norway, 2017, p. V010T09A051.
\newblock

\bibitem{Fukami2023sra}
K.~Fukami, K.~Fukagata, K.~Taira,
  \href{https://link.springer.com/10.1007/s00162-023-00663-0}{Super-resolution
  analysis via machine learning: a survey for fluid flows}, {\em Theoretical
  and Computational Fluid Dynamics}, 37:421--444 (2023).
\newblock

\bibitem{Hohlein2020acs}
K.~Höhlein, M.~Kern, T.~Hewson, R.~Westermann,
  \href{https://rmets.onlinelibrary.wiley.com/doi/10.1002/met.1961}{A
  comparative study of convolutional neural network models for wind field
  downscaling}, {\em Meteorological Applications}, 27:e1961 (2020).
\newblock

\bibitem{Stengel2020asr}
K.~Stengel, A.~Glaws, D.~Hettinger, R.~N. King,
  \href{https://pnas.org/doi/full/10.1073/pnas.1918964117}{Adversarial
  super-resolution of climatological wind and solar data}, {\em Proceedings of
  the National Academy of Sciences}, 117:16805--16815 (2020).
\newblock

\bibitem{Xie2018tat}
Y.~Xie, E.~Franz, M.~Chu, N.~Thuerey,
  \href{https://dl.acm.org/doi/10.1145/3197517.3201304}{{tempoGAN}: a
  temporally coherent, volumetric {GAN} for super-resolution fluid flow}, {\em
  ACM Transactions on Graphics}, 37:1--15 (2018).
\newblock

\bibitem{Werhahn2019amp}
M.~Werhahn, Y.~Xie, M.~Chu, N.~Thuerey,
  \href{https://dl.acm.org/doi/10.1145/3340251}{A {Multi}-{Pass} {GAN} for
  {Fluid} {Flow} {Super}-{Resolution}}, {\em Proceedings of the ACM on Computer
  Graphics and Interactive Techniques}, 2:1--21 (2019).
\newblock

\bibitem{Tran2020ges}
D.~T. Tran, H.~Robinson, A.~Rasheed, O.~San, M.~Tabib, T.~Kvamsdal,
  \href{https://iopscience.iop.org/article/10.1088/1742-6596/1669/1/012029}{{GANs}
  enabled super-resolution reconstruction of wind field}, {\em Journal of
  Physics: Conference Series}, 1669:012029 (2020).
\newblock

\bibitem{Larsen2020ota}
T.~N. Larsen, \href{https://hdl.handle.net/11250/2780978}{On the applicability
  of a perceptually driven generative-adversarial framework for
  super-resolution of wind fields in complex terrain}, Ph.D. thesis, NTNU,
  Trondheim, Norway (2020).

\bibitem{Gatski1996sam}
T.~B. Gatski, M.~Y. Hussaini, J.~L. Lumley (eds.),
  \href{https://academic.oup.com/book/40798}{{\em Simulation and {Modeling} of
  {Turbulent} {Flows}}}, Oxford University Press, 1996.
\newblock

\bibitem{Goodfellow2014gan}
I.~Goodfellow, J.~Pouget-Abadie, M.~Mirza, B.~Xu, D.~Warde-Farley, S.~Ozair,
  A.~Courville, Y.~Bengio,
  \href{https://proceedings.neurips.cc/paper_files/paper/2014/file/5ca3e9b122f61f8f06494c97b1afccf3-Paper.pdf}{Generative
  adversarial nets}, in Z.~Ghahramani, M.~Welling, C.~Cortes, N.~Lawrence,
  K.~Weinberger (eds.), {\em Advances in neural information processing
  systems}, vol.~27, Curran Associates, Inc., 2014.

\bibitem{Jolicoeur_Martineau2019trd}
A.~Jolicoeur-Martineau, \href{https://openreview.net/forum?id=S1erHoR5t7}{The
  relativistic discriminator: a key element missing from standard {GAN}}, {\em
  International Conference on Learning Representation} (2019).

\bibitem{Wang2019ees}
X.~Wang, K.~Yu, S.~Wu, J.~Gu, Y.~Liu, C.~Dong, Y.~Qiao, C.~C. Loy,
  \href{https://link.springer.com/10.1007/978-3-030-11021-5_5}{{ESRGAN}:
  {Enhanced} {Super}-{Resolution} {Generative} {Adversarial} {Networks}}, in
  L.~Leal-Taixé, S.~Roth (eds.), {\em Computer {Vision} – {ECCV} 2018
  {Workshops}}, vol. 11133, Springer International Publishing, Cham, 2019, pp.
  63--79, Series Title: Lecture Notes in Computer Science.
\newblock

\bibitem{Ledig2017prs}
C.~Ledig, L.~Theis, F.~Huszar, J.~Caballero, A.~Cunningham, A.~Acosta,
  A.~Aitken, A.~Tejani, J.~Totz, Z.~Wang, W.~Shi,
  \href{http://ieeexplore.ieee.org/document/8099502/}{Photo-{Realistic}
  {Single} {Image} {Super}-{Resolution} {Using} a {Generative} {Adversarial}
  {Network}}, in {\em 2017 {IEEE} {Conference} on {Computer} {Vision} and
  {Pattern} {Recognition} ({CVPR})}, IEEE, IEEE, Honolulu, HI, 2017, pp.
  105--114.
\newblock

\bibitem{Simonyan2015vdc}
K.~Simonyan, A.~Zisserman, \href{https://arxiv.org/abs/1409.1556}{Very {Deep}
  {Convolutional} {Networks} for {Large}-{Scale} {Image} {Recognition}}, {\em
  3rd International Conference on Learning Representations (ICLR2015)} (2015),
  Publisher: arXiv Version Number: 6.
\newblock

\bibitem{Rasheed2014amw}
A.~Rasheed, J.~K. Süld, T.~Kvamsdal,
  \href{https://dx.doi.org/10.1016/j.egypro.2014.07.238}{A {Multiscale} {Wind}
  and {Power} {Forecast} {System} for {Wind} {Farms}}, {\em Energy Procedia},
  53:290--299 (2014).
\newblock

\bibitem{Vesterkjaer2019e}
E.~Vesterkjær, \href{https://github.com/eirikeve/esrdgan/}{{ESRDGAN}},
  https://github.com/eirikeve/esrdgan/ (2019).

\bibitem{Liaw2018tar}
R.~Liaw, E.~Liang, R.~Nishihara, P.~Moritz, J.~E. Gonzalez, I.~Stoica,
  \href{https://arxiv.org/abs/1807.05118}{Tune: {A} {Research} {Platform} for
  {Distributed} {Model} {Selection} and {Training}}, {\em ICML 2018 AutoML
  Workshop} (2018), Publisher: arXiv Version Number: 1.
\newblock

\bibitem{Akiba2019oana}
T.~Akiba, S.~Sano, T.~Yanase, T.~Ohta, M.~Koyama,
  \href{https://dl.acm.org/doi/10.1145/3292500.3330701}{Optuna: {A}
  {Next}-generation {Hyperparameter} {Optimization} {Framework}}, in {\em
  Proceedings of the 25th {ACM} {SIGKDD} {International} {Conference} on
  {Knowledge} {Discovery} \& {Data} {Mining}}, acm, ACM, Anchorage AK USA,
  2019, pp. 2623--2631.
\newblock

\bibitem{Li2020asf}
L.~Li, K.~Jamieson, A.~Rostamizadeh, E.~Gonina, J.~Ben-tzur, M.~Hardt,
  B.~Recht, A.~Talwalkar,
  \href{https://proceedings.mlsys.org/paper_files/paper/2020/file/a06f20b349c6cf09a6b171c71b88bbfc-Paper.pdf}{A
  system for massively parallel hyperparameter tuning}, in I.~Dhillon,
  D.~Papailiopoulos, V.~Sze (eds.), {\em Proceedings of machine learning and
  systems}, vol.~2, 2020, pp. 230--246.

\bibitem{OPeNDAPh}
OPeNDAP, \href{https://www.opendap.org/}{Advanced software for remote data
  retrieval}, https://www.opendap.org/.

\bibitem{Ronneberger2015unc}
O.~Ronneberger, P.~Fischer, T.~Brox,
  \href{http://link.springer.com/10.1007/978-3-319-24574-4_28}{U-{Net}:
  {Convolutional} {Networks} for {Biomedical} {Image} {Segmentation}}, in
  N.~Navab, J.~Hornegger, W.~M. Wells, A.~F. Frangi (eds.), {\em Medical
  {Image} {Computing} and {Computer}-{Assisted} {Intervention} – {MICCAI}
  2015}, vol. 9351, Springer International Publishing, Cham, 2015, pp.
  234--241, Series Title: Lecture Notes in Computer Science.
\newblock

\bibitem{Sun2020pcb}
L.~Sun, J.-X. Wang,
  \href{https://linkinghub.elsevier.com/retrieve/pii/S2095034920300295}{Physics-constrained
  bayesian neural network for fluid flow reconstruction with sparse and noisy
  data}, {\em Theoretical and Applied Mechanics Letters}, 10:161--169 (2020).
\newblock

\bibitem{Maulik2020pnn}
R.~Maulik, K.~Fukami, N.~Ramachandra, K.~Fukagata, K.~Taira,
  \href{https://link.aps.org/doi/10.1103/PhysRevFluids.5.104401}{Probabilistic
  neural networks for fluid flow surrogate modeling and data recovery}, {\em
  Physical Review Fluids}, 5:104401 (2020).
\newblock

\bibitem{Stadtmann2023doa}
F.~Stadtmann, H.~G. Wassertheurer, A.~Rasheed,
  \href{https://arxiv.org/abs/2304.01093}{Demonstration of a {Standalone},
  {Descriptive}, and {Predictive} {Digital} {Twin} of a {Floating} {Offshore}
  {Wind} {Turbine}}, {\em arXiv preprint arXiv:2304.01093} (2023), Publisher:
  arXiv Version Number: 1.
\newblock

\bibitem{FMENorthWindnrc}
{FME NorthWind}, \href{https://www.northwindresearch.no/}{Norwegian {Research}
  {Centre} on {Wind} {Energy}}, https://www.northwindresearch.no/.

\end{thebibliography}

\clearpage
\onecolumn
\appendix
\section{Additional Tables and Figures}
% to have the distinct numbering scheme of tables and figures for the appendix:
\renewcommand{\thetable}{\Alph{section}\arabic{table}}
\renewcommand{\thefigure}{\Alph{section}\arabic{figure}}

\begin{table*}[h]
    \centering
    \caption{Average height in meters above the surface of the layers in the SIMRA simulation.  }
    \begin{tabular}{c|c| c| c| c| c| c| c| c| c| c| c| c| c| c}
        layer & 1& 2& 3& 4& 5& 6& 7& 8& 9& 10& 11& 12& 13& 14  \\\hline
        height above ground & 0& 2& 5& 8& 11& 15& 19& 24& 29& 35& 41& 49& 57& 67 \\\hline\hline
        
        layer & 15& 16& 17& 18& 19& 20& 21& 22& 23& 24& 25& 26& 27& 28  \\\hline
        height above ground & 77 & 90 & 104& 120& 138& 159& 183& 210& 241& 276& 317& 364& 417& 477  \\\hline\hline
        
        layer &29& 30& 31& 32& 33& 34& 35& 36& 37& 38& 39& 40 & 41 \\\hline
        height above ground & 546& 623& 711& 809& 918& 1037& 1167& 1305& 1448& 1590& 1722& 1878& 2049 
    \end{tabular}
    \label{tab:simra_vertical_spacing}
\end{table*}

\begin{table*}
	\caption{Details of the computational models, number of CPU, domain extent [km], number of mesh elements  [million], and total simulation time [minutes].}
	\begin{center}
		\label{table:compmodels}
		\begin{tabular}{c l l l l}
			& & & &\\ % put some space after the caption
			\toprule
			Model     & CORES    & Domain & N & Time\\
			\midrule
			HARMONIE  &1840    & $1875\times2400\times26$ & 46 & 87 \\
			SIMRA     &  48    & $  30\times30\times3.0$  & 1.6& 13 \\
			\bottomrule
		\end{tabular}
	\end{center}
\end{table*}

 \begin{figure*}
        \centering
        \begin{subfigure}[b]{0.475\linewidth}
            \includegraphics[width=\linewidth]{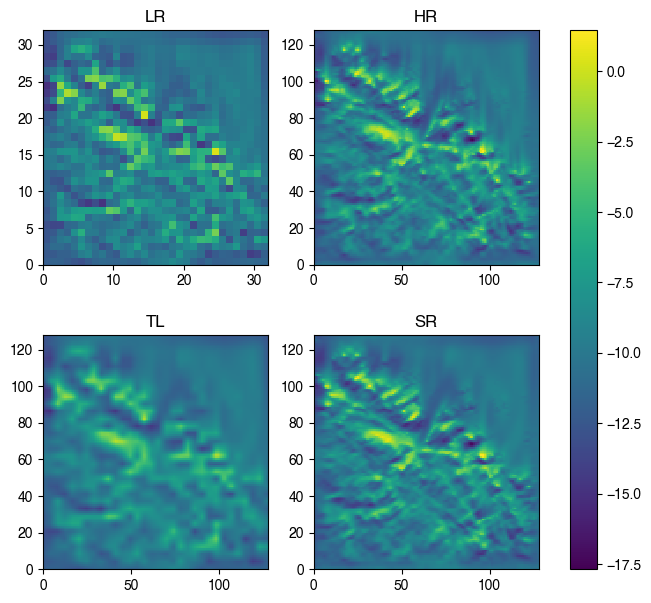}
            \caption{Wind velocity along the $x$-axis}
        \end{subfigure}
        \begin{subfigure}[b]{0.475\linewidth}
            \includegraphics[width=\linewidth]{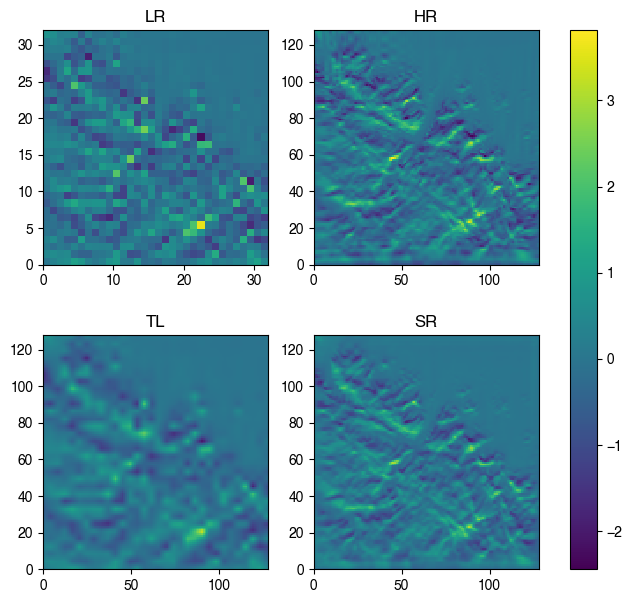}
            \caption{Wind velocity along the $z$-axis}
        \end{subfigure} \hfill
        \begin{subfigure}[b]{0.475\linewidth}
            \includegraphics[width=\linewidth]{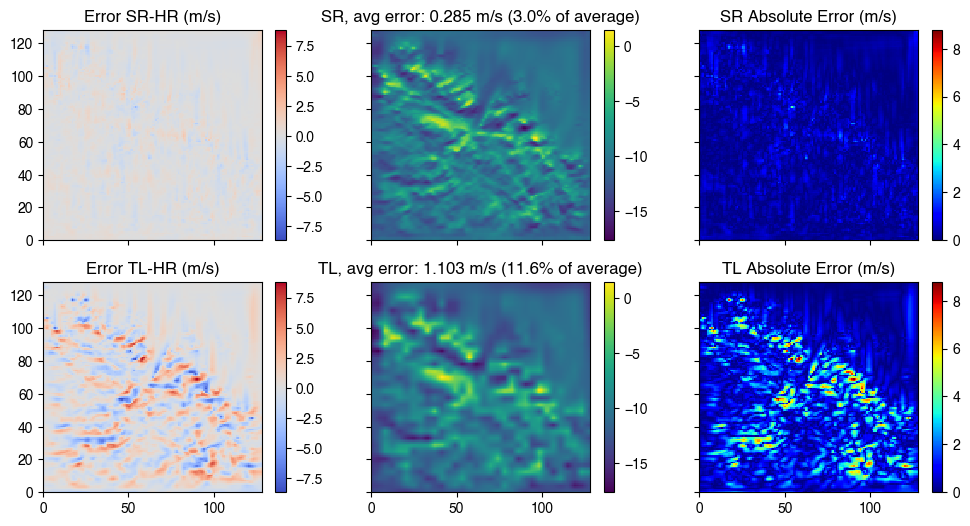}
            \caption{Error for wind velocity along the $x$-axis }
        \end{subfigure}\hfill
        \begin{subfigure}[b]{0.475\linewidth}
            \includegraphics[width=\linewidth]{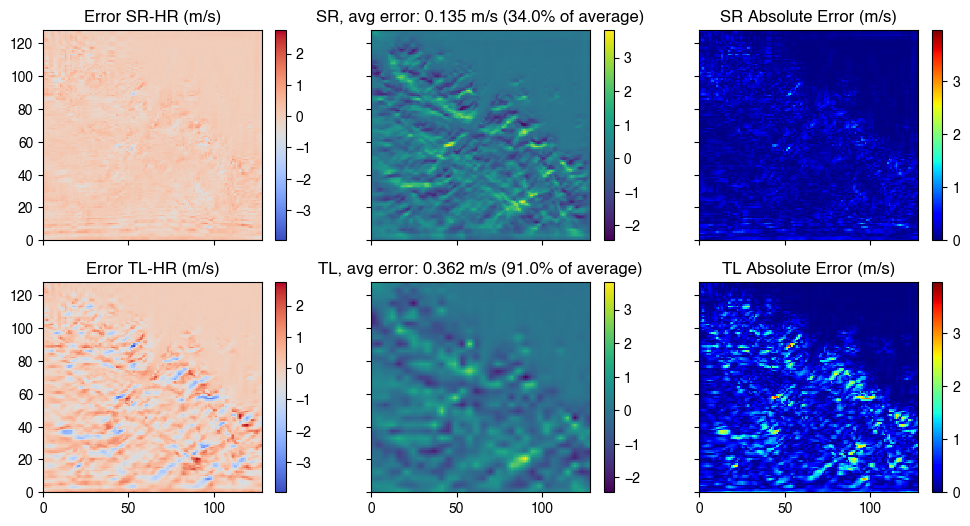}
            \caption{Error for wind velocity along the $z$-axis}
        \end{subfigure}
        \caption[2D Best Model Results 2]{Comparison of the second highest horizontal 2D slice of the wind field for a randomly selected wind field. LR means low resolution, the input of the model, TL means trilinear interpolation of the LR data, SR means the super-resolved wind field generated by the trained network and HR means the true high resolution wind field. "Avg error" refers to average absolute error for the displayed wind component in the displayed slice, the "\% of average" means this value divided by the average of value of that wind component in that slice for the HR wind field.}
        \label{fig:u9w9}
    \end{figure*}

\clearpage
\begin{longtable}{ p{4.2cm} p{4cm} p{9cm} }
    \caption{Full list of standard hyperparameters with explanation}\label{tab:full-fyperparameters}\\%\begin{tabular}{ |p{4.2cm}|p{4cm}|p{9cm}| }
    \toprule
    \multicolumn{3}{c}{\textbf{General Parameters}} \\
    \midrule
    \textbf{Parameter} & \textbf{Value} & \textbf{Meaning/Comments} \\
    \midrule
    \textit{include\_pressure} & True & \textbf{G} input channel\\
    \textit{include\_z\_channel} & True & \textbf{G} input channel\\
    \textit{included\_z\_layers} & 1-10 & 10 of the 11 layers closest to the ground (dropping the one furthest down) \\
    \textit{conv\_mode} & 3D & Alternatives: 3D, 2D, Horizontal3D (separate kernel for each horizontal layer)\\
    \textit{interpolate\_z} & False & Interpolate as specified in Section~\ref{subsec:data_processing}\\
    \textit{include\_altitude\_channel} & False & include $z_{alt}$ as G input channel \\
    \textit{load\_G\_from\_save} & False & Initialise \textbf{G} with a model trained without adversarial cost  \\
    \midrule
    \multicolumn{3}{c}{\textbf{Data-related parameters}} \\
    \midrule
    \textbf{Parameter} & \textbf{Value} & \textbf{Meaning/Comments} \\
    \midrule
    \textit{enable\_slicing} & True & Slice 64x64 slices as specified in Section~\ref{subsec:data_processing} \\
    \textit{scale} & 4 & $xy$ dimension scale difference between LR and SR, e.g. 16x16 $\rightarrow$ 64x64 \\
    \textit{batch\_size} & 32 & \\
    \textit{data\_augmentation\_flipping} & True & \\
    \textit{data\_augmentation\_rotation} & True & \\
    \midrule
    \multicolumn{3}{c}{\textbf{Architecture}} \\
    \midrule
    \textbf{Parameter} & \textbf{Value} & \textbf{Meaning/Comments} \\
    \textbf{G/both} \hfill \textbf{D} & \textbf{G/both} \hfill \textbf{D} & \\
    \midrule
    \textit{num\_features} & 128 \hfill 32 & D features double for every DownConv block except the last \\
    \textit{terrain\_number\_of\_features} & 16 & Number of features in the terrain feature extractor part of \textbf{G}\\
    \textit{num\_RRDB} & 16 & \\
    \textit{num\_RDB\_convs} & 5 & RDBs per RRDB\\
    \textit{RDB\_res\_scaling} & 0.2 & Residual scaling, $\alpha$  in the pink part of Figure~\ref{fig:newESRGAN}\\
    \textit{RRDB\_res\_scaling} & 0.2 &  Residual scaling, $\alpha$ in the green part of Figure~\ref{fig:newESRGAN}\\
    \textit{hr\_kern\_size} & 5 & Kernel size after upscaling in the generator (5x5x5)\\
    \textit{kern\_size} & 3 & Standard kernel size (3x3x3)\\
    \textit{weight\_init\_scale} & 0.1 \hfill 0.2 & Smaller kaiming weight initialization \\
    \textit{lff\_kern\_size} & 1 & local feature fusion layer of RDB\\
    \textit{dropout\_probability} & 0.1 \hfill 0.2 & \\
    \textit{G\_max\_norm} & 1.0 & Maximum gradient due to gradient clipping \\
    \midrule
    \multicolumn{3}{c}{\textbf{Training}} \\
    \midrule
    \textbf{Parameter} & \textbf{Value} & \textbf{Meaning/Comments} \\
    \textbf{G}/both \hfill \textbf{D} & \textbf{G}/both \hfill \textbf{D} & \\
    \midrule
    \textit{learning\_rate} & 1e-5 \hfill 1e-5 & \\
    \textit{adam\_weight\_decay\_G} & 0 \hfill 0 & \\
    \textit{adam\_beta1\_G} & 0.9 \hfill 0.9 & \\
    \textit{adam\_beta2\_G} & 0.999 \hfill 0.999 & \\
    \textit{multistep\_lr} & True & Reduce learning rate during training\\
    \textit{multistep\_lr\_steps} & [10k, 30k, 50k, 70k, 100k] & schedule for reducing learning rate during training\\
    \textit{lr\_gamma} & 0.5 & factor of reduction when lr is adjusted\\
    \textit{gan\_type} & relativisticavg & Section~\ref{subsec:GAN}\\
    \textit{adversarial\_loss\_weight} & 0.005 & $\eta_6$, as specified in Equation \eqref{eq:g_loss_total}\\
    \textit{gradient\_xy\_loss\_weight} & 1.0 & $\eta_2$, as specified in Equation \eqref{eq:g_loss_total}\\
    \textit{gradient\_z\_loss\_weight} & 0.2 & $\eta_3$, as specified in Equation \eqref{eq:g_loss_total}\\
    \textit{xy\_divergence\_loss\_weight} & 0.25 & $\eta_5$, as specified in Equation \eqref{eq:g_loss_total}\\
    \textit{divergence\_loss\_weight} & 0.25 & $\eta_4$, as specified in Equation \eqref{eq:g_loss_total}\\
    \textit{pixel\_loss\_weight} & 0.15 & $\eta_1$, as specified in Equation \eqref{eq:g_loss_total}\\
    \textit{d\_g\_train\_ratio} & 1 & How often D is trained relative to G \\
    \textit{use\_noisy\_labels} & False & Add noise to labels when training D\\
    \textit{use\_label\_smoothing} & True & Set HR labels to $0.9 + 0.1 \frac{it}{it_{max}}$ instead of 1.0 when training D \\
    \textit{flip\_labels} & False & Randomly flip (set HR\_label=false, SR\_label=True) some labels when training D \\
    \textit{use\_instance\_noise} & True & Add instance noise when training \textbf{D}, $\mathbf{D}(\mathbf{x}) \rightarrow \mathbf{D}(\mathbf{x} + \bm{\epsilon})$ \\
    & & $\bm{\epsilon}_i \sim \mathcal{N}\left(0,\,2.0 \left(1- \frac{it}{niter}\right)\right)$ \vspace{1mm}\\
    \textit{niter} & 90k & Number of training iterations\\
    \textit{train\_eval\_test\_ratio} & 0.8 & Ratio for training, remaining fraction split equally between validation and testing\\
    \bottomrule
%\end{tabular}\end{table*}
\end{longtable}

\newpage
\begin{table*}
    \caption[Experiment 1 Results]{Test results of \textit{Experiment~1}. \textbf{name} references the combinations in Table~\ref{tab:exp1}. 1 and 2 refer to the two different seeds used for training. 
    Since the models working with interpolated low-resolution data are trained to produce interpolated high-resolution data, the generated data is interpolated back to the original coordinates and evaluated against the original high-resolution data during testing. Those results are to the right, and evaluation results against the interpolated high-resolution data are to the left in the "interp" rows. 
    Trilinear interpolation performance is included for comparison.}
    \label{tab:results-exp1}
    \centering
    \begin{tabular}{lrrrr}
        \toprule
        \textbf{name} & \textbf{PSNR1 (db)} & \textbf{PSNR2 (db)} & \textbf{pix1 (m/s)} &  \textbf{pix2 (m/s)} \\ \midrule
        \textit{only\_wind} & 39.35 & 39.32 & 0.283 & 0.285\\ 
        \textit{$z$\_channel} & 39.96 & 40.52 & 0.268 & 0.246\\ 
        \textit{$p$\_channel} & 39.59 & 39.65 & 0.278 & 0.276\\ 
        \textit{$p\_z$\_channels} & 40.27 & 40.05 & 0.253 & 0.261\\ 
        \textit{$z_{ground}$\_channels} & 40.6 & 40.61 & 0.242 & 0.244\\ 
        \textit{$p\_z_{ground}$\_channels} & 40.54 & 40.00 & 0.245 & 0.264\\ 
        \textit{only\_wind\_interp} & 39.06 / 39.10 & 39.25 / 39.22  & 0.293 / 0.292 & 0.288 / 0.289 \\ 
        \textit{$z$\_channel\_interp} & 40.59 / 40.53 & 40.65 / 40.61 & 0.246 / 0.247 & 0.243 / 0.243 \\ 
        \textit{$p$\_channel\_interp} & 39.47 / 39.49 & 40.03 / 39.99 & 0.283 / 0.289 & 0.265 / 0.267 \\ 
        \textit{$z\_p$\_channels\_interp} & 39.73 / 39.75 & 39.77 / 39.77 & 0.274 / 0.274 & 0.276 / 0.273 \\ 
        \textit{trilinear} & 36.53 &  & 0.377 & \\ 
        \textit{trilinear\_interp} & 36.35 / 36.42  &  & 0.385 / 0.382 & \\ 
        \bottomrule
    \end{tabular}
\end{table*}

\begin{table*}
    \caption[Experiment 2 Results]{Test results of \textit{Experiment~2}. \textbf{name} references the combinations in Table~\ref{tab:exp2}, \textbf{PSNR} and  \textbf{pix} are metrics defined in Section~\ref{subsec:EvaluationMetrics}, and 1 and 2 refer to the two different seeds used for training.}
    \label{tab:results-exp2}
    \centering
    \begin{tabular}{lrrrr}
    \toprule
        \textbf{name} & \textbf{PSNR1 (db)} & \textbf{PSNR2 (db)} & \textbf{pix1 (m/s)} &  \textbf{pix2 (m/s)} \\ \hline
        \textit{only\_pix\_cost} & 38.00 & 37.37 & 0.335 & 0.344\\ 
        \textit{grad\_cost} & 39.67 & 39.73 & 0.276 & 0.273\\ 
        \textit{div\_cost} & 39.65 & 36.62 & 0.276 & 0.378\\ 
        \textit{$xy$\_cost} & 38.84 & 31.73 & 0.309 & 0.519\\ 
        \textit{std\_cost} & 40.27 & 40.05 & 0.253 & 0.261\\ 
        \textit{large\_grad\_cost} & 41.05 & 40.94 & 0.234 & 0.239\\ 
        \textit{large\_div\_cost} & 41.42 & 41.35 & 0.225 & 0.227\\ 
        \textit{large\_$xy$\_cost} & 41.55 & 40.72 & 0.223 & 0.248\\ \bottomrule
    \end{tabular}
\end{table*}

\begin{table*}
    \caption[Modified hyperparameters of the final model]{Modified hyperparameters of the final model from the standard setup.}
    \label{tab:best-setup} 
    \centering
    \begin{tabular}{ p{4.0cm} p{1.8cm} p{5.5cm} }
    \toprule
    \multicolumn{3}{c}{\textbf{Input}} \\
    \midrule
    \textbf{Parameter} & \textbf{Value} & \textbf{Meaning/Comments} \\
    \midrule
    \textit{include\_pressure} & False & \textbf{G} input channel\\
    \textit{interpolate\_z} & True & Interpolate input in vertical direction\\
    \midrule 
    \multicolumn{3}{c}{\textbf{Training}} \\
    \midrule
    \textbf{Parameter} & \textbf{Value} & \textbf{Meaning/Comments} \\
    \midrule
    \textit{learning\_rate} & 8e-5 & \\
    niter & 150k & Number of training iterations\\
    \textit{adversarial\_loss\_weight} & 0 & $\eta_6$, as specified in Equation \eqref{eq:g_loss_total}\\
    \textit{gradient\_xy\_loss\_weight} & 3.064 & $\eta_2$, as specified in Equation \eqref{eq:g_loss_total}\\
    \textit{gradient\_z\_loss\_weight} & 0 & $\eta_3$, as specified in Equation \eqref{eq:g_loss_total}\\
    \textit{xy\_divergence\_loss\_weight} & 0.721 & $\eta_5$, as specified in Equation \eqref{eq:g_loss_total}\\
    \textit{divergence\_loss\_weight} & 0.366 & $\eta_4$, as specified in Equation \eqref{eq:g_loss_total}\\
    \textit{pixel\_loss\_weight} & 0.136 & $\eta_1$, as specified in Equation \eqref{eq:g_loss_total}\\
    \textit{pixel\_loss\_weight\_pre-train} & 0.034 & $\eta_1$, as specified in Equation \eqref{eq:g_loss_total}, used in pre-training\\
    \textit{G\_max\_norm} & $\infty$ & Maximum gradient due to gradient clipping\\
    \bottomrule
    \end{tabular}
\end{table*}

\end{document}